\documentclass[final,3p,times]{elsarticle}
\usepackage{lineno,hyperref}
\usepackage{amssymb}
\usepackage{subfigure}
\usepackage{amsmath}
\usepackage{amsthm}
\usepackage{mathrsfs}
\usepackage{slashbox}
\usepackage[ruled,vlined]{algorithm2e}
\include{pythonlisting}
\usepackage{setspace}

\DeclareMathOperator*{\argminA}{arg\,min}
\modulolinenumbers[200]
\journal{Journal}

\begin{document}

\begin{frontmatter}

\title{Adaptive activation functions accelerate convergence in deep and physics-informed neural networks}


\author{Ameya D. Jagtap$^a$, George Em Karniadakis$^{a,b,*}$}
\cortext[mycorrespondingauthor]{Corresponding author Emails:  ameya$\_$jagtap@brown.edu (A.D.Jagtap), george$\_$karniadakis@brown.edu (G.E.Karniadakis)}

\address{$^a$Division of Applied Mathematics, Brown University, 182 George Street, Providence, RI 02912, USA. \linebreak $^b$Pacific Northwest National Laboratory, Richland, WA 99354, USA.}

\begin{abstract}
 We employ adaptive activation functions for regression in deep and physics-informed neural networks (PINNs) to approximate smooth and discontinuous functions as well as solutions of linear and nonlinear partial differential equations. In particular, we solve the nonlinear Klein-Gordon equation, which has smooth solutions, the nonlinear Burgers equation, which can admit high gradient solutions, and the Helmholtz equation. We introduce a scalable hyper-parameter in the activation function, which can be optimized to achieve best performance of the network as it changes dynamically the topology of the loss function involved in the optimization process. The adaptive activation function has better learning capabilities than the traditional one (fixed activation) as it improves greatly the convergence rate, especially at early training, as well as the solution accuracy. To better understand the learning process, we plot the neural network solution in the frequency domain to examine how the network captures successively different frequency bands present in the solution. We consider both forward problems, where the approximate solutions are obtained, as well as inverse problems, where parameters involved in the governing equation are identified. Our simulation results show that the proposed method is a very simple and effective approach to increase the efficiency, robustness and accuracy of the neural network approximation of nonlinear functions as well as solutions of partial differential equations, especially for forward problems.
\end{abstract}


\end{frontmatter}

\linenumbers
\section{Introduction}


Neural networks (NNs) have found applications in the context of numerical solution of partial differential equations, integro-differential equations and dynamical systems. Since, a neural network is an universal approximator, thus it is natural to consider the neural network space as an ansatz space of the solution of governing equation. In \cite{RK, RK1, LLF}, NNs are successfully used to obtain the approximate solution of partial diferential equations (PDEs).
One can also construct the physics-informed learning machine, which makes use of systematically structured prior information about the solution. In the earlier study, Owhadi \cite{OW} showed the promising approach of exploiting such prior information to construct physics-informed learning machines for the numerical homogenization problem. Raissi et al \cite{RK2, RK3} employed Gaussian process regression to obtain representation of functionals of linear operators, which not only infer the solution accurately but also provide uncertainty estimates for many physical problems. Further, their method was extended to nonlinear problems by Raissi and Karniadakis \cite{RK} and Raissi et al \cite{RK0} in the context of solution inference and system identification. Data-driven turbulence modeling has been developed by Wang \textit{et al.} \cite{Wa} and Duraisamy \& co-workers in a series of papers \cite{ZD,PD,DZS}.
Physics-informed neural networks (PINNs) can accurately solve both forward problems, where the approximate solutions of governing equations are obtained, as well as inverse problems, where parameters involved in the governing equation are obtained from the training data \cite{RK1}.  
In the PINN algorithm, along with the contribution from the neural network the loss function is enriched by the addition of residual term from the governing equation(s), which act as a penalizing term that constrains the space of admissible solutions. 
\begin{figure} [htpb] 
\centering 
\includegraphics[trim=0cm 2.2cm 0cm 0cm, clip=true, scale=0.48, angle = 0]{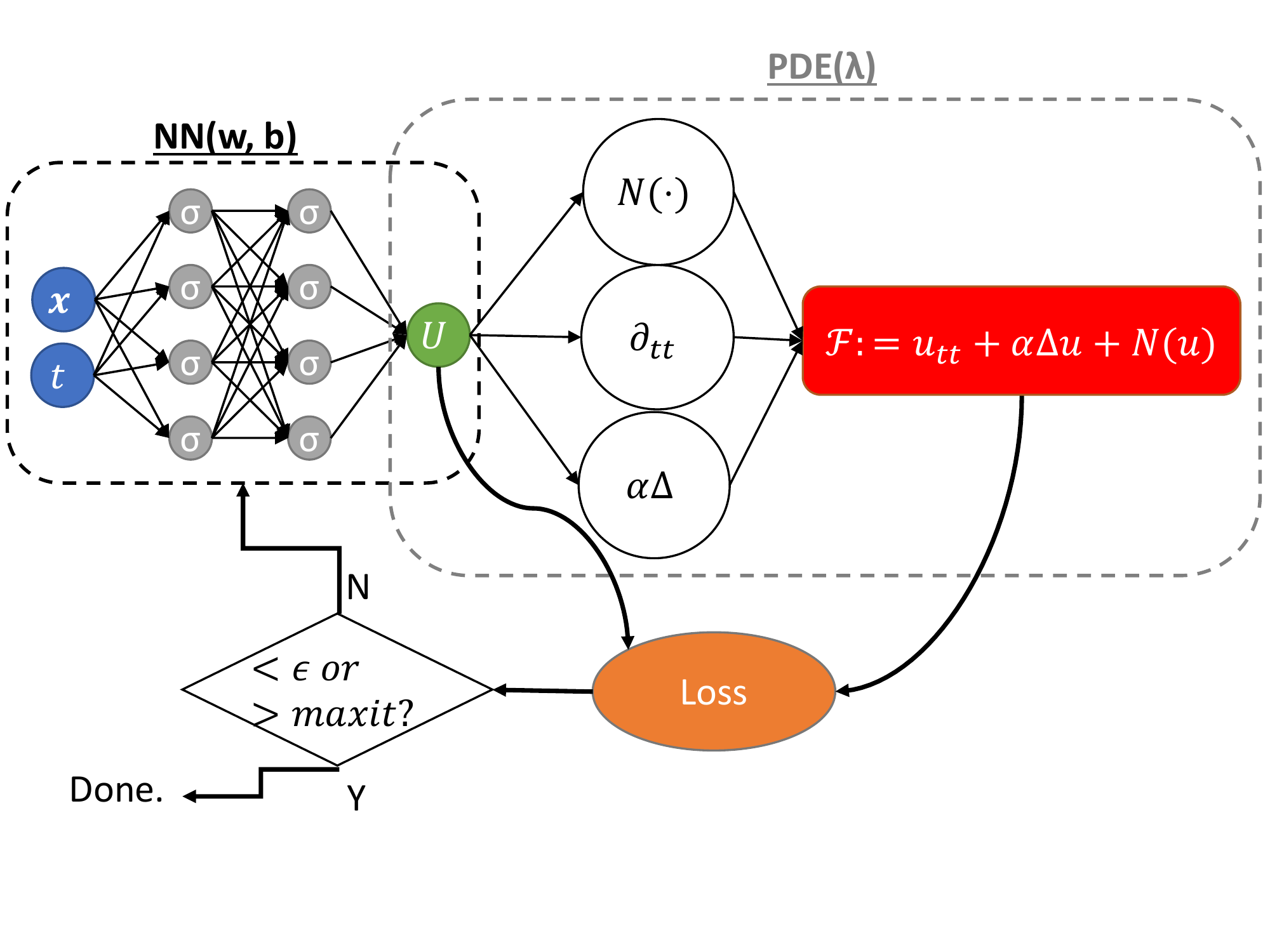}
\caption{Schematic of PINN for the Klein-Gordon equation. The left NN is the uninformed network while the right one induced by the governing equation
is the informed network. The two NNs share hyper-parameters and they both contribute to the loss function.}
\label{fig:PINN}
\end{figure}
Figure \ref{fig:PINN} gives a sketch of a PINN algorithm for the Klein-Gordon equation where one can see the neural network along with the supplementary physics-informed part. The loss function is evaluated using the contribution from the neural network part as well as the residual from the governing equation given by physics-informed part. Then, one seeks the optimal values of weights $(w)$ and biases $(b)$ in order to minimize the loss function below certain tolerance $\epsilon$ or until a prescribed maximum number of iterations.

The activation function plays an important role in such training process due to the dependence of the derivative of loss function on optimization parameters, which, in turn, depends on the derivative of the activation function. In the PINN algorithm various activation functions such as \textit{tanh}, \textit{sin} \textit{etc} are used to solve various problems, see \cite{RK1, VIV} for more details. There is no obvious choices for the activation function since it solely depends on the problem at hand.
To tackle this issue, various methods are proposed in the literature like adaptive sigmoidal activation function for multilayer feedforward NNs proposed by Yu et al \cite{Yu}, while Qian et al \cite{Qian} focuses on learning activation functions in convolutional NNs via combining basic activation functions in a data-driven way. Dushkoff and Ptucha \cite{MiRa} proposed multiple activation functions per neuron, where individual neurons select between a multitude of activation functions. A tunable activation function is proposed by Li \textit{et al.} \cite{LLR}, where only a single hidden layer is used and the activation function is tuned. In \cite{SWCC}, Shen \textit{et al.} used a similar idea of tunable activation function but with multiple outputs.

In this paper the activation function is tuned for any number of hidden layers by introducing an adaptable hyper-parameter with a scaling factor. Along with the deep neural network problem, PINN-based forward and inverse problems involving smooth solutions (like, nonlinear Klein-Gordon equation and Helmholtz equation) as well as steep-gradient solution (Burgers equation) are solved using the proposed method and compared with the fixed activation function. One can clearly see the advantages like increase in the accuracy and fast convergence rate,  using the proposed adaptive activation function, especially in the early training period.

This paper is organized as follows: After the introduction in section 1, section 2 gives a brief discussion of the proposed methodology, where we also discuss about training data, loss function, optimization methods and the proposed method. Section 3 gives the results and detailed discussions on deep neural network approximations of smooth and discontinuous functions as well as PINN based solutions of forward/inverse problems, including the Burgers, Klein-Gordon and Helmholtz equations. Finally, in section 4, we summarize the conclusions of our work.

\section{Methodology}

We consider a NN of depth $D$ corresponding to a network with an input layer, $D-1$ hidden layers and an output layer. In the $k^{th}$ hidden layer, $N_k$ number of neurons are present. Each hidden layer of the network receives an output $x^{k-1} \in\mathbb{R}^{N_{k-1}}$ from the previous layer where an affine transformation of the form 
\begin{equation}\label{afft}
\mathcal{L}_k (x^{k-1}) \coloneqq w^k x^{k-1} + b^k,
\end{equation}
is performed. The network weights $w^k$ and bias term $b^k$ associated with the $k^{th}$ layer are chosen from  \textit{independent and identically distributed (iid)} samplings. The nonlinear activation function $\sigma(\cdot)$ is applied to each component of the transformed vector before sending it as an input to the next layer. The activation function is an identity function after an output layer. Thus, the final neural network representation is given by the composition 
$$ u_{\boldsymbol{\Theta}}(x) = (\mathcal{L}_k \circ \sigma \circ \mathcal{L}_{k-1} \circ \ldots \circ \sigma \circ \mathcal{L}_1 )(x),$$
where the operator $\circ$ is the composition operator, $\boldsymbol{\Theta} = \{w^k, b^k\}_{k=1}^D$ represents the trainable parameters in the network,  $u$ is the output and $x^0 = x$  is the input.


\subsection{Training data}
In the supervised learning, training data is important to train the neural network, which can be obtained from the exact solution (if available) or from high-resolution numerical solution using methods like spectral method, discontinuous Galerkin method \textit{etc}, as per the problem at hand. Here we select the training points either randomly from the uniform/normal distribution or one can also choose these points depending upon the small length and time scales (high gradient regions) present in the solution space. Training data can also be obtained from carefully performed experiments, which may yield both high- and low-fidelity data sets.


\subsection{Loss function and optimization algorithm}
We aim to find the optimal weights for which the suitably defined loss function is minimized. In PINN the loss function is defined as
\begin{equation}\label{loss}
 J(\boldsymbol{\Theta}) =  MSE_{\mathcal{F}} + MSE_u 
\end{equation}
where the mean squared error (MSE) is given by
\begin{align*}
MSE_{\mathcal{F}} &=  \frac{1}{N_f}\sum_{i=1}^{N_f} |\mathcal{F}(x^i_f,y^i_f,t^i_f)|^2, ~~~~~~  MSE_u = \frac{1}{N_u}\sum_{i=1}^{N_u} |u^i - u(x^i_u,y^i_u,t^i_u)|^2.
\end{align*}
 Here $\{x_f^i,y_f^i,t^i_f\}_{i=1}^{N_f}$ denotes the residual training points in the space-time domain and $\{x_u^i,y_u^i, t^i_u\}_{i=1}^{N_u}$ denotes the boundary/initial training data. The aim of including the first term is that the neural network solution must satisfy the governing equation at randomly chosen points in the domain, which constitutes the physics-informed part of neural network whereas the second term includes the known boundary/initial conditions, which must be satisfied by the neural network solution.
The resulting optimization problem leads to finding the minimum of a loss function by optimizing the parameters, \textit{i.e.}, we seek to find
\begin{align*}
  w^* &= \argminA_{w \in  \boldsymbol{\Theta}} ~(J(w)); \ \ \ \ \ \   b^* = \argminA_{b \in  \boldsymbol{\Theta}} ~(J(b)).
\end{align*}

One can approximate the solutions to this minimization problem iteratively by one of the forms of gradient descent algorithm. The stochastic gradient descent (SGD) algorithm is widely used in machine learning community, see \cite{Rud} for a survey. In SGD the weights are updated as
$$w^{m+1} = w^m - \eta_l \nabla_{w} J^m(w)$$
where $\eta_l>0$ is the learning rate and $J^m$ is the loss function at $m^{th}$ iteration. SGD methods can be initialized with some starting value $w^0$. In this work, the ADAM optimizer \cite{ADAM}, which is a variant of the SGD method is used in all problems.

\subsection{Adaptive activation functions}
In the literature, various activation functions are available such as sigmoid or logistic, tanh. ReLU, Leaky-ReLU \textit{etc} \cite{Hay}. 
The role of activation function is to decide whether particular neuron should fire or not. When the activation function is absent the weights and bias would simply do a linear transformation, which is a case of a linear regression model. Such linear model is simple to solve but is limited in its capacity to solve complex problems. The nonlinear activation function performs the nonlinear transformation to the input data making it capable to learn and perform more complex tasks. In the back-propagation algorithm, the evaluation of gradient of the loss function involves the gradient of the activation function. Activation functions make the back-propagation possible since the gradients are supplied to update the weights and biases; hence, without the differentiable nonlinear activation function, this would not be possible. Thus, one must choose an activation function which is less prone to the \textit{vanishing} and the \textit{exploding} gradient problem \cite{Ben}. In this work, we have tried different activation functions, and their performance is compared in the results and discussion section.

\begin{figure} [htpb] 
\centering
\includegraphics[trim=3.8cm 0.2cm 3.5cm 0.2cm, clip=true, scale=0.5, angle = 0]{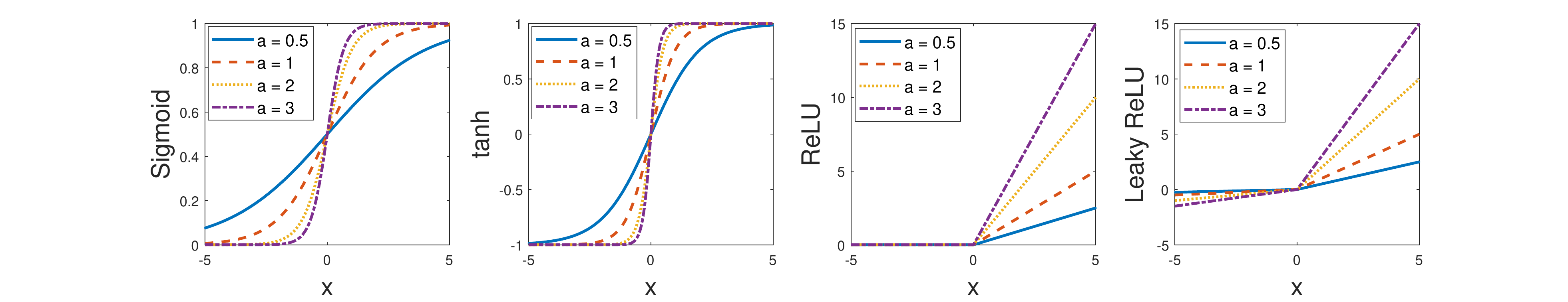}
\caption{(Left to right) Sigmoid or logistic, tanh, ReLU and Leaky-ReLU activation functions for various values of $a$.}
\label{fig:Af}
\end{figure}

The size of a network is correlated to its capacity to reproduce complicated functions. A deep network is required to solve complex problems, which on the other hand is difficult to train. In most cases, a suitable architecture is selected based on the researcher's experience, which is a trial and error based method. One can think of tuning the network to get the best performance out of it. To achieve this, we introduce the hyper-parameter $a$ in the activation function as $$\sigma(a~\mathcal{L}_k (x^{k-1})),$$
which needs to be optimized. The resulting optimization problem leads to finding the minimum of a loss function by optimizing $a$ along with the weights and biases, \textit{i.e.}, we seek to find
$$ a^* = \argminA_{a \in  \mathbb{R}^+\text{\textbackslash}\{0\}} ~(J(a)).$$
The parameter $a$ is  updated as
$$a^{m+1} = a^m - \eta_l \nabla_{a} J^m(a).$$
Mathematically, such hyper-parameter can change the slope of the activation function, which is one of the important aspects of the neural network training. Figure ~\ref{fig:Af} shows the sigmoid, tanh, ReLU and Leaky-ReLU activation functions with different values of hyper-parameter $a$, where we can see the changes in slope of the activation function with $a$. The corresponding expressions of these activation function are given by
\begin{align*}
 \text{Sigmoid}: &~~\frac{1}{1+e^{-ax}},
 \\ \text{Hyperbolic tangent}: & ~~ \frac{e^{ax}-e^{-ax}}{e^{ax}+e^{-ax}},
\\ \text{ReLU}: & ~~ \text{max}(0,ax),
\\ \text{Leaky ReLU}: & ~~ \text{max}(0,ax)-\nu~{max}(0,-ax).
\end{align*} 

The learning rate has a great impact while searching for global  minima. Large learning rate can over-shoot the global minima whereas small learning rate can increase the computational cost although it slowly moves towards global minima.
The common practice is to use small learning rate for such optimization problem, which gives slow variation or say slow convergence towards the optimized value. Intuitively, one can think of some scale factor $n \geq 1$ multiplied by $a$, which accelerates convergence towards global minima. Thus, the final form of the activation function is given by $$ \sigma(na~\mathcal{L}_k (x^{k-1})).$$
It is important to note that the introduction of the scalable hyper-parameter does not change the structure of loss function defined previously. 
Then, the final adaptive activation function based neural network representation is given by
$$ u_{\tilde{\boldsymbol{\Theta}}}(x) = (\mathcal{L}_k \circ \sigma \circ na\mathcal{L}_{k-1} \circ \sigma \circ na\mathcal{L}_{k-2} \circ \ldots \circ \sigma \circ na\mathcal{L}_1 )(x).$$
In this case, the trainable parameters are $\tilde{\boldsymbol{\Theta}} = \{w^k, b^k, a\}_{k=1}^D$. Compared to the original neural network, the adaptive activation function based PINN has one additional scalable hyper-parameter $a$ to train. The adaptive activation function based PINN algorithm is summarized as follows.

\vspace{5mm}

\begin{algorithm}[H]
\SetAlgoLined
 \textbf{Step 1} : Specification of training set\\
Training data : $u_{NN}$ network $\{x_u^i,y_u^i, t^i_u\}_{i=1}^{N_u}.$  \\
Residual training points :  $\mathcal{F}$ network $\{x_f^i,y_f^i,t^i_f\}_{i=1}^{N_f}.$ \\
 \textbf{Step 2} : Construct neural network $u_{NN}(\tilde{\boldsymbol{\Theta}})$ with random initialization of parameters $\tilde{\boldsymbol{\Theta}}$. \\
  \textbf{Step 3} : Construct the residual neural network $\mathcal{F}$ by
  substituting surrogate $u_{NN}$ into the governing equations using 
automatic differentiation \cite{AD}  and other arithmetic operations.\\
\textbf{Step 4}: Specification of loss function:\\
$$J(\tilde{\boldsymbol{\Theta}}) = \frac{1}{N_f}\sum_{i=1}^{N_f} |\mathcal{F}(x^i_f,y^i_f,t^i_f)|^2+ \frac{1}{N_u}\sum_{i=1}^{N_u} |u^i - u(x^i_u,y^i_u,t^i_u)|^2.$$\\
\textbf{Step 5}: Find the best parameters using suitable optimization method for minimizing the loss function 
$$\tilde{\boldsymbol{\Theta}}^* =  \text{arg min}~(\tilde{\boldsymbol{\Theta}}).$$
 \caption{Adaptive activation function based PINN algorithm}
\end{algorithm}

\section{Results and discussion}
In this section, first we shall approximate nonlinear smooth and discontinuous functions using neural networks with fixed and adaptive activations. Subsequently, we will solve Burgers, Klein-Gordon and Helmholtz equations, which can admit both continuous as well as high gradient solutions using PINNs with fixed and adaptive activations. Both forward problems, where the solution is inferred, as well as inverse problems, where the parameters involved in the governing equation are obtained using the proposed adaptive activation function based PINN. To show the effectiveness of the proposed method, various comparisons are made. 
In this study, the optimal value of learning rate is found by experiments to be 0.0008.

\subsection{Neural network approximation of nonlinear smooth and discontinuous functions}
In this test case we are using a standard NN (without physics-informed part) to approximate smooth and discontinuous functions. In this case the loss function consists of just a neural network part given by
$$J(\tilde{\boldsymbol{\Theta}}) = \frac{1}{N_u}\sum_{i=1}^{N_u} |u^i - u(x^i_u,y^i_u,t^i_u)|^2.$$
The activation function is \textit{tanh} and the number of hidden layers is four with 50 neurons in each layer.

\begin{figure}  [htpb]
\centering
\includegraphics[scale=0.65]{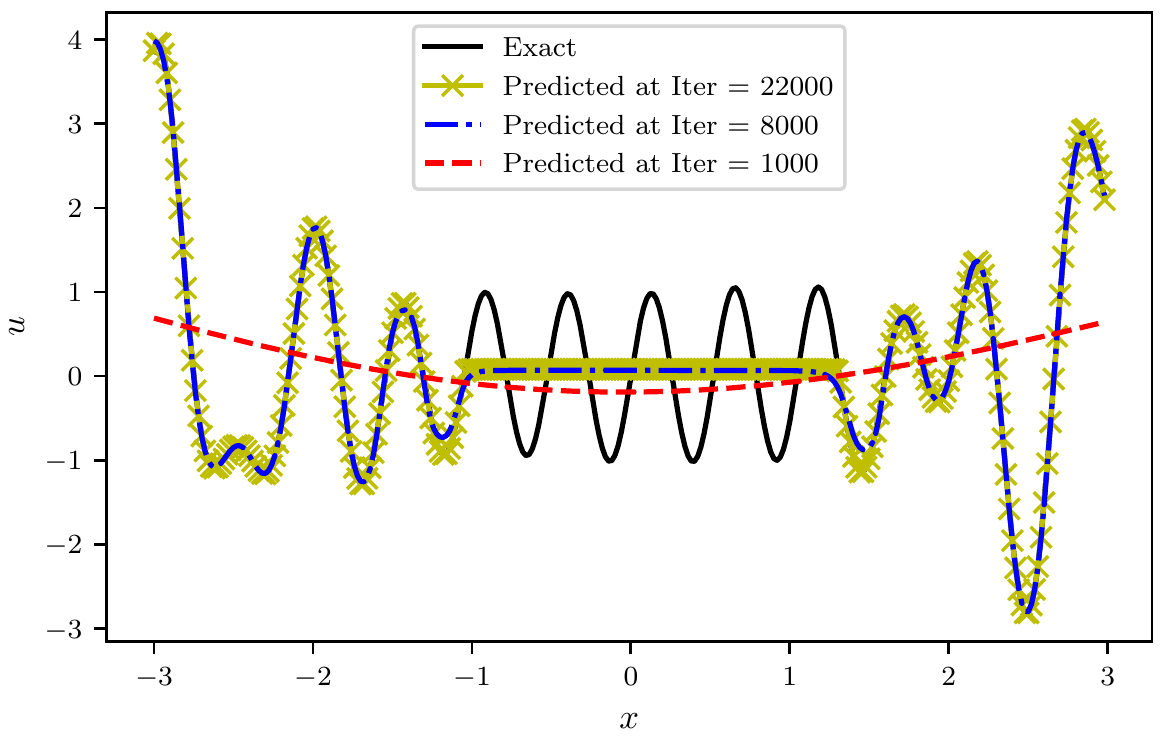}
\includegraphics[scale=0.6, angle = 0]{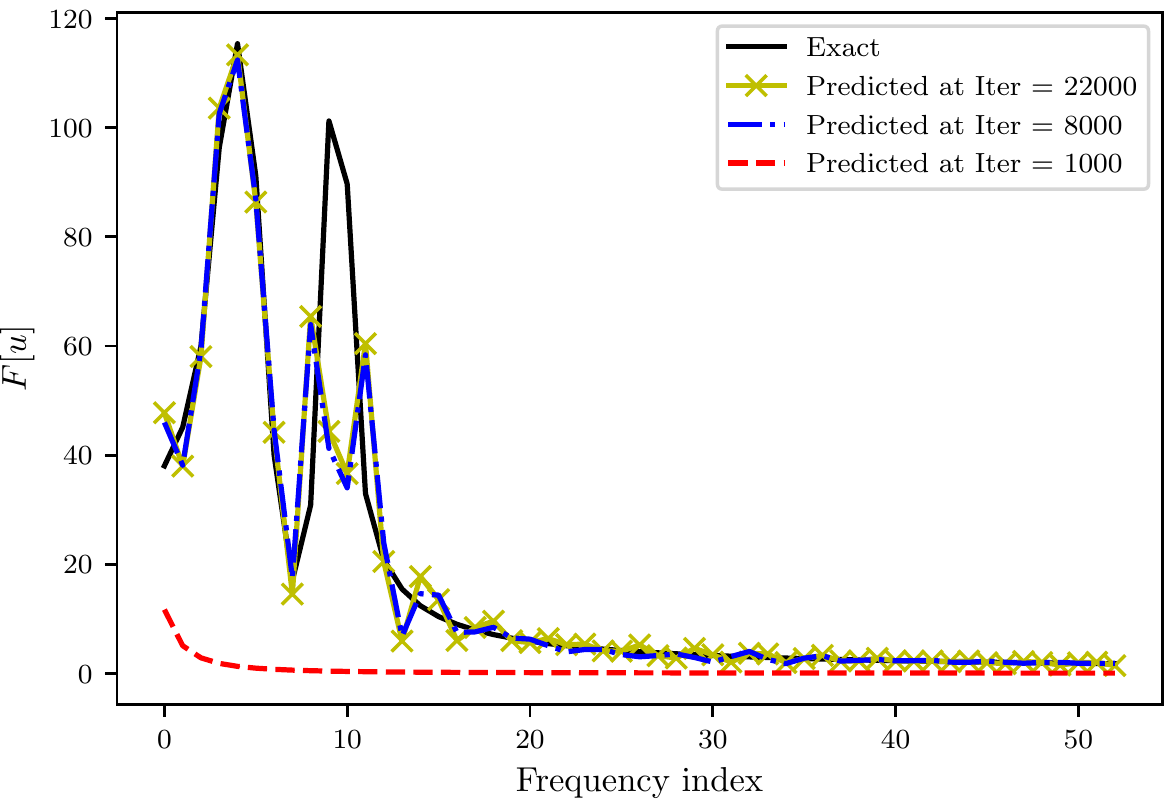}
\includegraphics[scale=0.65, angle = 0]{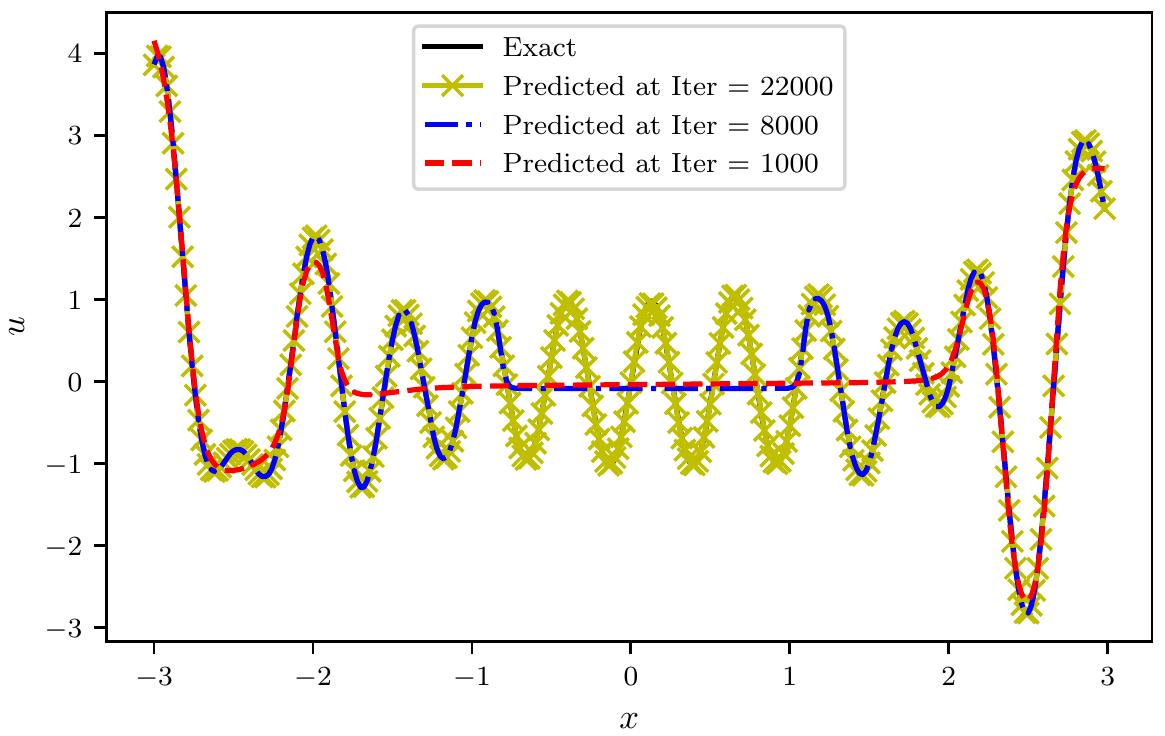}
\includegraphics[scale=0.6, angle = 0]{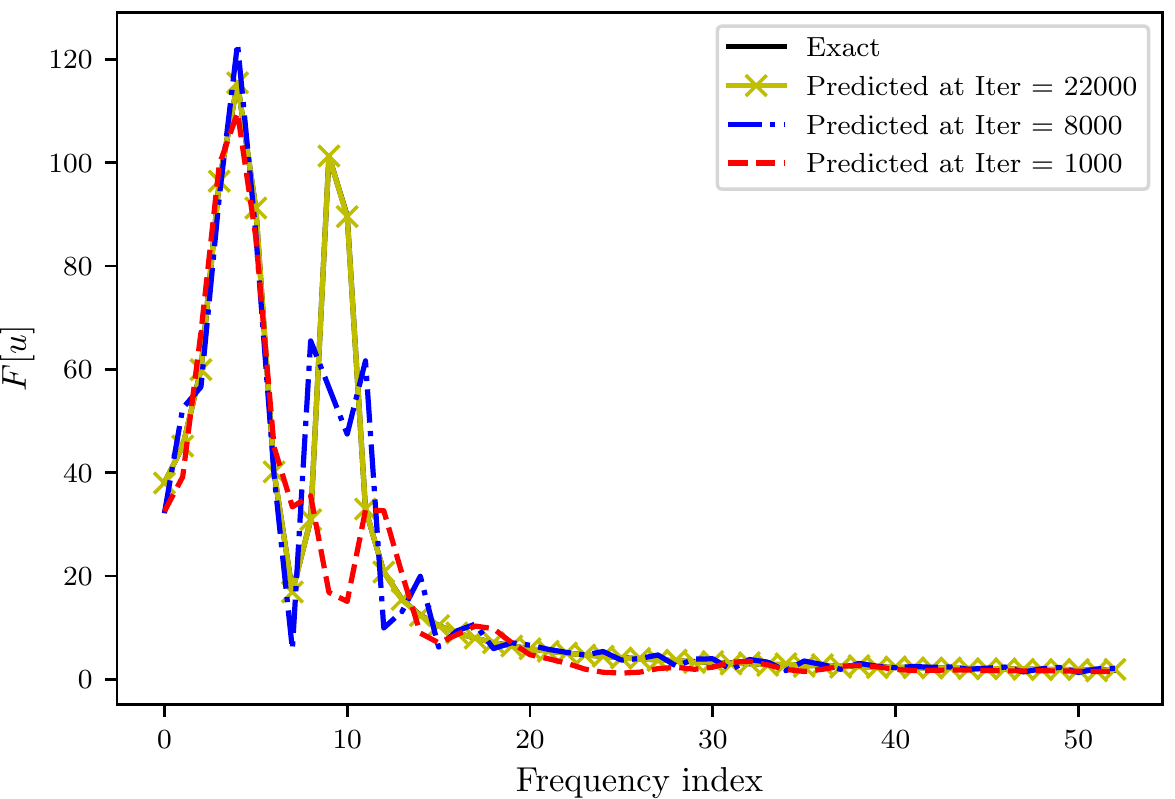}
\includegraphics[scale=0.65, angle = 0]{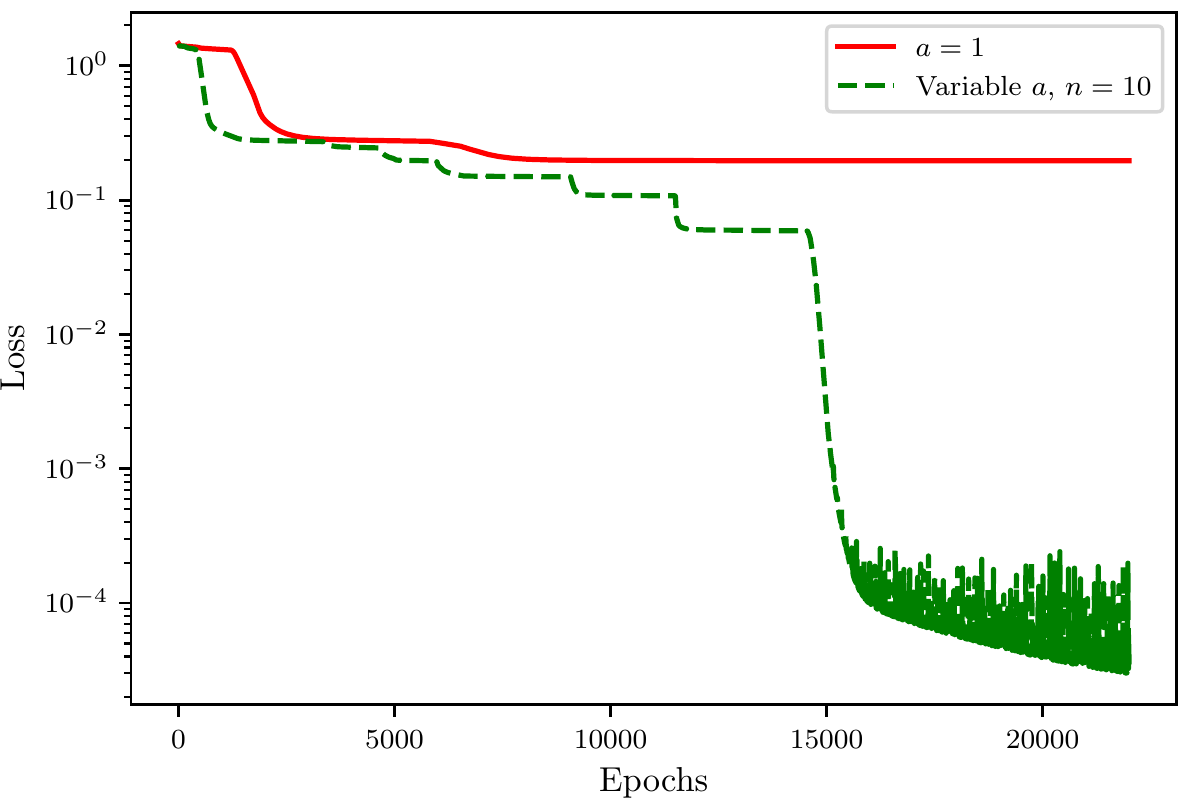}
\includegraphics[scale=0.62, angle = 0]{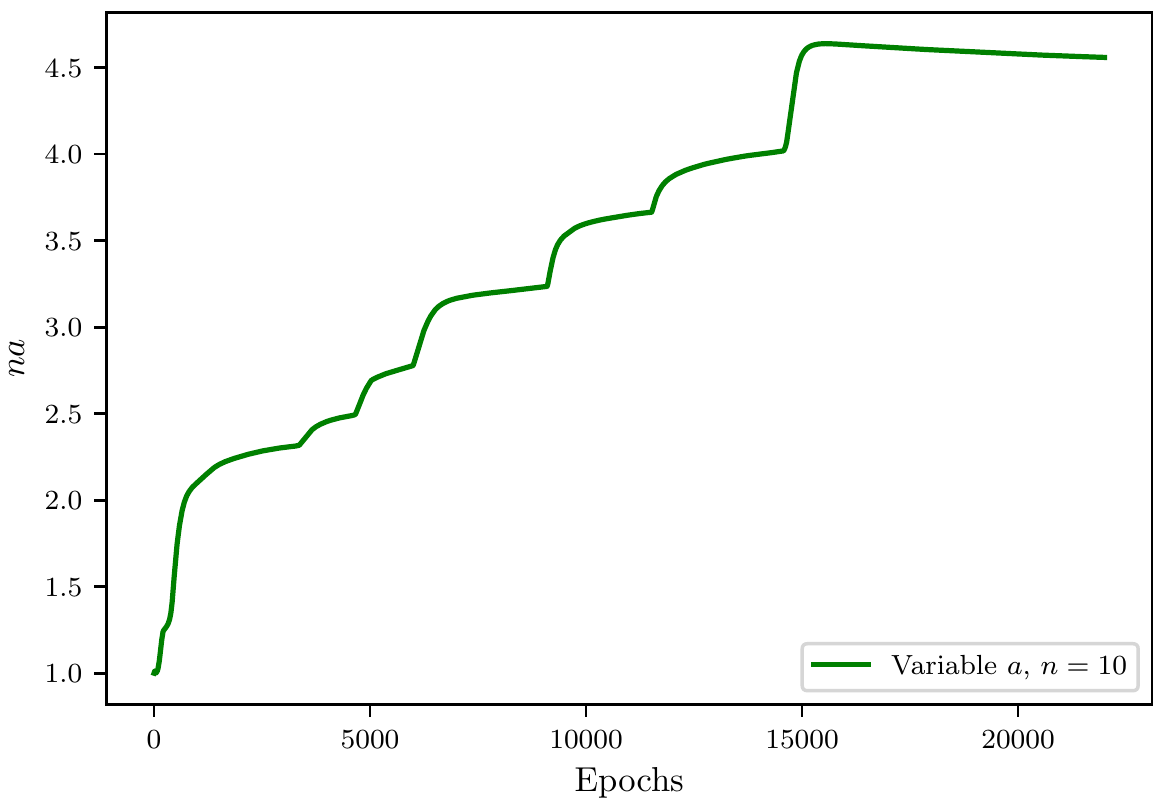}
\caption{Smooth function: Neural network solution of fixed (top row) and adaptive activation (middle row) functions. First column (top and middle row) shows the solution, which is also plotted in frequency domain (zoomed-view) as shown by the corresponding second column. Bottom row shows the loss function comparison for fixed and adaptive activation (left) and variation in $a$ with $n = 10$ (right).}
\label{fig:NNsmt}
\end{figure}

\begin{figure}  [htpb]
\centering
\includegraphics[scale=0.65]{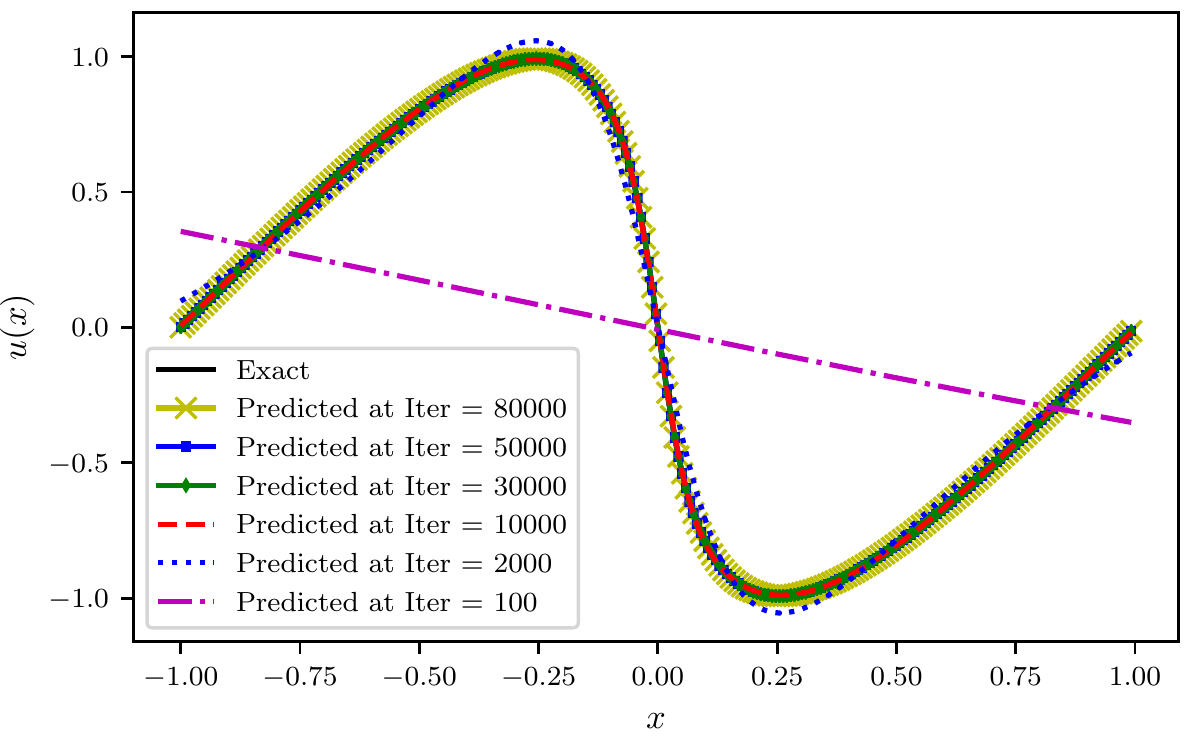}
\includegraphics[scale=0.6, angle = 0]{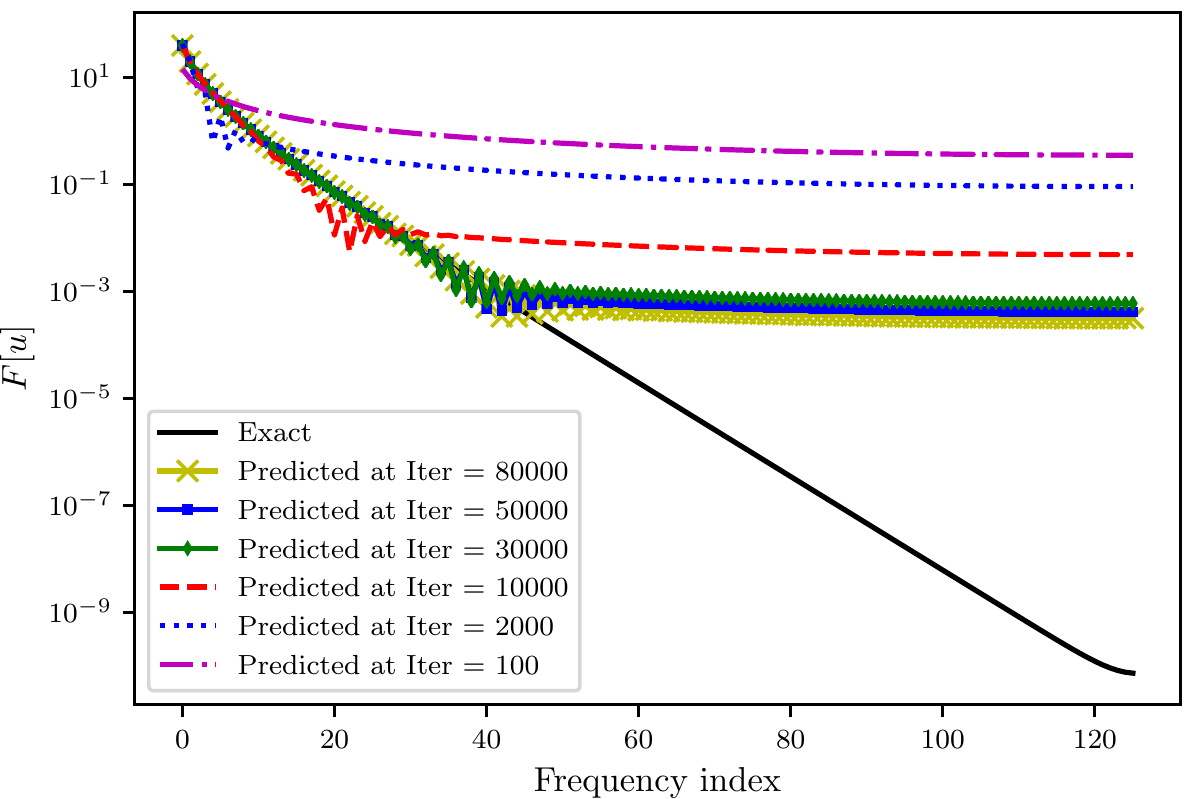}
\includegraphics[scale=0.65, angle = 0]{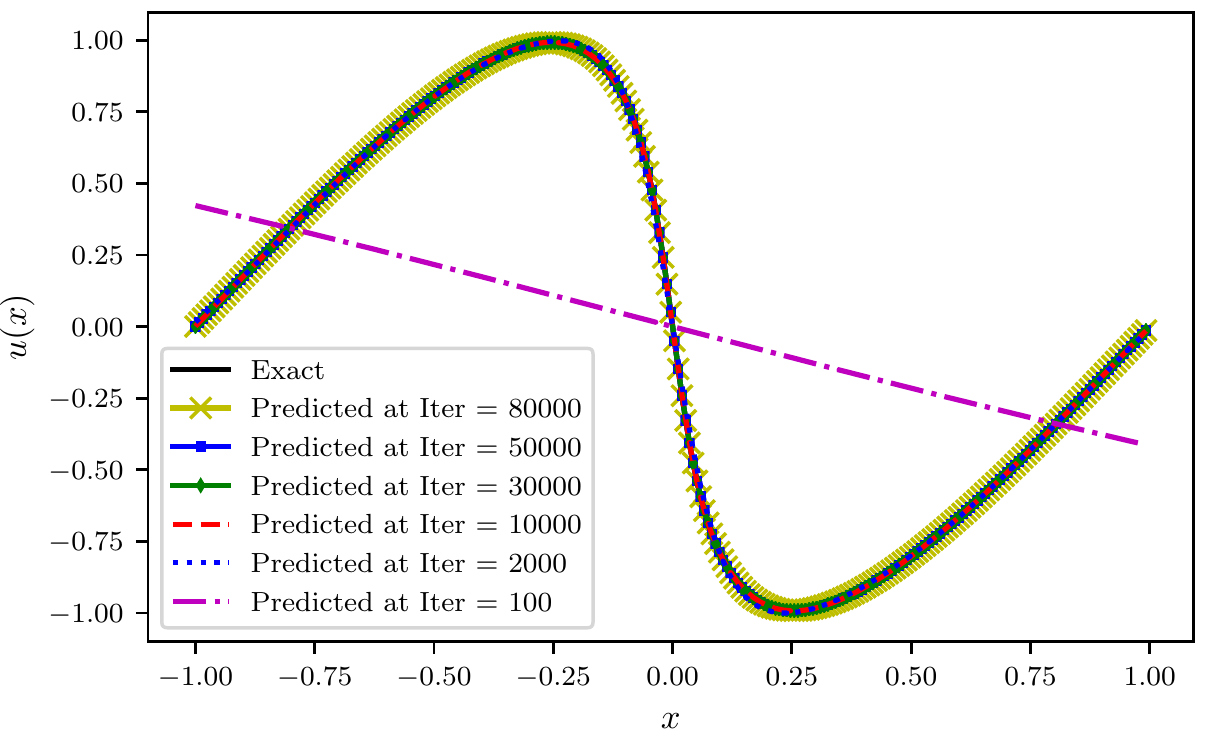}
\includegraphics[scale=0.6, angle = 0]{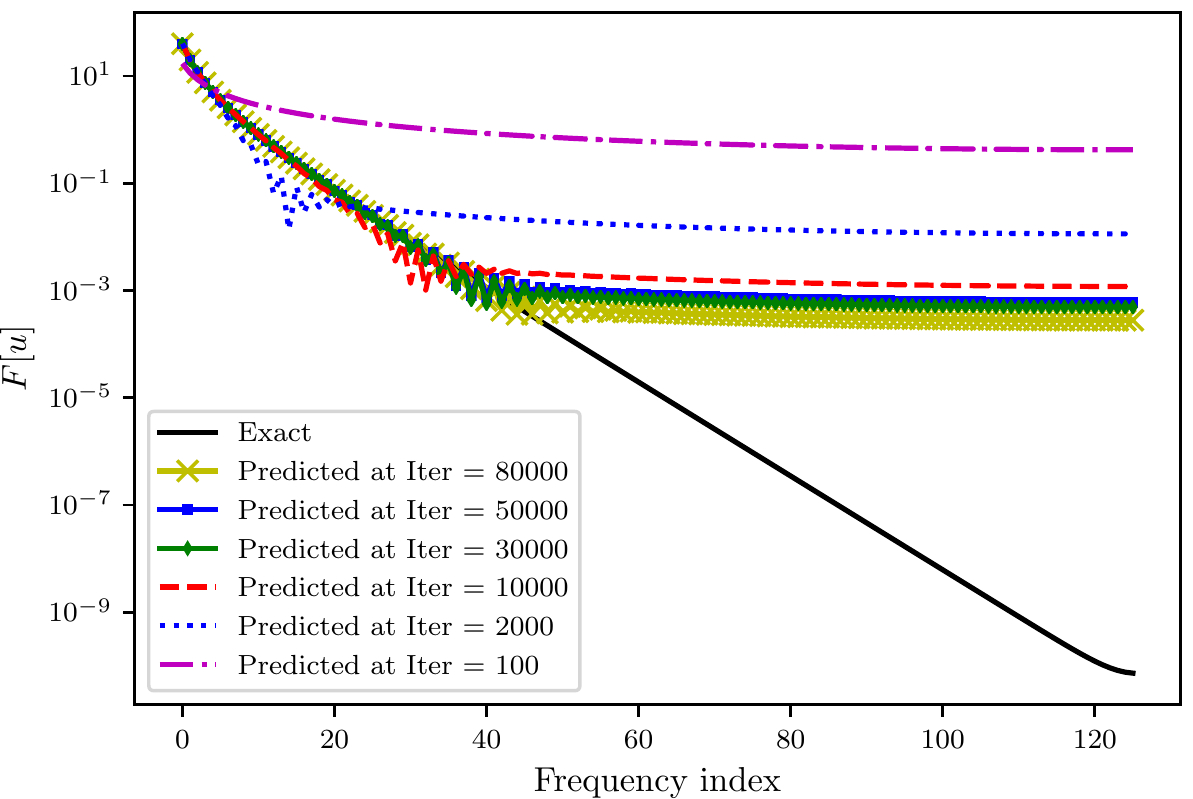}
\includegraphics[scale=0.65, angle = 0]{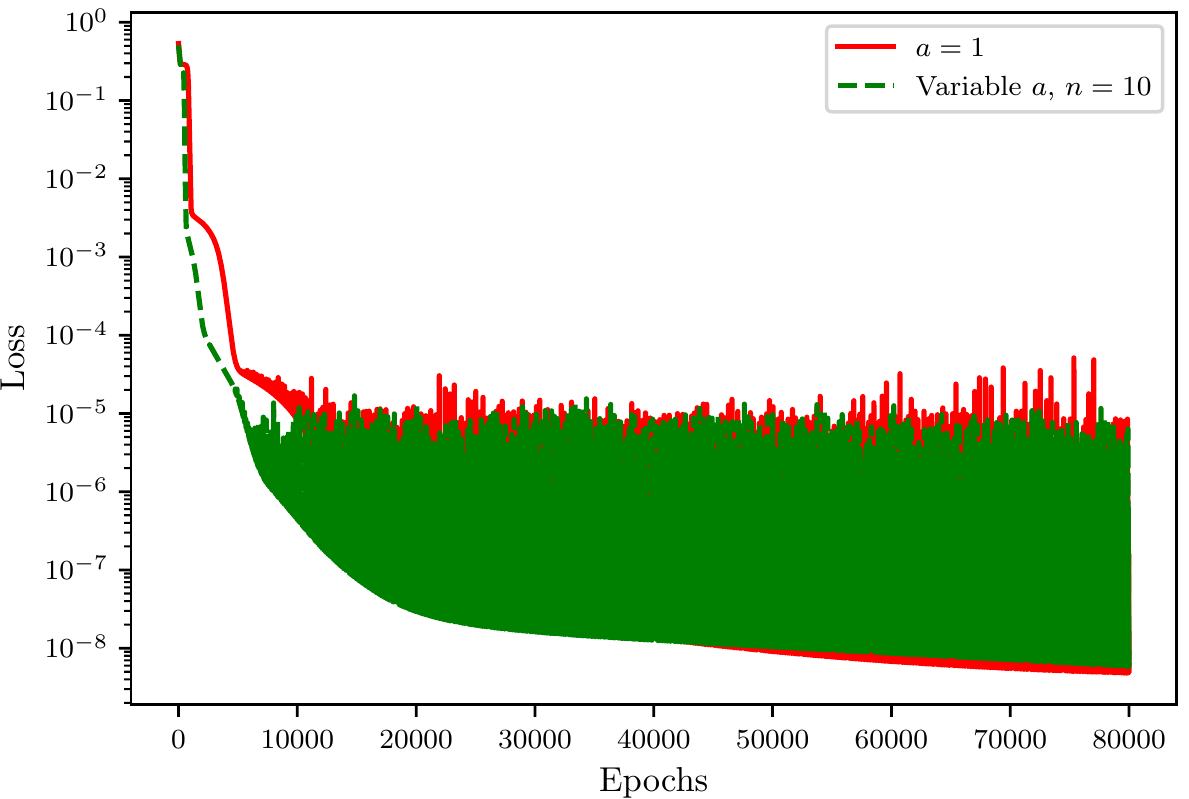}
\includegraphics[scale=0.62, angle = 0]{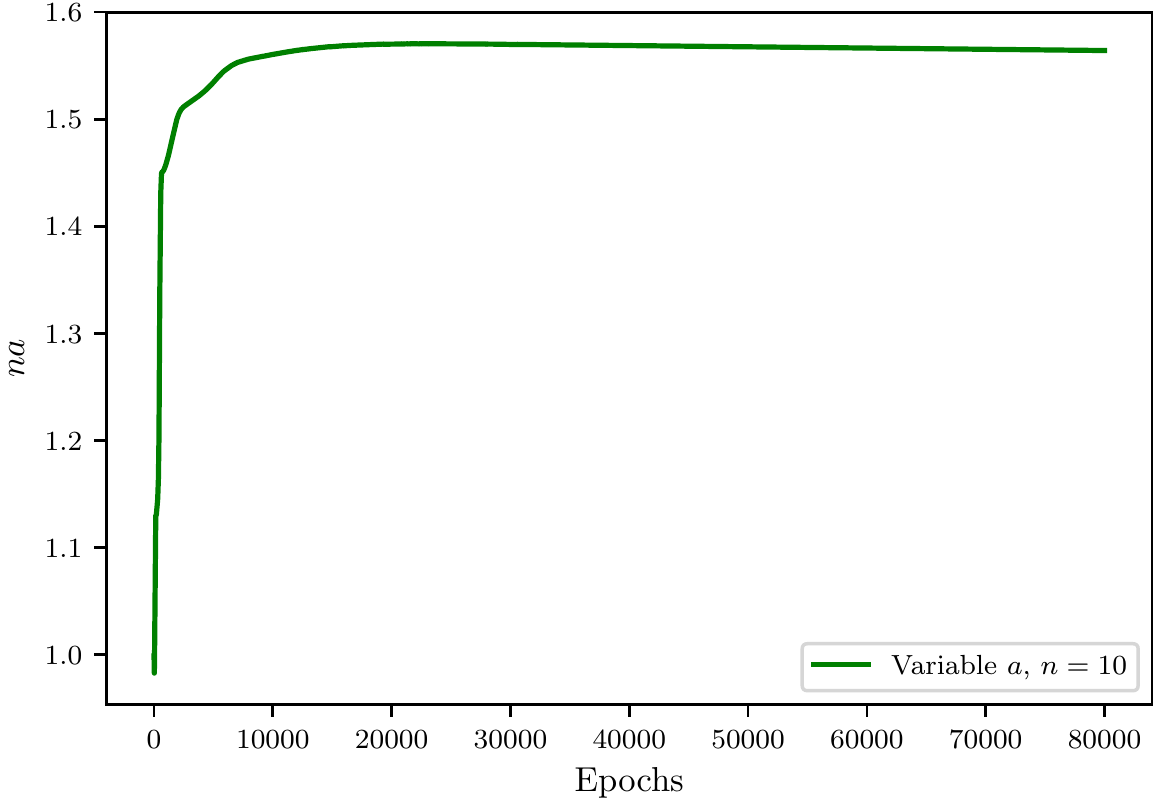}
\caption{Smooth Burgers solution: Neural network solution of fixed (top row) and adaptive activation (middle row) functions. First column (top and middle row) shows the solution which is also plotted in frequency domain as shown by the corresponding second column. Bottom row shows the loss function comparison for fixed and adaptive activation (left) and variation in $a$ with $n = 10$ (right).}
\label{fig:NNsmt2}
\end{figure}
Here, we shall consider two smooth functions and one discontinuous function. First, the smooth function is given by
\begin{equation}
 u(x) = (x^3-x)~\frac{\sin(7x)}{7} + \sin(12x),~~ x \in [-3,~3].
\end{equation}
In this function both high and low frequency components are present. The number of training points $N_u$ used is 300.
Next, we consider the Burgers equation given by
\begin{equation}\label{Bur}
u_{t} + u u_x  = \tilde{\epsilon} u_{xx}, \ \ x \in [-1,~1],\,\, t>0
\end{equation}
with initial condition $ u(x,0) = -\sin(\pi x)$, boundary conditions $u(-1,t) = u(1,t) = 0$ and $\tilde{\epsilon} = 0.01/\pi$. The analytical solution can be obtained using the Hopf-Cole transformation, see Basdevant et al., \cite{Bes} for more details. We consider the smooth solution of Burgers equation at time $t = 0.25$ to be approximated by a deep neural network. The number of training points used is 256.

Finally, the following discontinuous function with discontinuity at $x = 0$ location is approximated by a deep neural network.
\begin{equation}
 u(x) = \begin{cases}
         0.2~\sin(6x)  & \text{If}~ x\leq 0 \\ 1+ 0.1~x\cos(12x)  & \text{Otherwise.} 
        \end{cases}
\end{equation}
Here, the domain is $[-4,~3.75]$. In this case, the number of training points used is 300.
\begin{figure}  [htpb]
\centering
\includegraphics[scale=0.65]{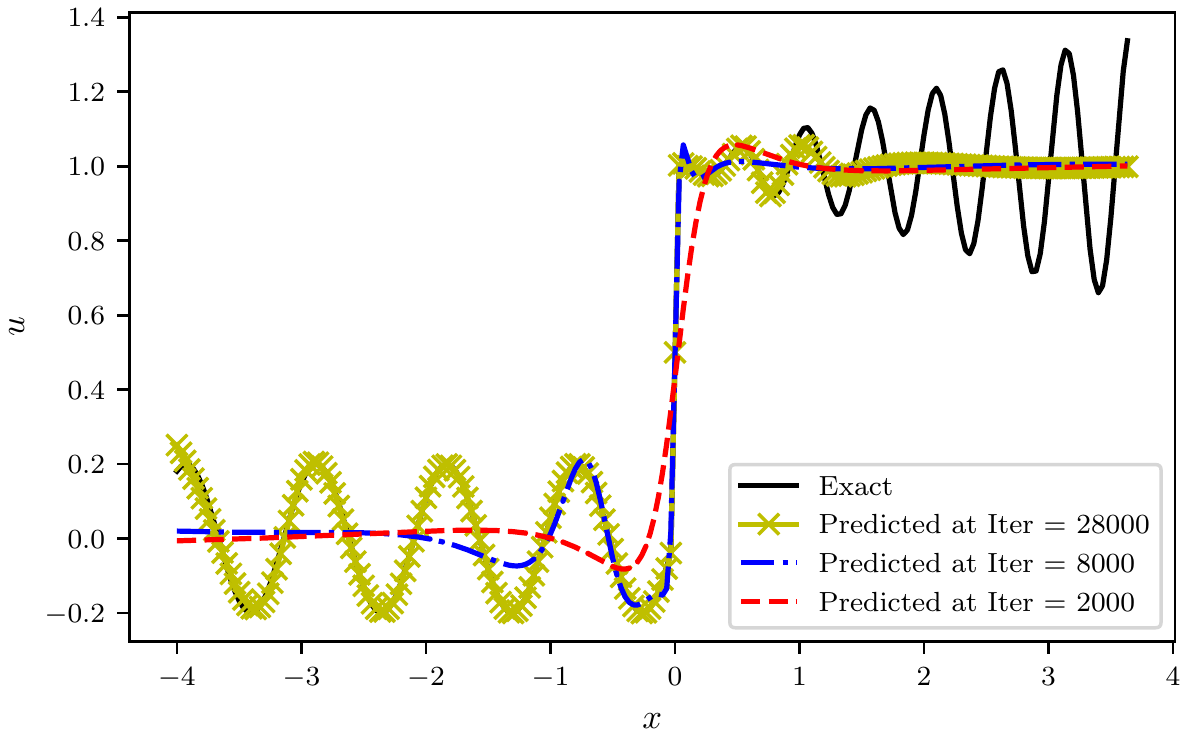}
\includegraphics[scale=0.6, angle = 0]{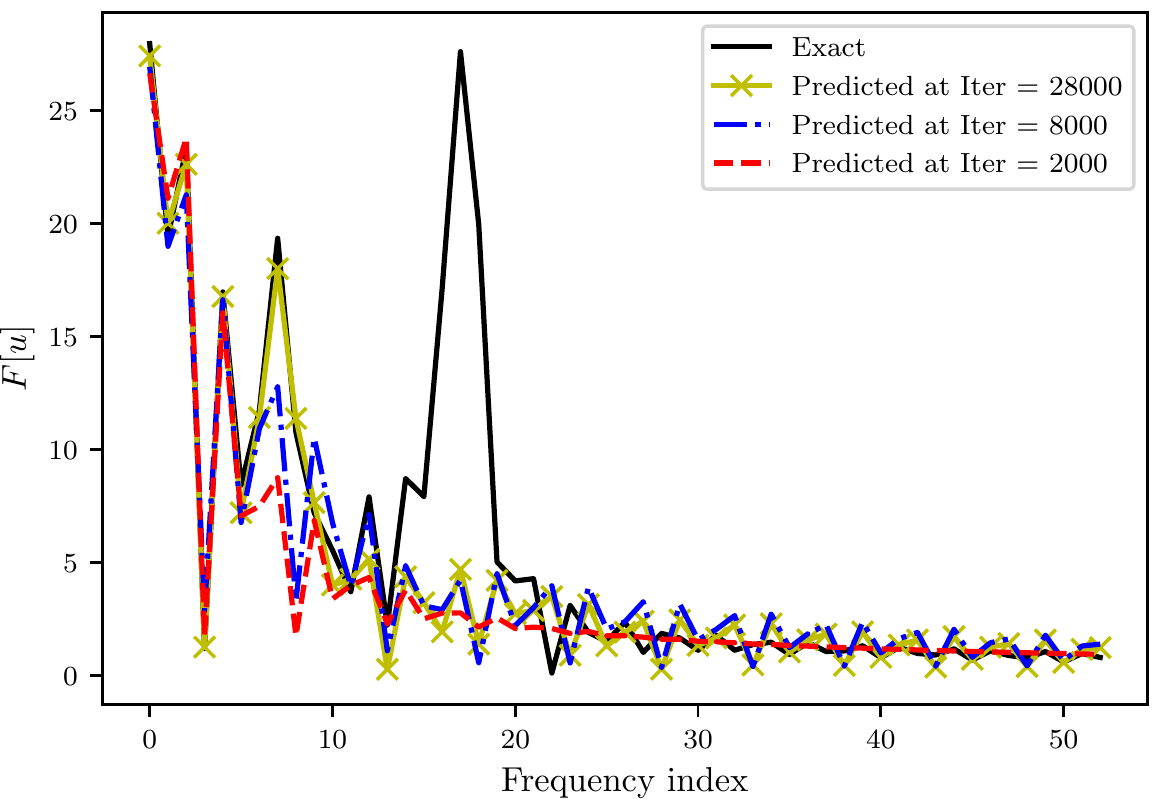}
\includegraphics[scale=0.65, angle = 0]{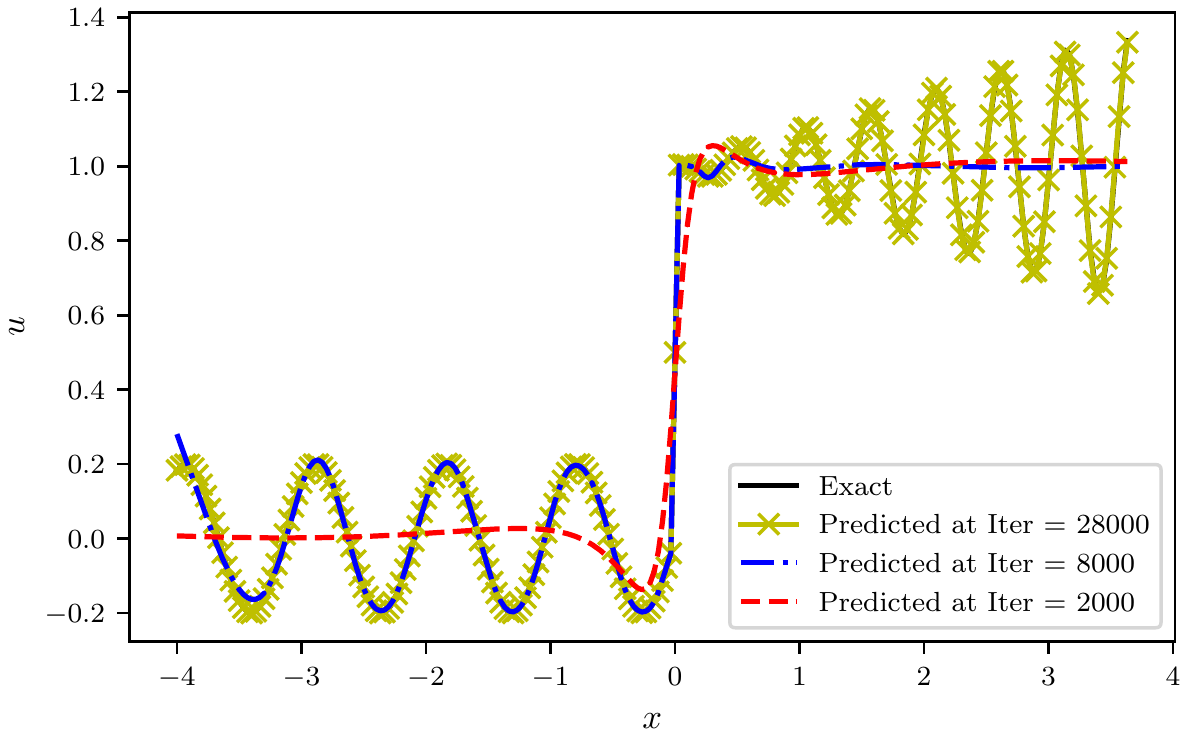}
\includegraphics[scale=0.6, angle = 0]{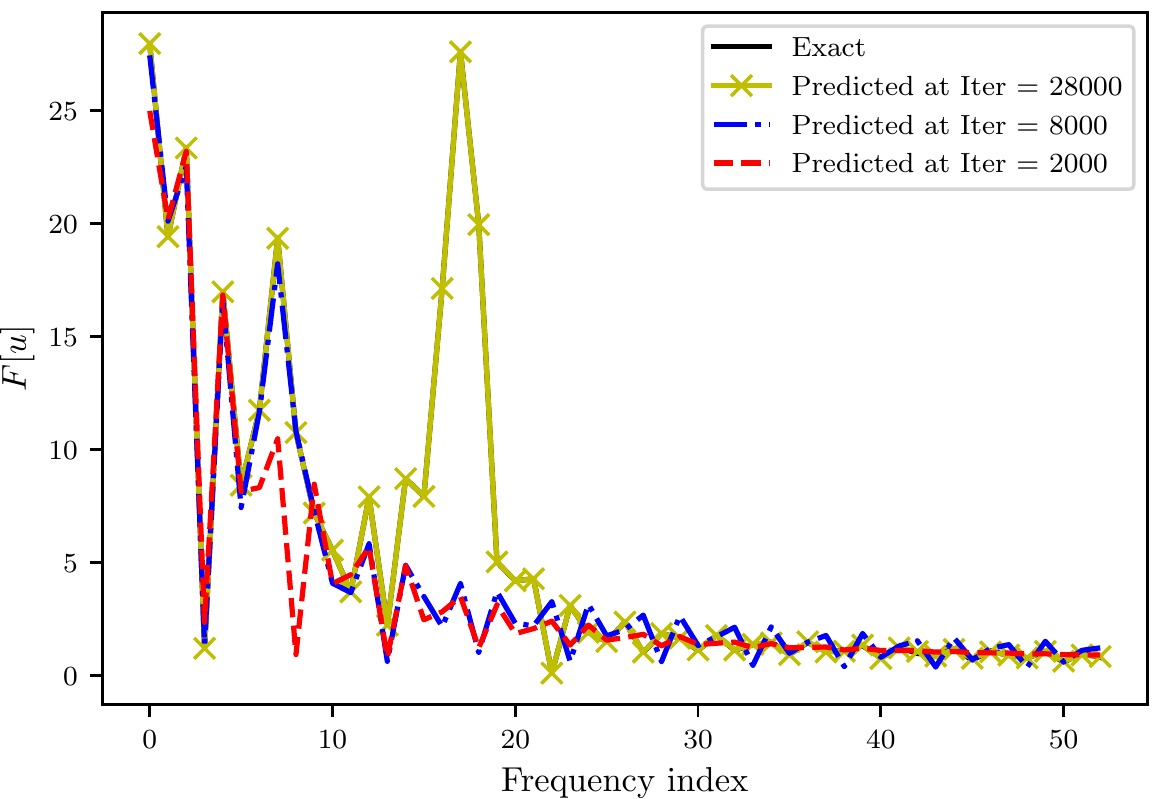}
\includegraphics[scale=0.65, angle = 0]{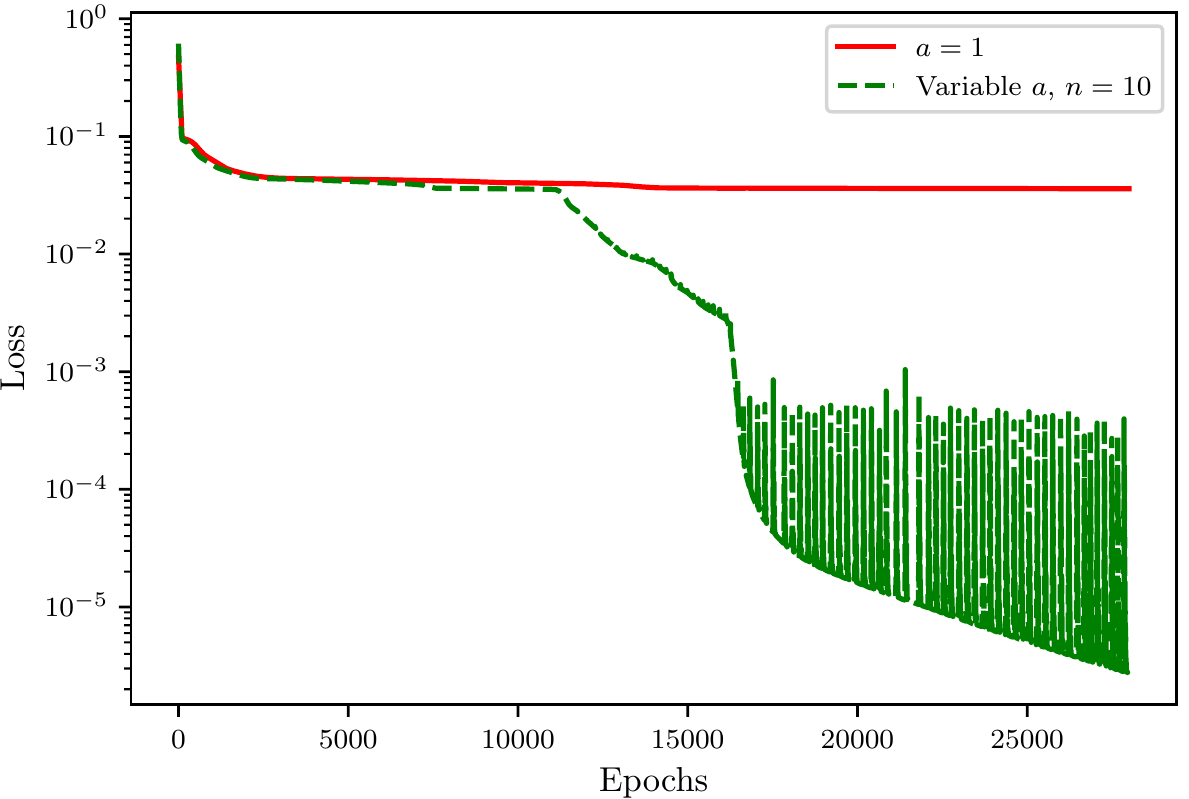}
\includegraphics[scale=0.62, angle = 0]{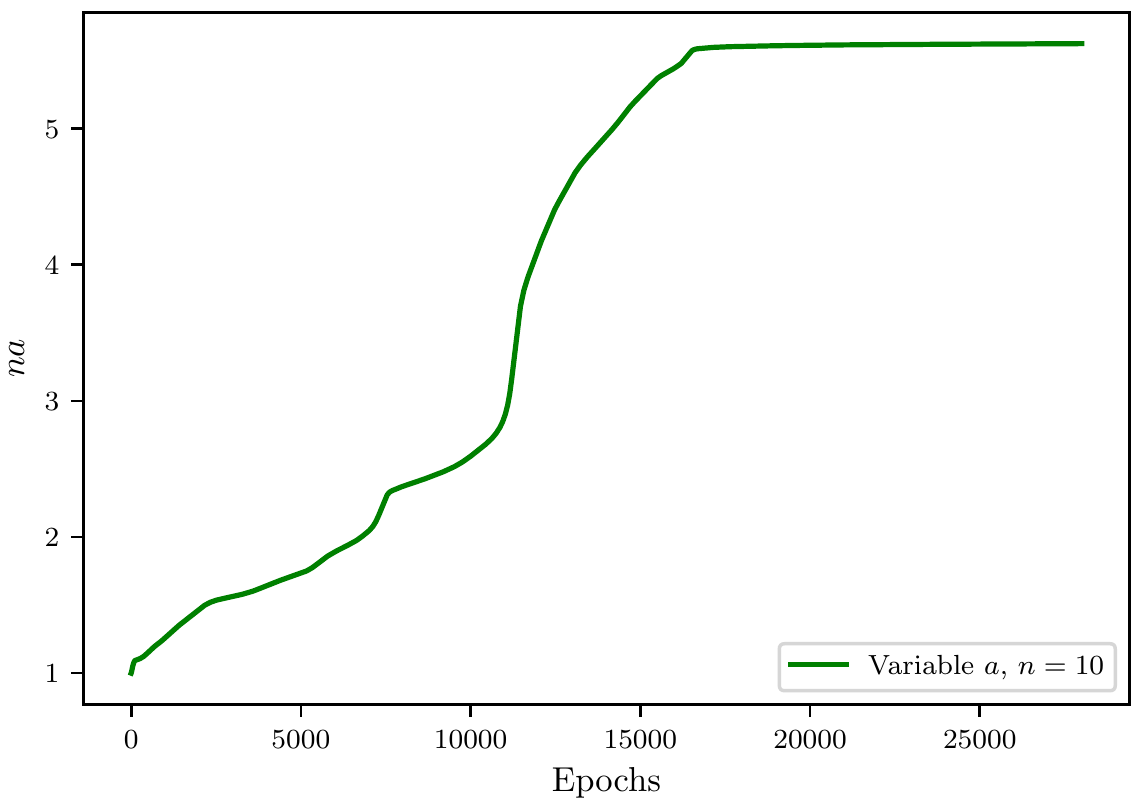}
\caption{Discontinuous function: Neural network solution of fixed (top row) and adaptive activation (middle row) functions. First column (top and middle row) shows the solution which is also plotted in frequency domain (zoomed-view) as shown by the corresponding second column. Bottom row shows the loss function comparison for fixed and adaptive activation (left) and variation in $a$ with $n = 10$ (right).}
\label{fig:NN}
\end{figure}

Knowledge of training process of neural network is important in order to optimize this process. In the earlier study, Arpit, \textit{et al.} \cite{ARP} suggested that the neural network learns simple pattern first before memorizing. Similar results have been found by Rahaman, \textit{et al.} \cite{RAH}, who suggested that the neural network learns low frequencies first. They showed that the amplitude of each frequency component of the network output is controlled by the spectral norm of the small-size network, which increases gradually during the training process. Thus, longer training time allows the network to learn complex functions by allowing it to capture the high frequencies components in the solution. Another such tool is the \textit{frequency principle} or \textit{F-principle} proposed by Xu, \textit{et al.},  see \cite{FPrin} and the references therein. In this case, the evolution of solution is observed in frequency domain by taking the Fourier transform $F[u]$ of a solution, where one can see that the neural network captures the low frequency first during the training process and then captures high frequency components. In this work, we shall use the F-principle to observe the performance of adaptive activation function.

Figure \ref{fig:NNsmt} shows the smooth solution given by fixed (top row) and adaptive activation (middle row) functions. The first column (top and middle rows) shows the solutions whereas the second column (top and middle rows) shows the solution in frequency domain. The adaptive activation function captures all frequencies in 22000 iterations as opposed to fixed activation. This behavior is also reflected in their solution plots. The bottom left figure shows the loss function comparison for fixed and adaptive activation with scaling factor 10. The loss is decreasing faster in case of adaptive activation and the optimal value of $a$ is around 4.5 as shown in the bottom right figure. The neural network tries to capture the low frequency first and then high frequency components, which can be seen from the solution as well as from the corresponding frequency plots.

Figure \ref{fig:NNsmt2} shows the smooth Burgers solution given by fixed (top row) and adaptive activation (middle row) functions. The first column (top and middle rows) shows the solutions whereas the second column (top and middle rows) shows the solution in frequency domain. We observe that even after 80000 iterations we are unable to capture all frequencies in the solution. Nonetheless, the adaptive activation function captures the frequencies faster than the fixed activation. The bottom left figure shows the loss function comparison for fixed and adaptive activation with scaling factor 10. The loss is decreasing faster in the case of adaptive activation and the optimal value of $a$ is around 1.56 as shown in the bottom right figure.

Finally, figure \ref{fig:NN} shows the discontinuous solution given by fixed (top row) and adaptive activation (middle row) functions. The first column (top and middle rows) shows the solutions whereas the second column (top and middle rows) shows the solution in frequency domain. The adaptive activation function captures all frequencies in 28000 iterations whereas the fixed activation function fails to capture these frequencies, which can be also be seen from their solution plots. The bottom left figure shows the loss function comparison for fixed and adaptive activation with $n = 10$. The loss is decreasing faster in the case of adaptive activation. Finally, the bottom right figure gives the variation of $na$, which gives the optimal value around 5.86. In this case, the neural network first captures the discontinuity and then low frequency components present in $x<0$ region and finally captures the high frequencies present in $x>0$ region. 

\subsection{Burgers equation}
The Burgers equation is one of the fundamental partial differential equation arising in various fields such as nonlinear acoustics, gas dynamics, fluid mechanics \textit{etc}, see Whitham \cite{WH} for more details.
The Burgers equation was first introduced by H. Bateman \cite{Bat} and later studied by J.M. Burgers \cite{Bur} in the context of theory of turbulence. The inviscid Burgers equation is a nonlinear first-order hyperbolic partial differential equation, which can develop discontinuities even when the initial condition is sufficiently smooth. Here, in the case of vanishing viscosity, we shall consider Burgers equation given by equation \eqref{Bur} along with its initial and boundary conditions.
The nonlinearity in the convection term develops very steep solution due to small $\tilde{\epsilon}$ value.
\begin{figure} [htpb] 
\centering
\includegraphics[trim=0cm 5cm 0cm 0cm, clip=true, scale=0.99, angle = 0]{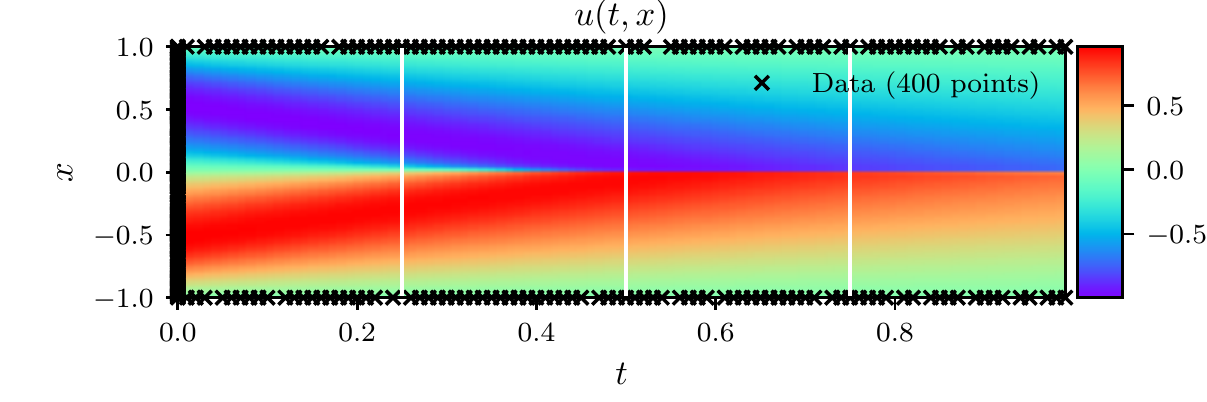}
\caption{Solution of Burgers equation (after 2000 iterations) on $x-t$ domain with variable $a, n = 5$. The three vertical white lines indicate the locations where the exact solution is compared with the solution given by PINN in figure \ref{fig:Bur333}}
\label{fig:Bur33}
\end{figure}
\begin{figure}  
\centering
\includegraphics[trim=0cm 0cm 0cm 4cm, clip=true, scale=0.9, angle = 0]{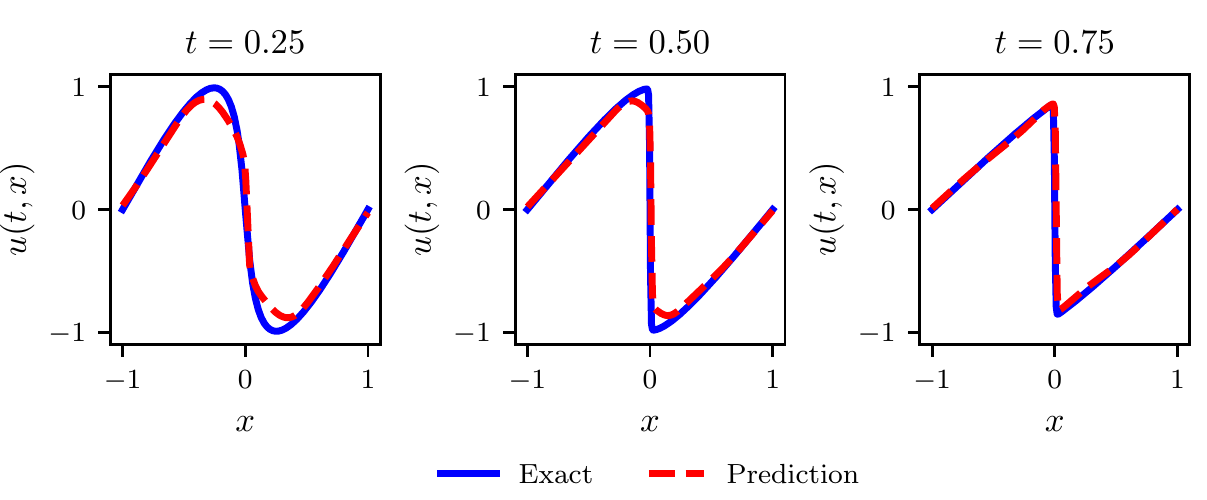}
\includegraphics[trim=0cm 0cm 0cm 4cm, clip=true, scale=0.9, angle = 0]{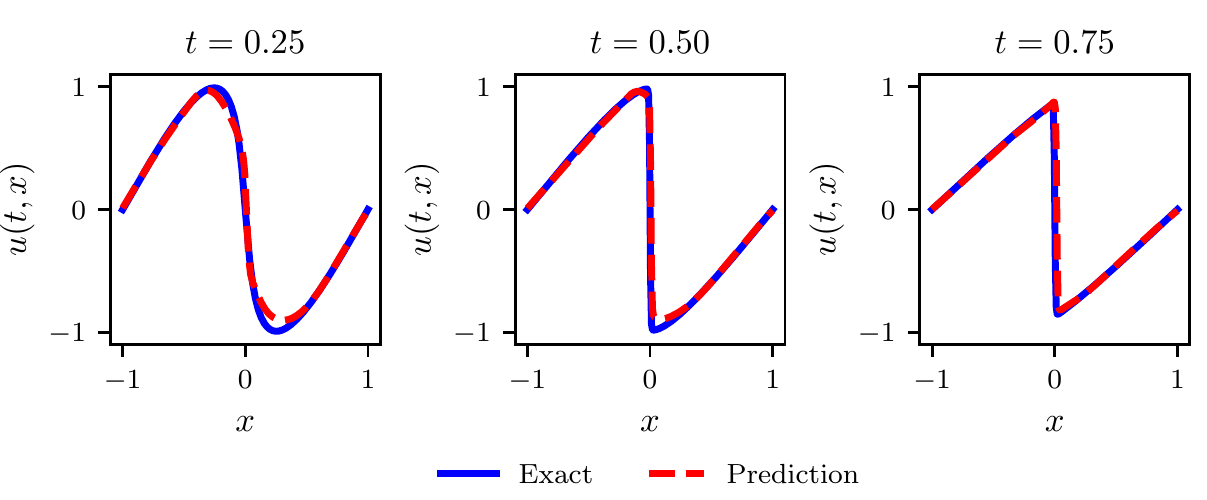}
\includegraphics[trim=0cm 0cm 0cm 4cm, clip=true, scale=0.9, angle = 0]{BurgersN5.pdf}
\caption{Burgers equation: Comparison of the exact solution with the solution given by PINN using $a = 1$ (top), variable $a, n = 1$ (middle) and variable $a, n = 5$ (bottom) obtained after 2000 iterations.}
\label{fig:Bur333}
\end{figure}
\begin{figure} [htpb] 
\centering
\includegraphics[scale=0.65]{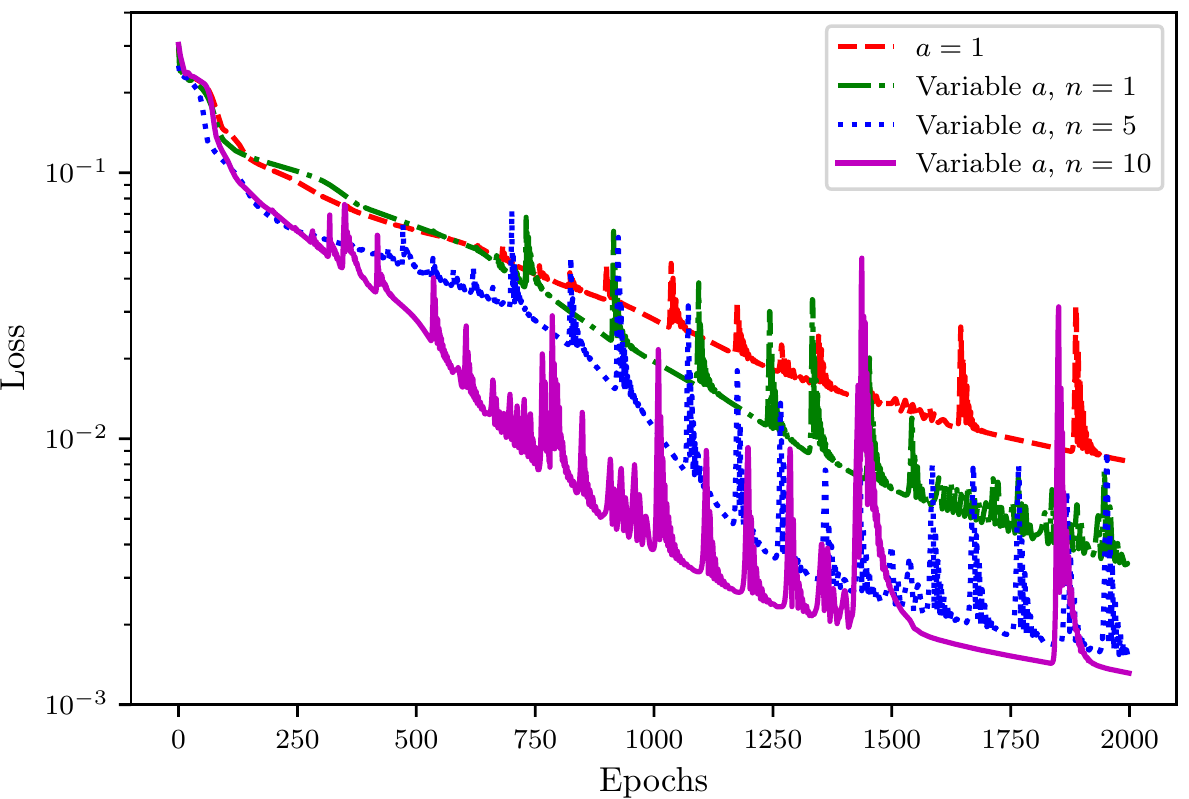}
\includegraphics[scale=0.65]{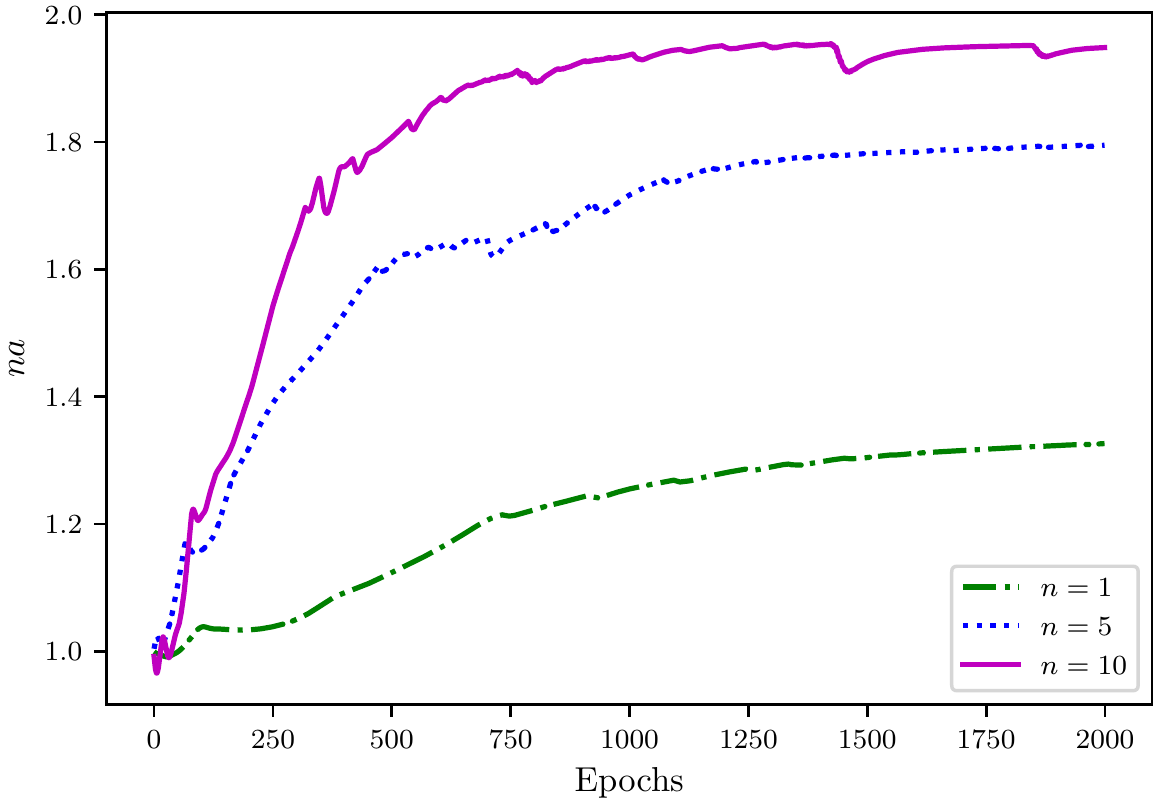}
\caption{Burgers equation: Loss vs. epochs for fixed and variable $a$ with different values of $n$ (left) and corresponding variation in $na$ with epochs (right).}
\label{fig:BurMSE}
\end{figure}
The number of training data points on the boundary is 400, the number of residual training points is  10000 and $$ \mathcal{F}\coloneqq (u_{NN})_{t} + u_{NN} (u_{NN})_x  -\tilde{\epsilon} (u_{NN})_{xx},$$
where $u_{NN}$ represents solution given by NN. The neural network architecture used for the computation consists of six hidden layers with 20 neurons in each layer.

Figure \ref{fig:Bur33} shows the contour plot of the solution of the Burgers equation and figure \ref{fig:Bur333} shows the comparison of exact and PINN solutions of Burgers equation with $\tilde{\epsilon} = 0.01/\pi$. In the top figure, the fixed activation function without any tuning parameters is used where the value of $a$ is unity. The middle and bottom figures present the results of adaptive activation function, where $a$ is a variable and scaling factors $n = 1$ and 5 are used in the respective figures. By introducing the adjustable parameter, the accuracy of the solution improves. Figure \ref{fig:BurMSE} (left) shows the variation of the loss function with epochs, where we can see the effect of hyper-parameter introduced in the activation function. For variable $a$, the loss function converges faster with increasing scaling factor as compared to fixed activation function.

Figure \ref{fig:BurMSE} (right) shows the optimization process for variable $a$. For unity scaling factor the optimization process of $a$ is slow due to small learning rate. One can increase the speed of tuning process further by increasing the scaling factor as shown in figure.
\begin{table}[htpb]
\begin{center}
\small \begin{tabular}{|c|c|c|c|c|} \hline 
 &  $a = 1$&
 Variable $a, (n = 1)$&Variable $a, (n = 5)$ & Variable $a, (n = 10)$
 \\ \hline
 Relative $L_2$ error &  1.913973e-01& 1.170261e-01 & 9.928848e-02 & 9.517134e-02 
 \\
 \hline 
 \end{tabular}
 \vspace{0.2 cm}
\caption{Burgers equation: Relative $L_2$ error after 2000 iterations with different values of $a$ with clean data.}\label{Table1}
\end{center}
\end{table}
Table \ref{Table1} gives the relative $L_2$ error for fixed as well as adaptive activation function, where it can be observed that the error decreases with increasing scaling factor.

\begin{figure} [htpb] 
\centering
\includegraphics[scale=0.44]{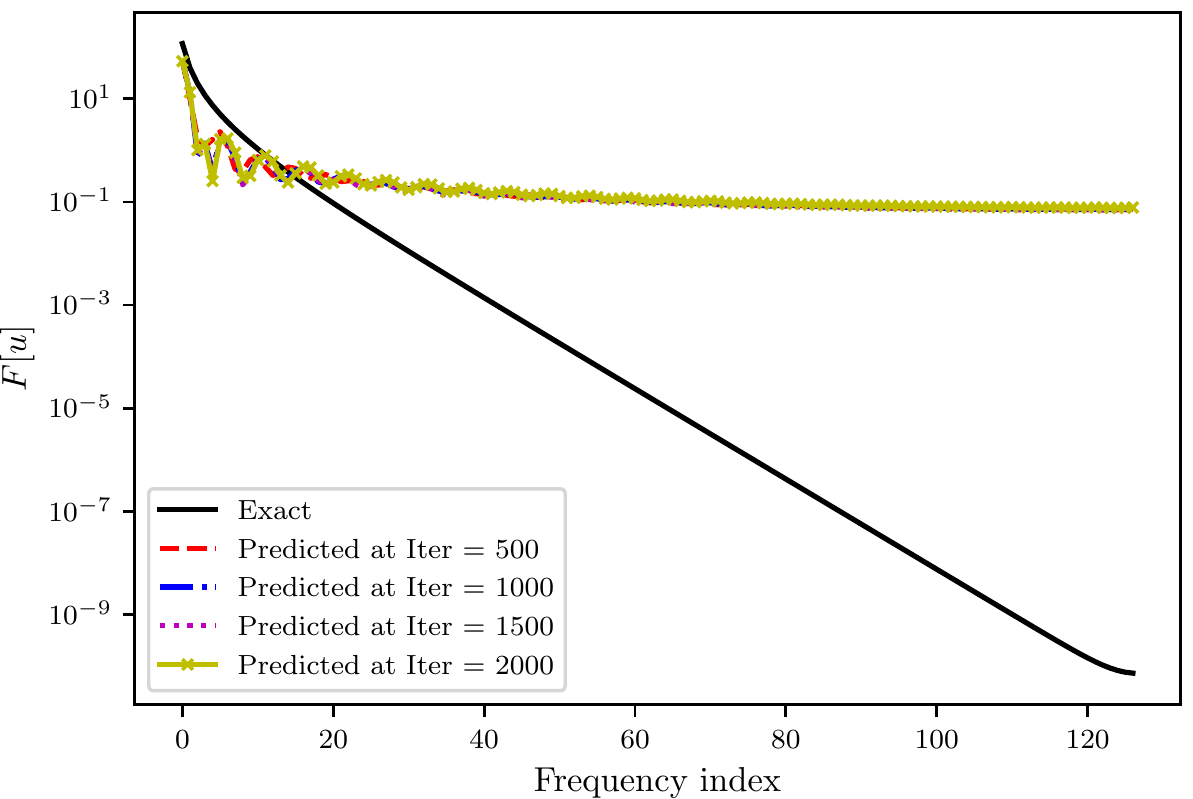}
\includegraphics[scale=0.44]{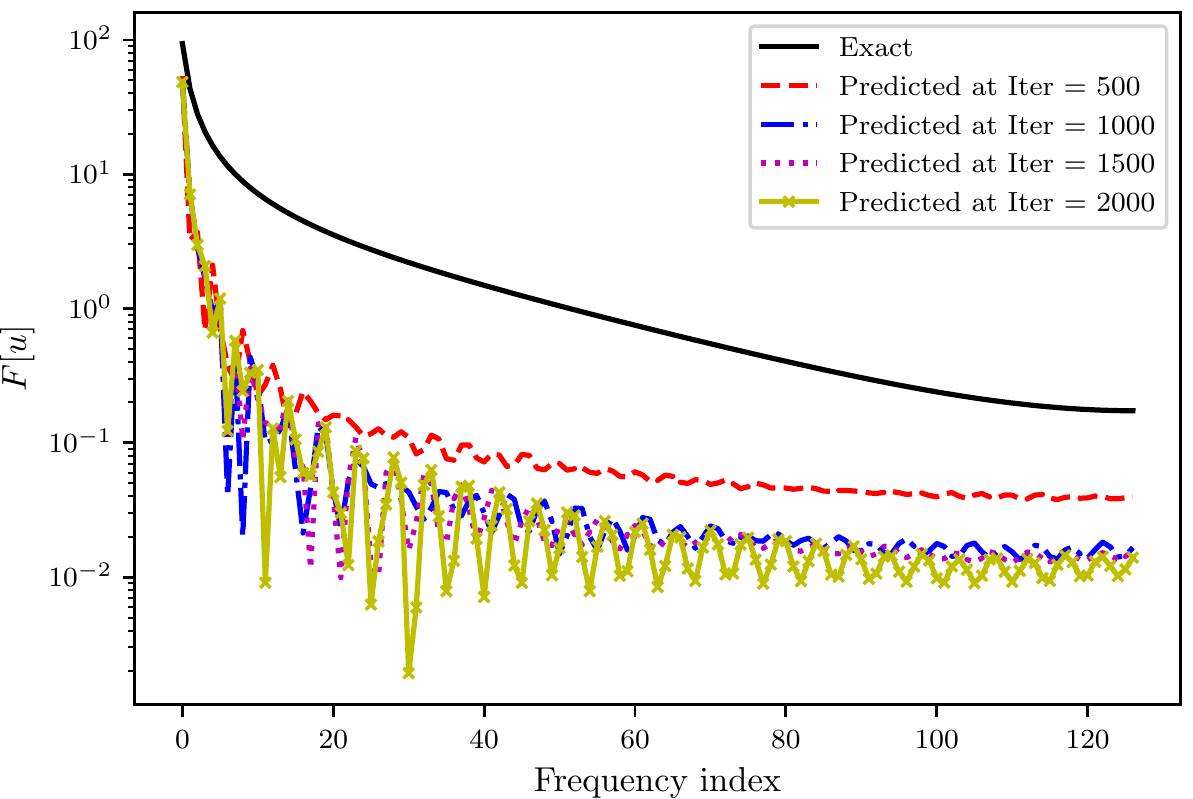}
\includegraphics[scale=0.44]{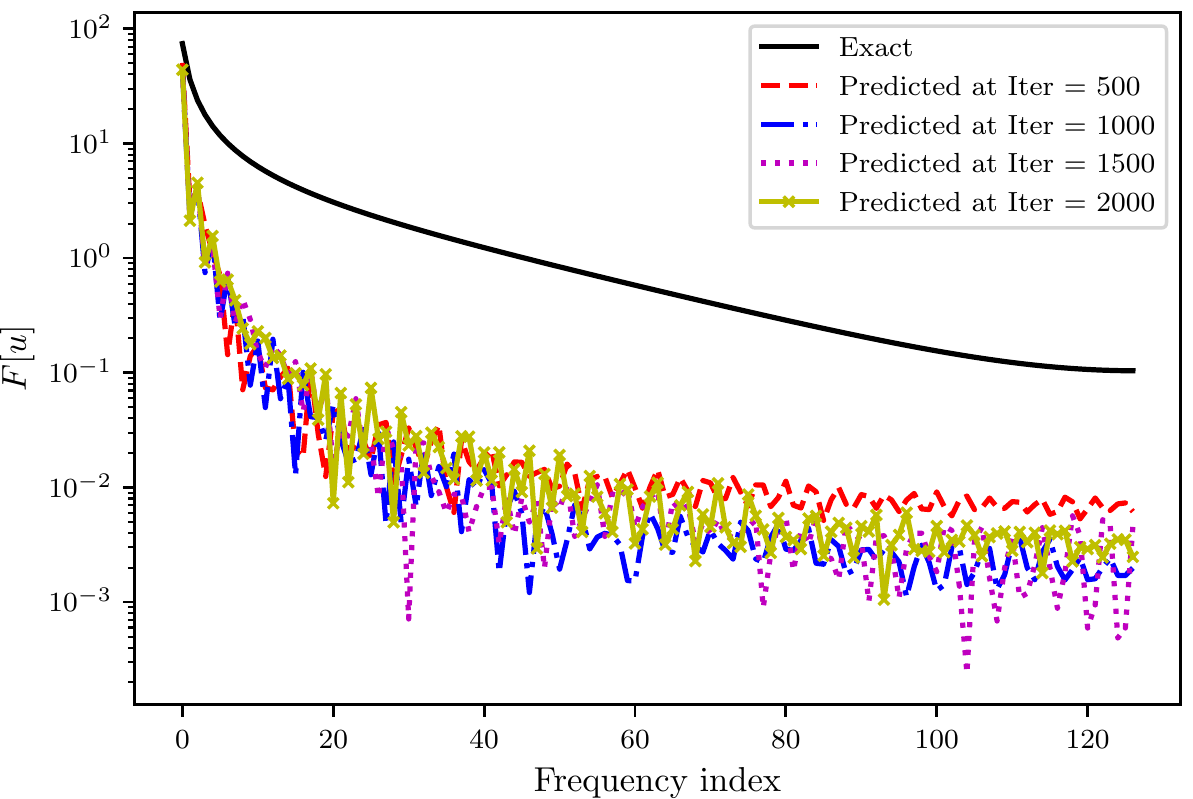}

\includegraphics[scale=0.44]{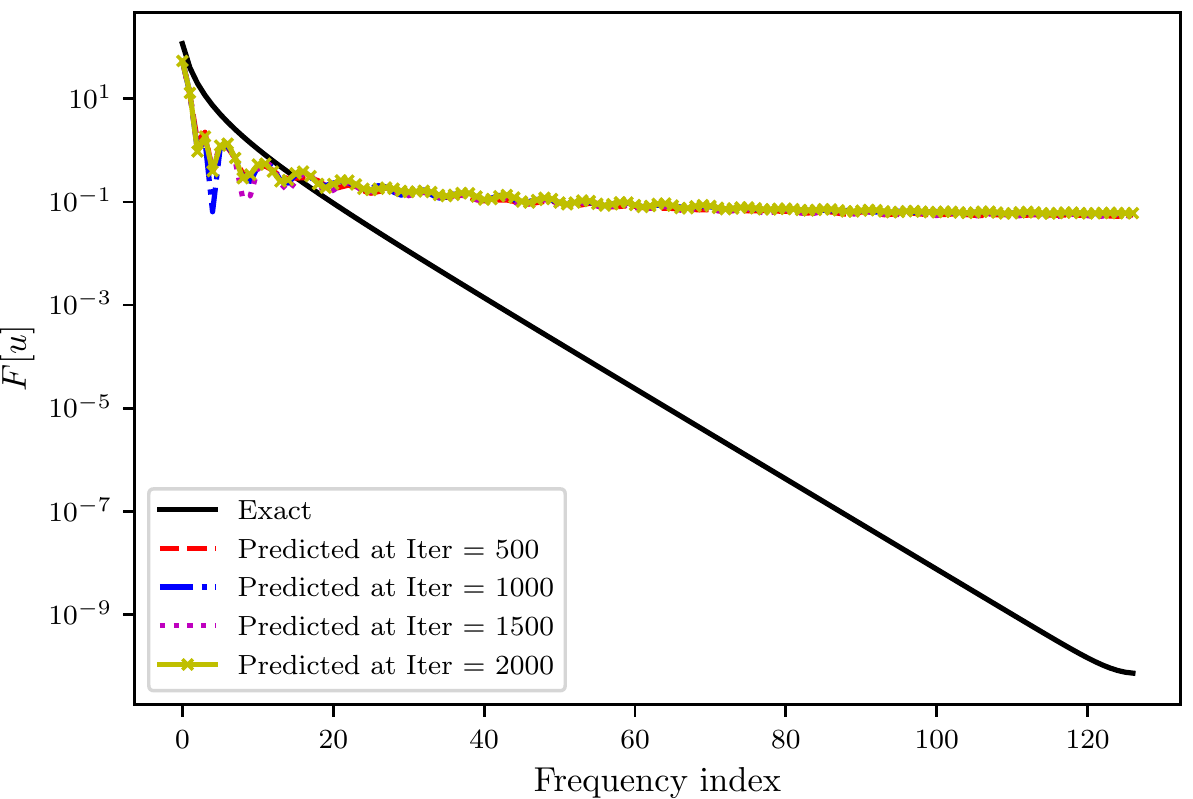}
\includegraphics[scale=0.44]{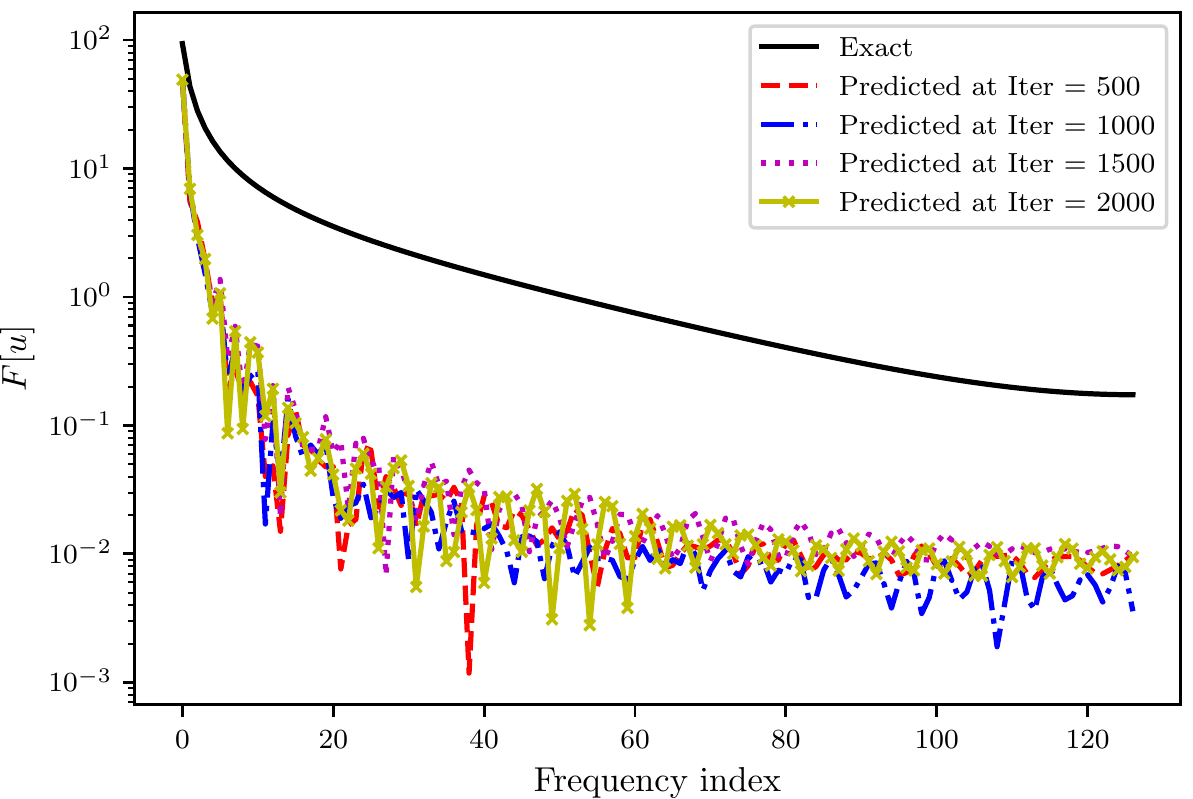}
\includegraphics[scale=0.44]{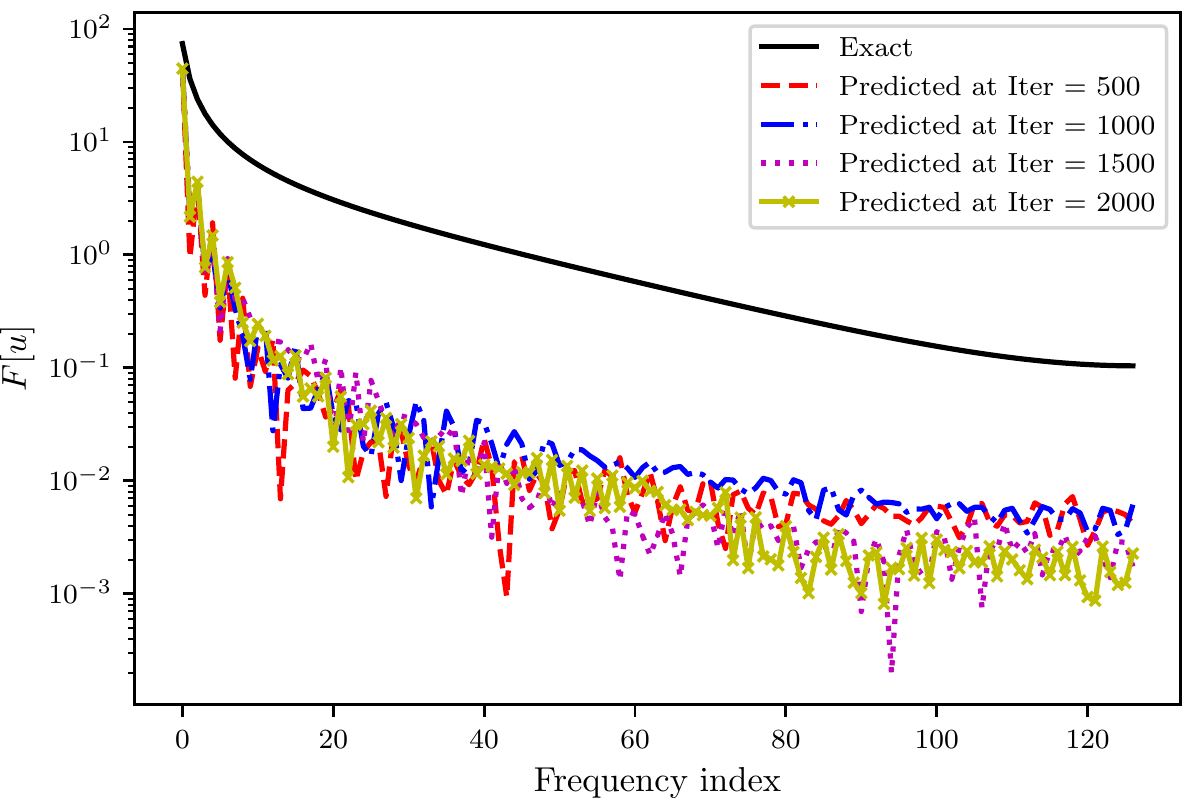}
\caption{ReLU activation: Comparison of solution of Burgers equation in frequency domain with fixed (1st row) and variable $a, n = 5$ (2nd row) \textit{'ReLU'} activation function. Columns (left to right) represent the solution in frequency domain at $t = 0.25, 0.5$ and 0.75, respectively.}
\label{fig:BurR}
\end{figure}
\begin{figure} [htpb] 
\centering
\includegraphics[scale=0.44]{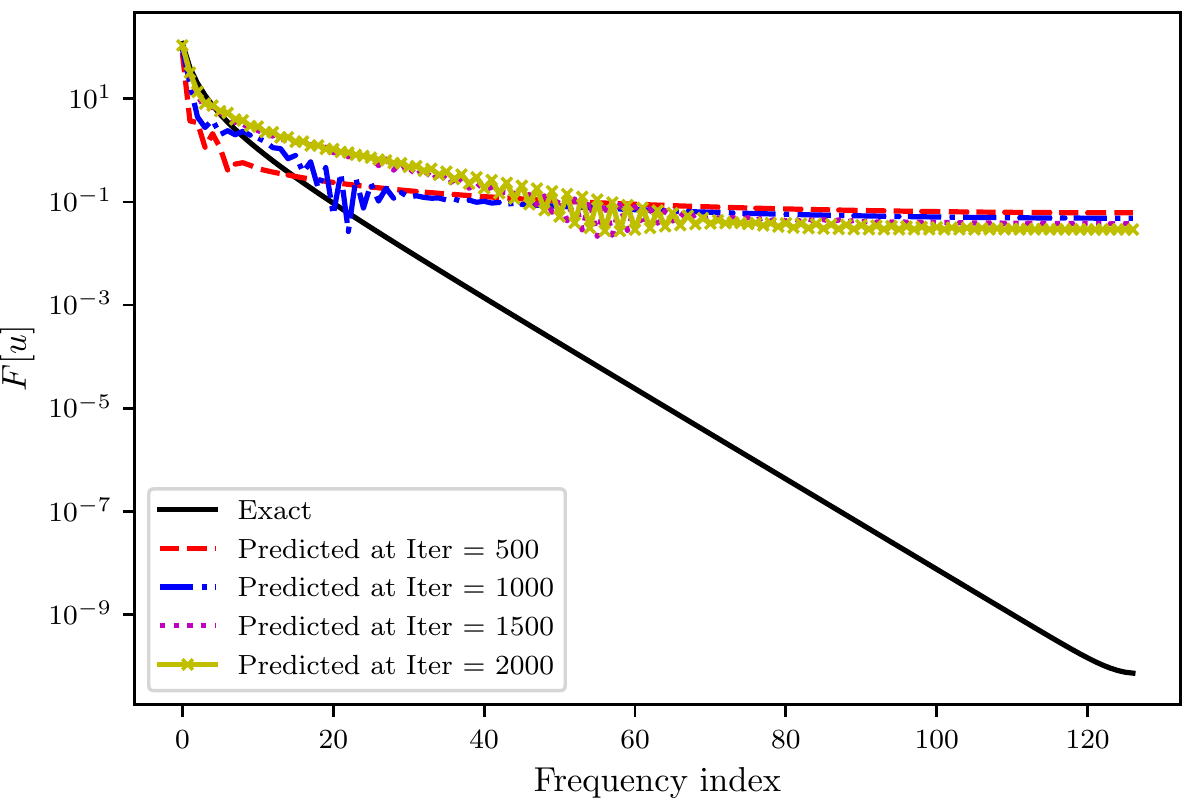}
\includegraphics[scale=0.44]{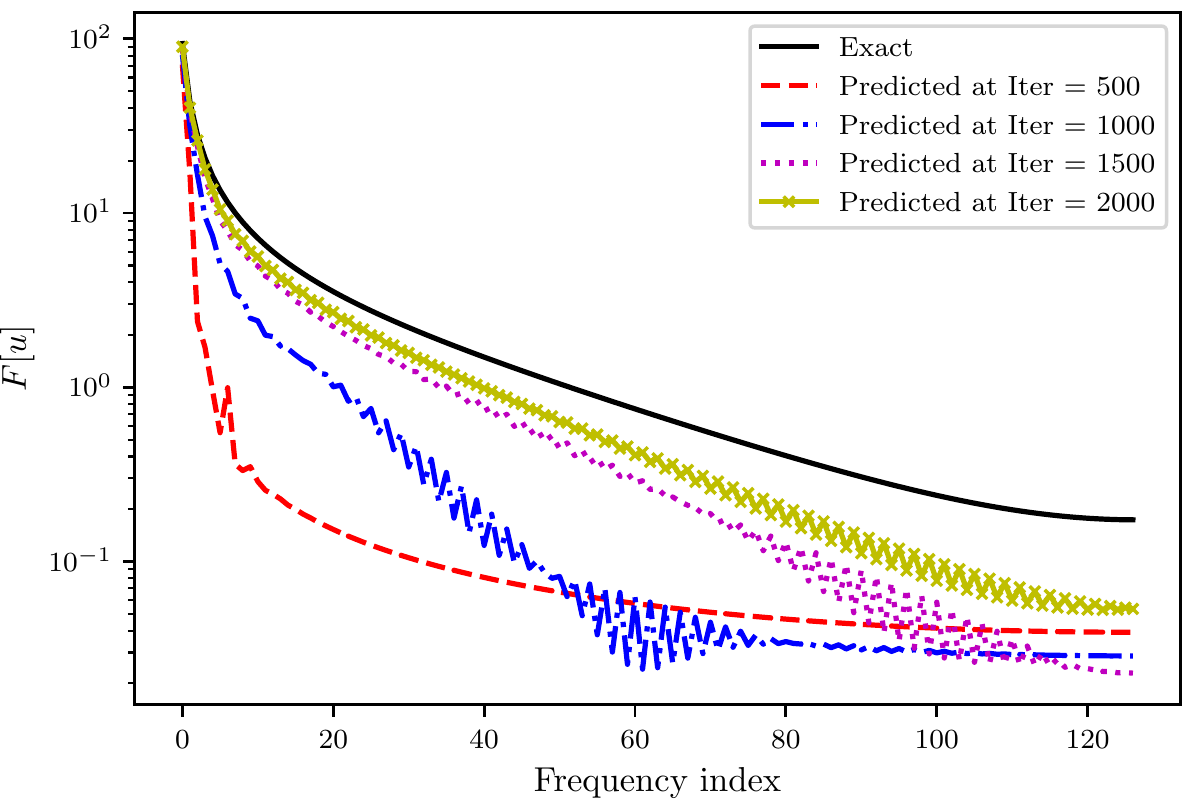}
\includegraphics[scale=0.44]{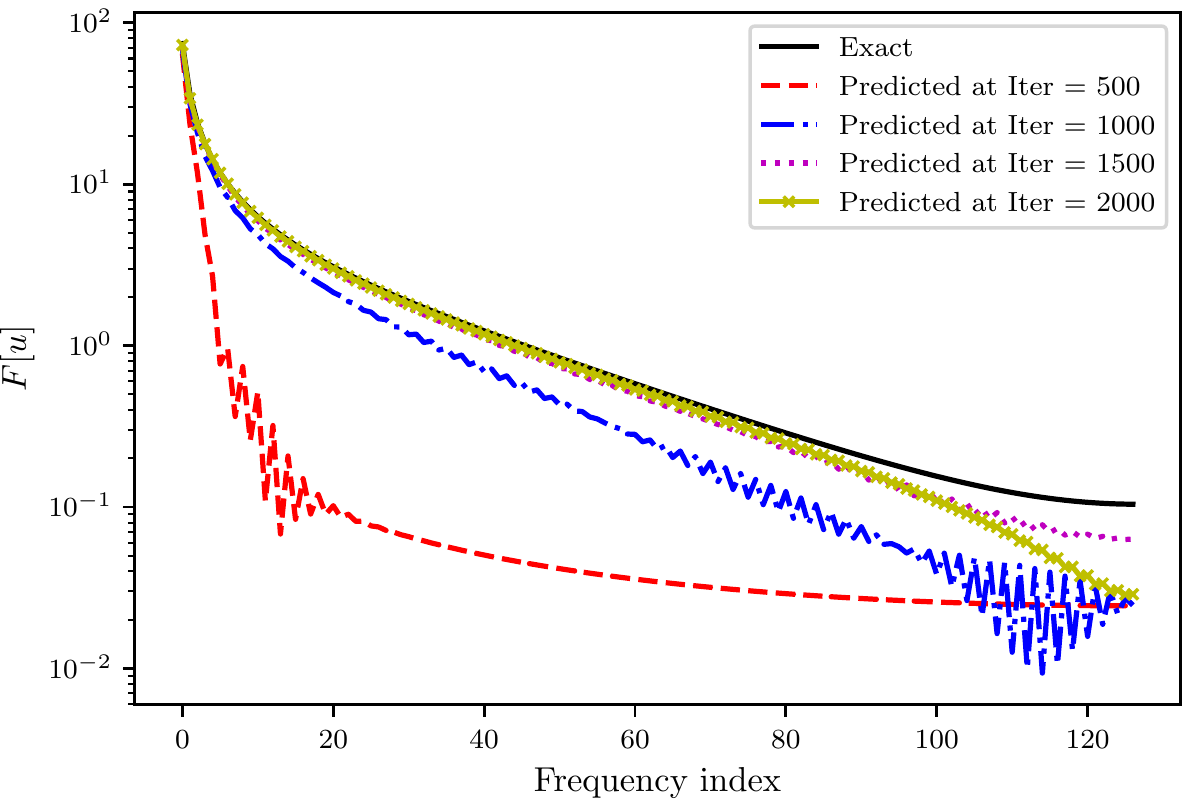}

\includegraphics[scale=0.44]{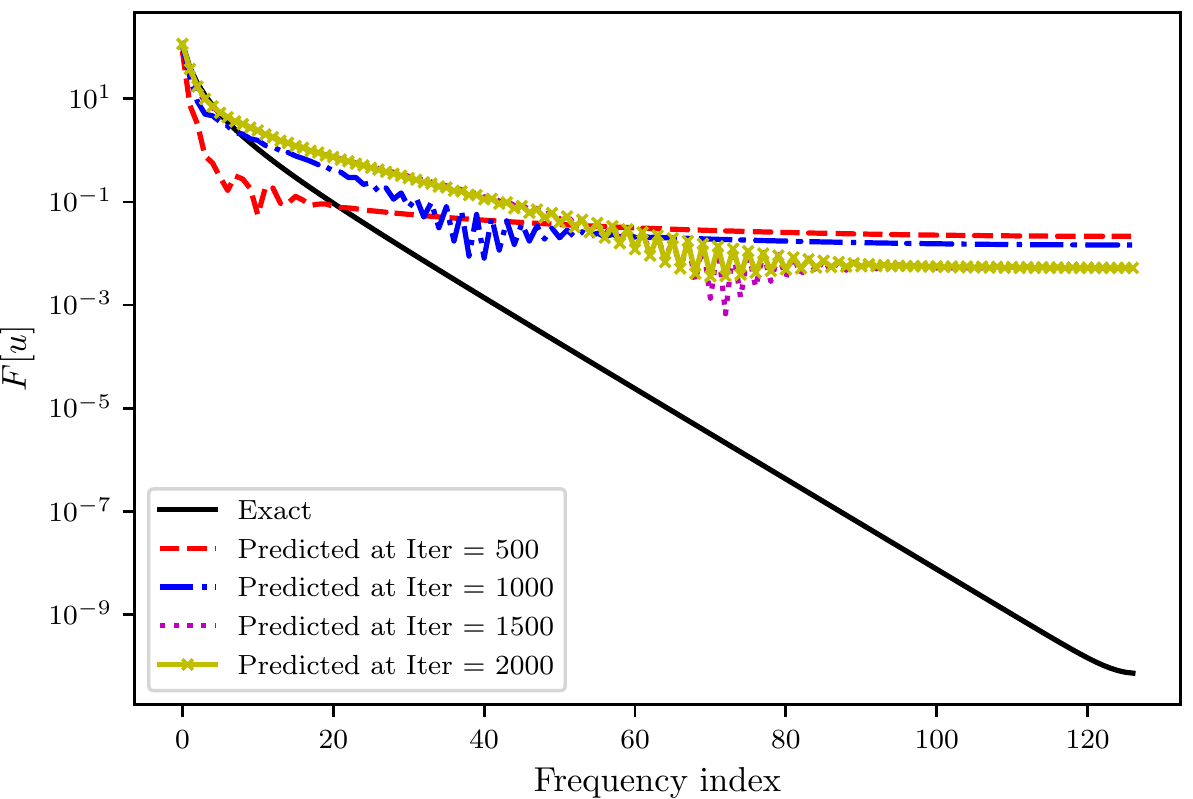}
\includegraphics[scale=0.44]{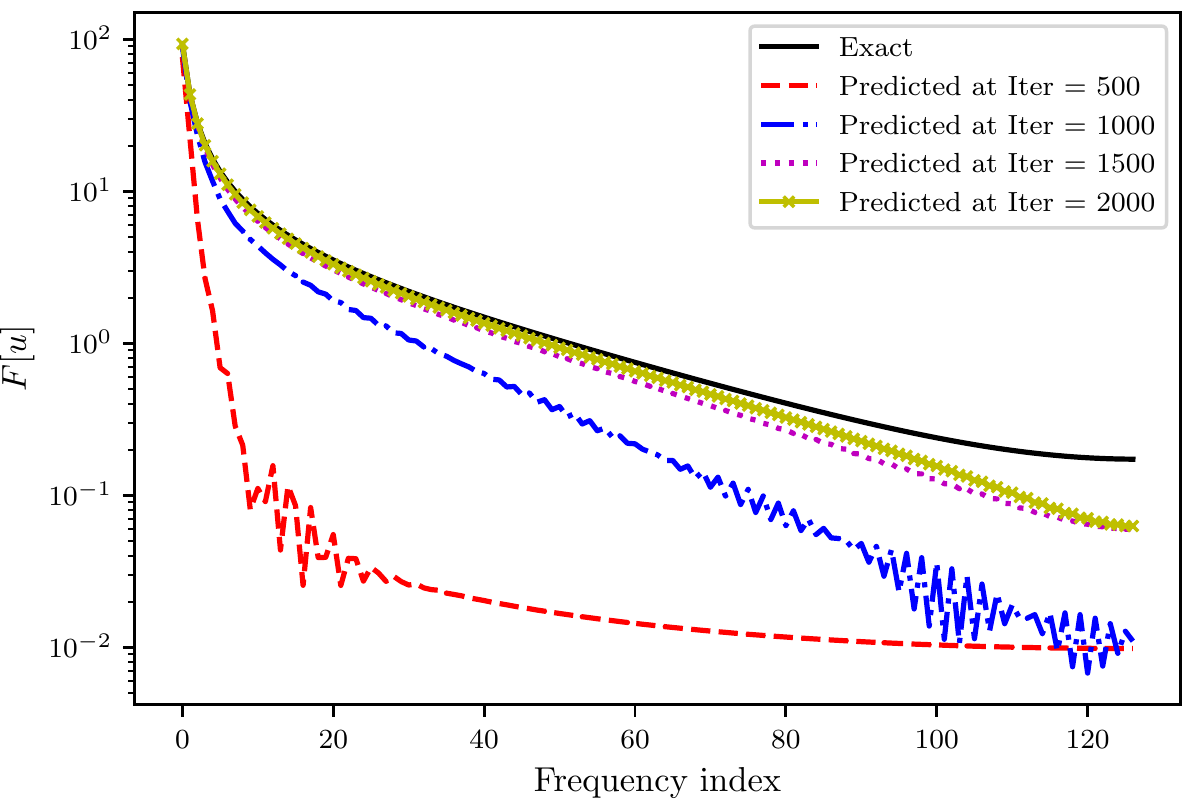}
\includegraphics[scale=0.44]{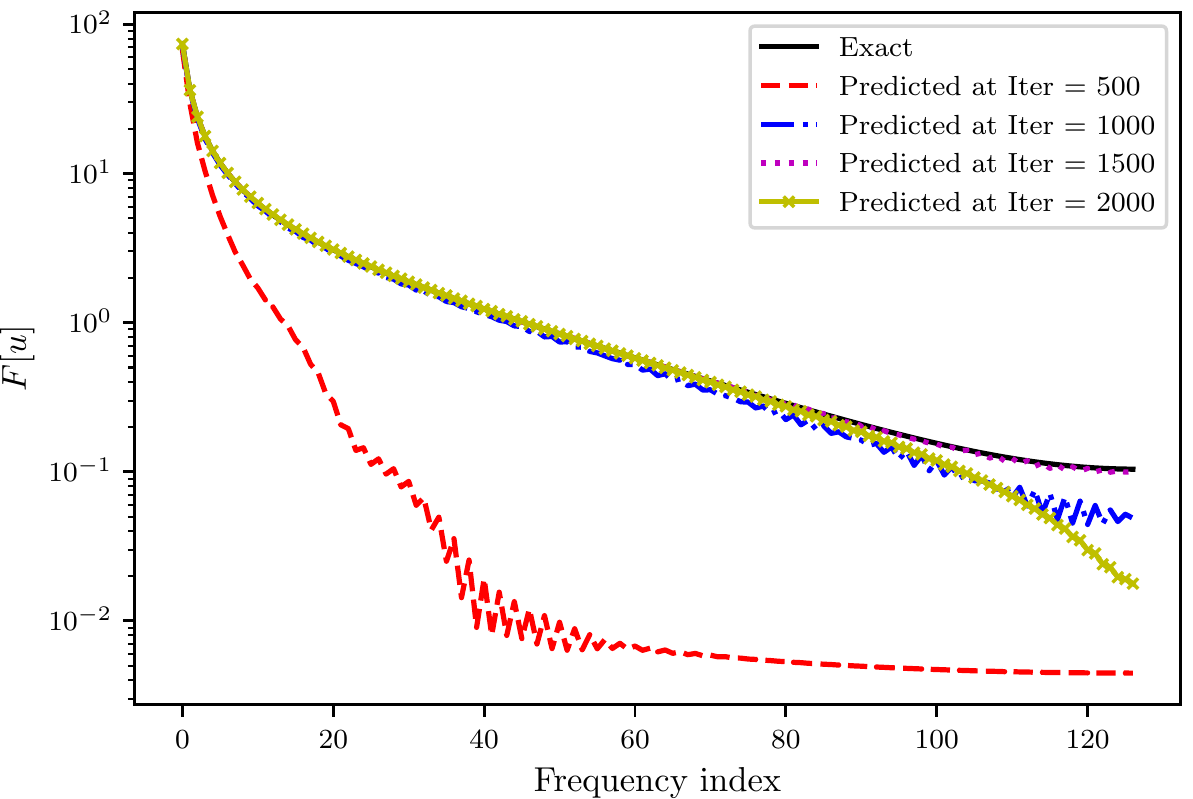}
\caption{\textit{'Tanh'} activation: Comparison of solution of Burgers equation in frequency domain with fixed (1st row) and variable $a, n = 5$ (2nd row) \textit{'tanh'} activation function. Columns (left to right) represent the solution in frequency domain at $t = 0.25, 0.5$ and 0.75, respectively.}
\label{fig:BurT}
\end{figure}
\begin{figure} [htpb] 
\centering
\includegraphics[scale=0.44]{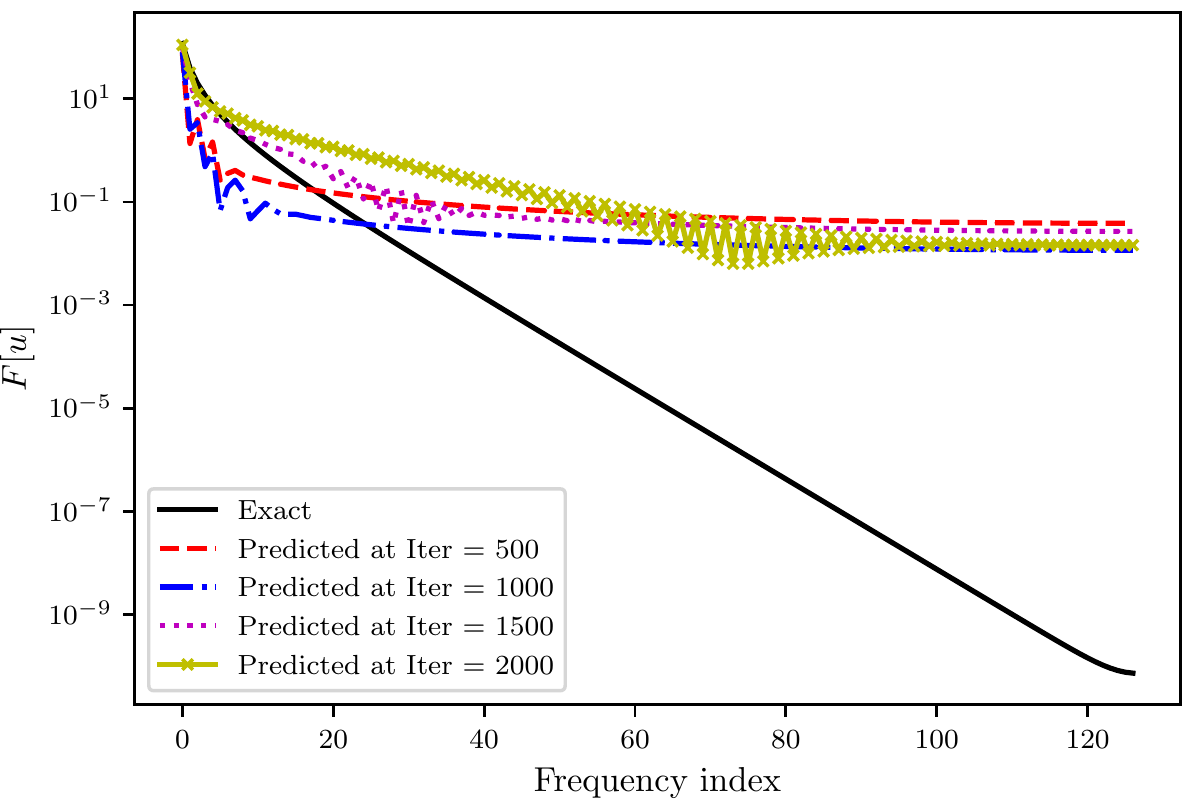}
\includegraphics[scale=0.44]{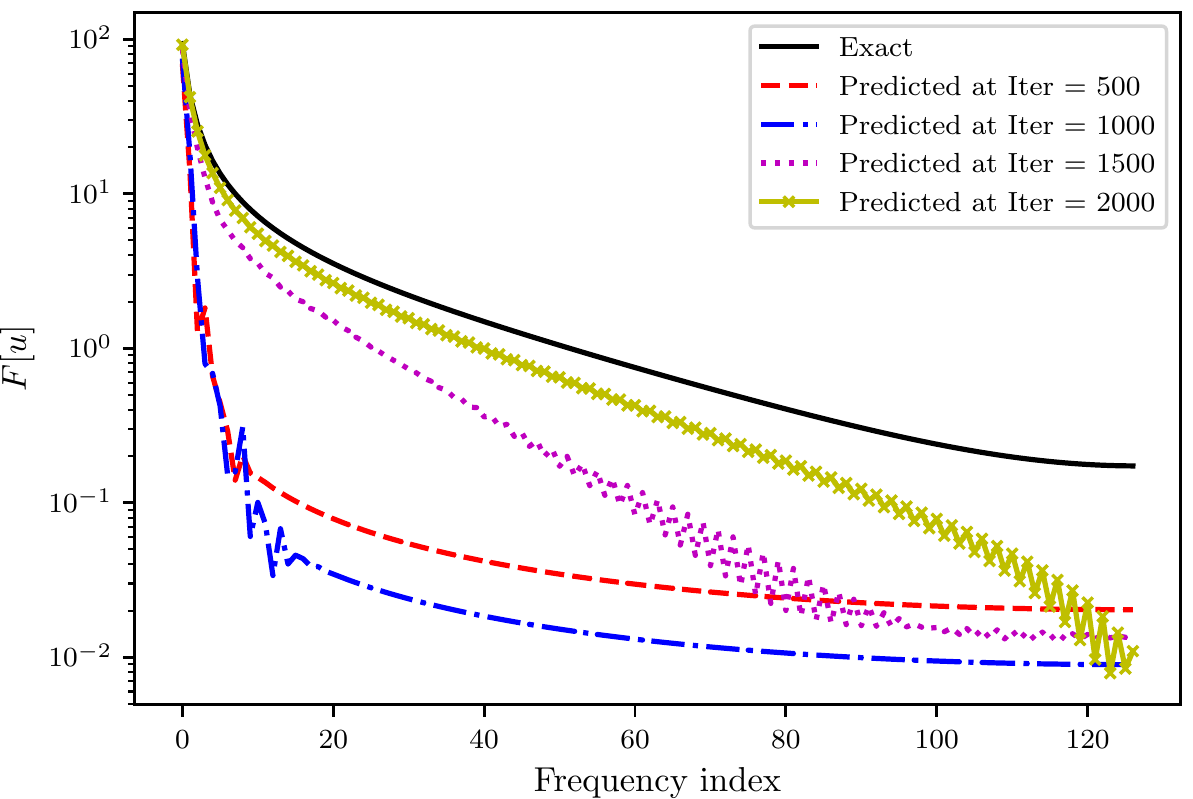}
\includegraphics[scale=0.44]{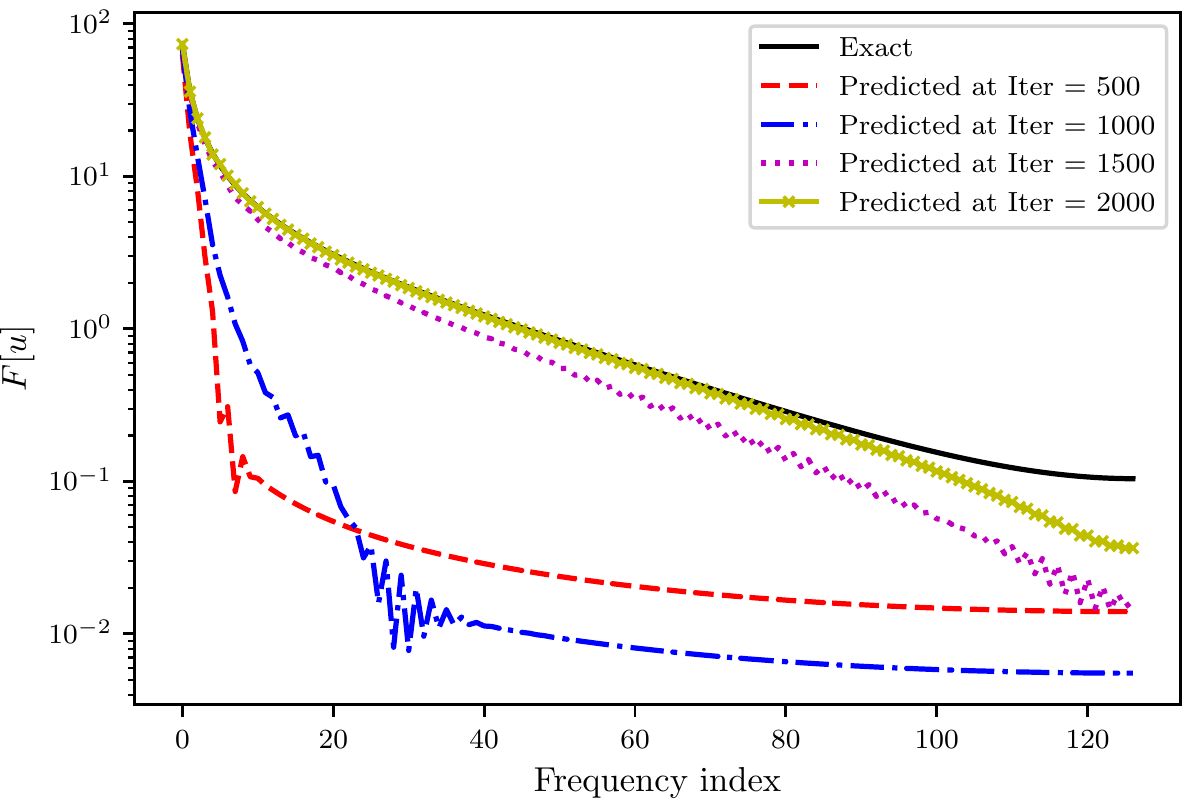}

\includegraphics[scale=0.44]{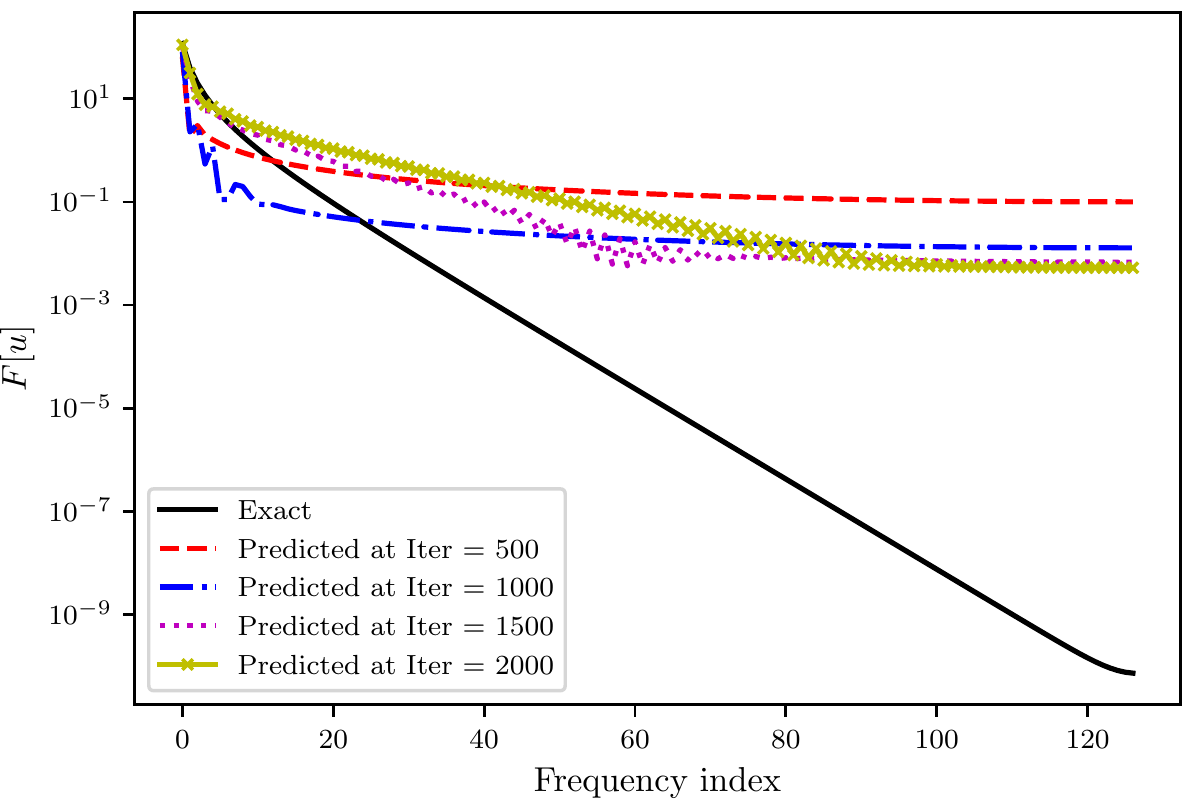}
\includegraphics[scale=0.44]{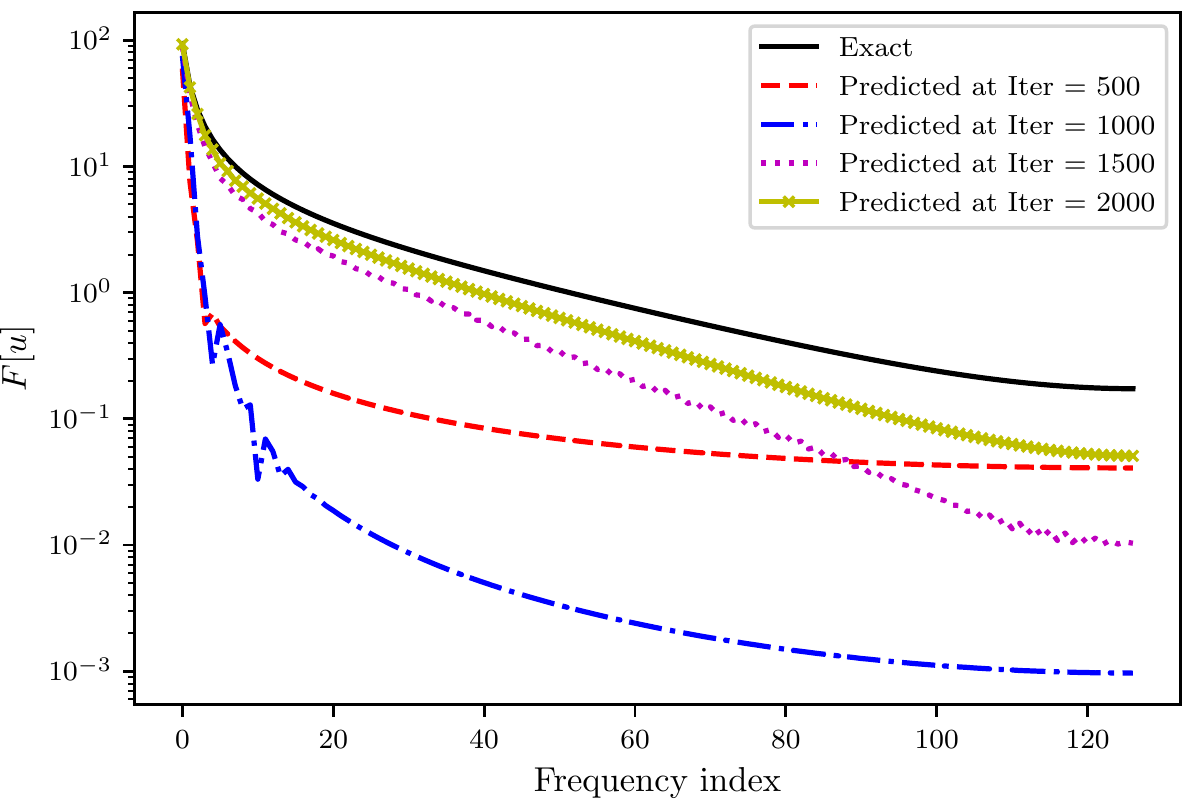}
\includegraphics[scale=0.44]{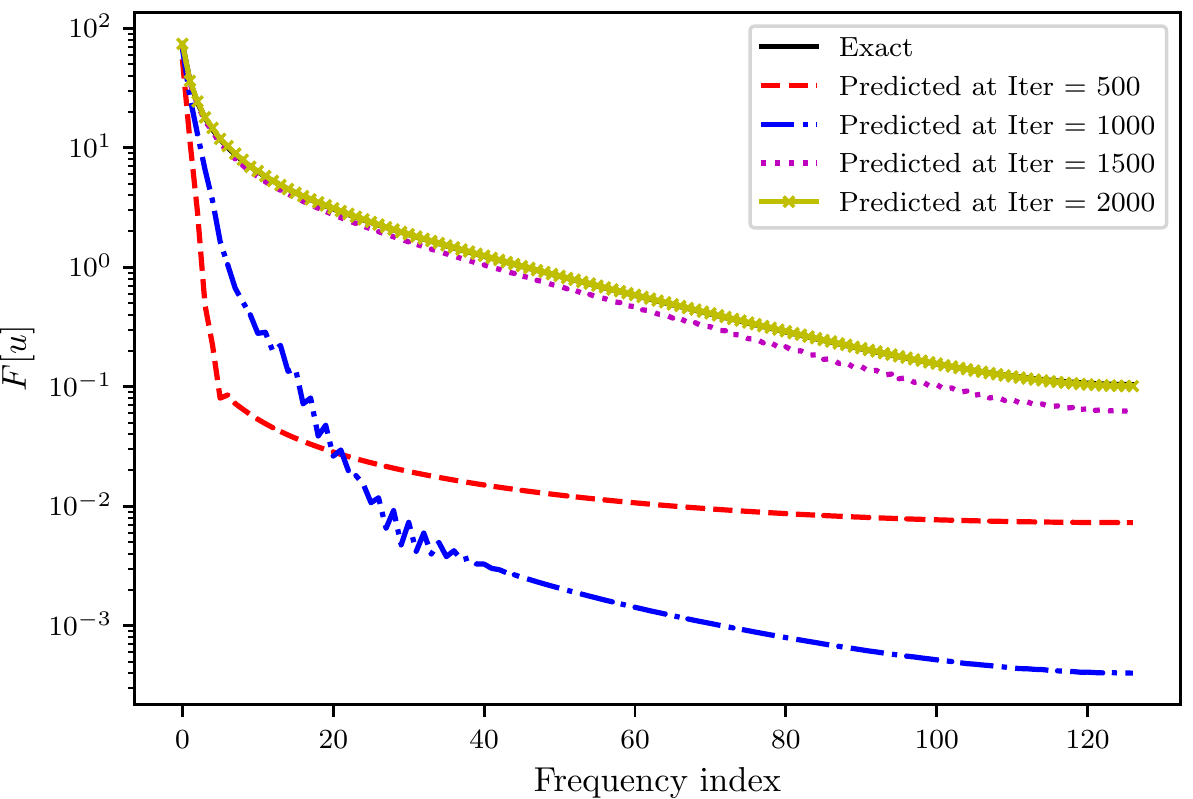}
\caption{\textit{'Sin'} activation: Comparison of solution of Burgers equation in frequency domain with fixed (1st row) and variable $a, n = 5$ (2nd row) \textit{'sin'} activation function. Columns (left to right) represent the solution in frequency domain at $t = 0.25, 0.5$ and 0.75, respectively. }
\label{fig:BurS}
\end{figure}

To show the comparison of fixed and adaptive activation functions we plotted the neural network solution of Burgers equation at $t = 0.25, 0.5$ and $0.75$ (column-wise) in frequency domain and compared their results. Figure \ref{fig:BurR} shows the results for ReLU activation function. The first row shows the fixed activation function whereas the second row gives result for the adaptive activation. In both cases, activation is unable to capture the frequencies present in the solution. Thus, ReLU is not a good activation function for this problem.  Next, we used \textit{'tanh'} activation and figure \ref{fig:BurT} shows the corresponding results. In the case of adaptive activation function, the process of capturing the correct frequency is faster than the fixed activation function for $t = 0.5$ and $0.75$. From the middle figures (top and bottom), the adaptive activation function captures almost all frequencies compared to fixed activation in a short duration. Similar behavior can be observed from the right figures (top and bottom) where dominant frequencies are captured in just 1000 iterations using the adaptive activation shown by blue curve (dash-dot line). Figure \ref{fig:BurS} shows the results for \textit{'sin'} activation function where a similar trend can be seen in the solution for both activation functions at $t = 0.5$ and $0.75$. In the case of smooth solution at $t = 0.25$ all activation functions fail to capture the frequencies present in the solution. This is clear from our previous example of approximating the smooth Burgers solution at $t = 0.25$ using NN requiring a large number of iterations to capture all frequencies present in the solution.

\begin{figure} [htpb] 
\centering
\includegraphics[scale=0.6]{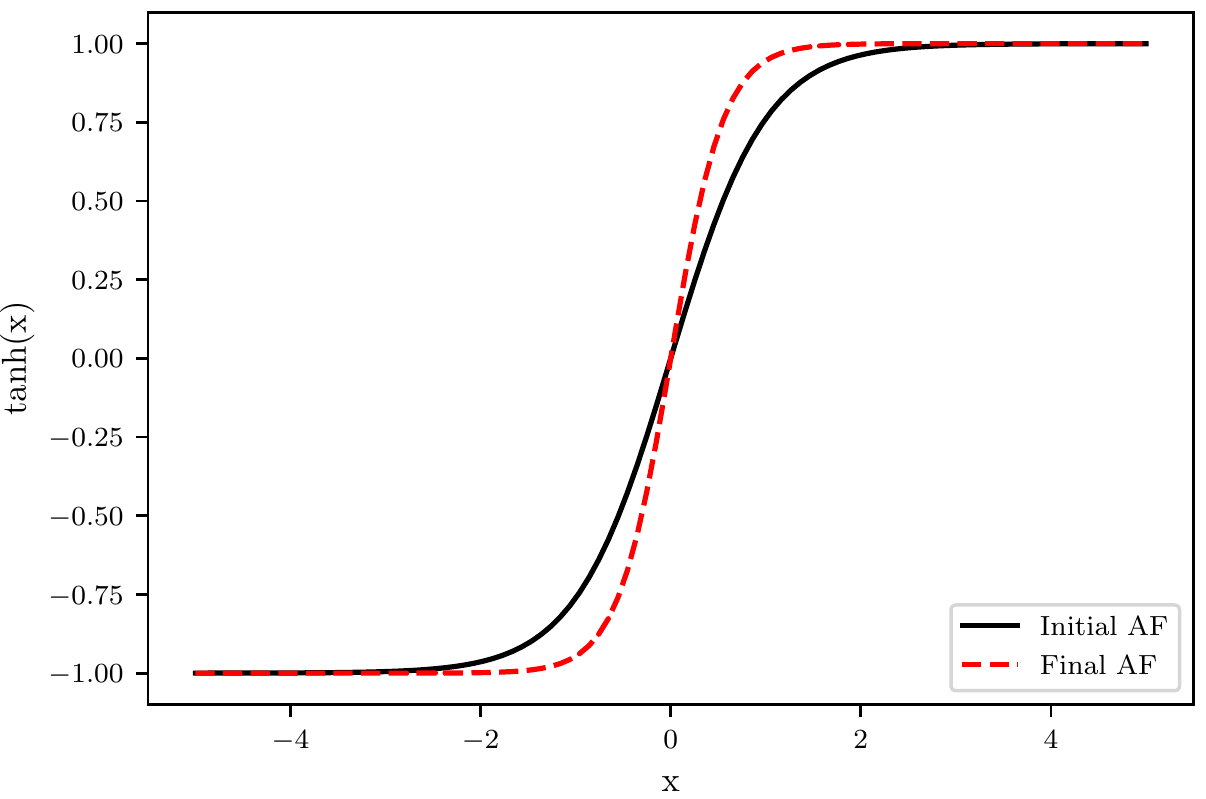}
\includegraphics[scale=0.6]{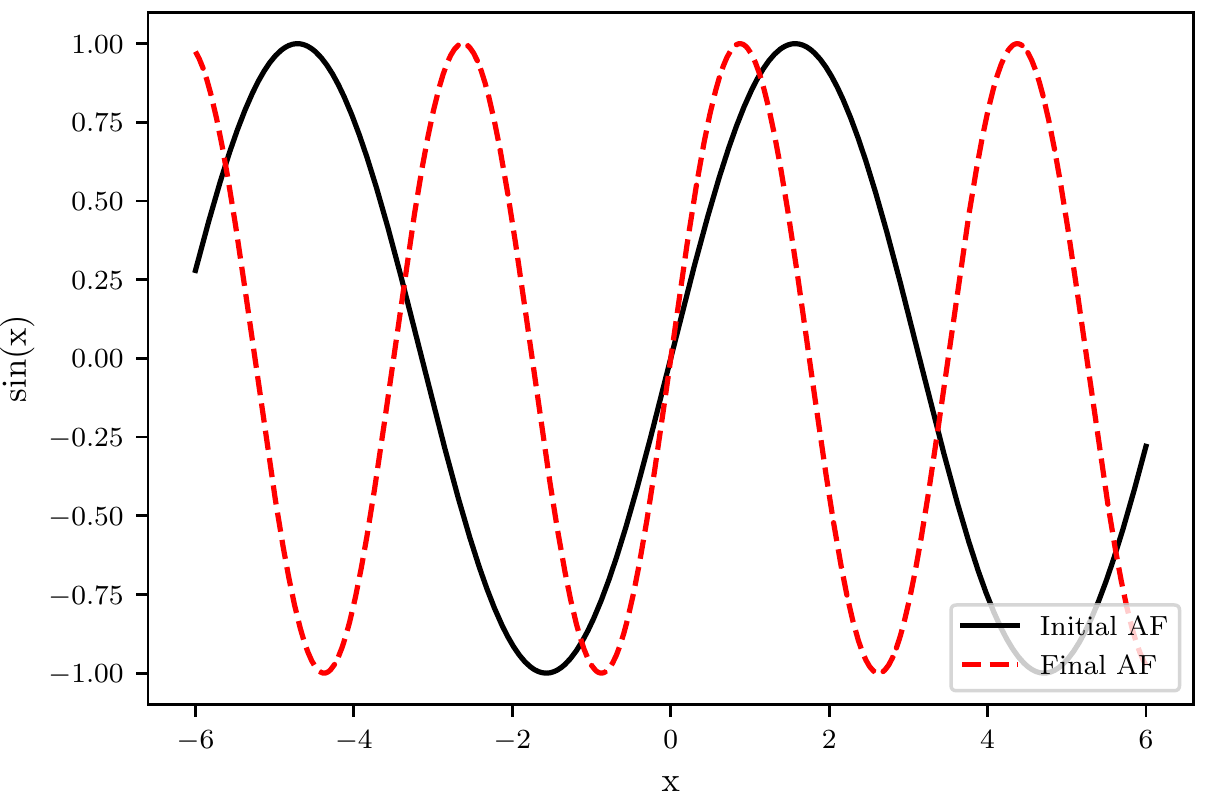}
\includegraphics[scale=0.54]{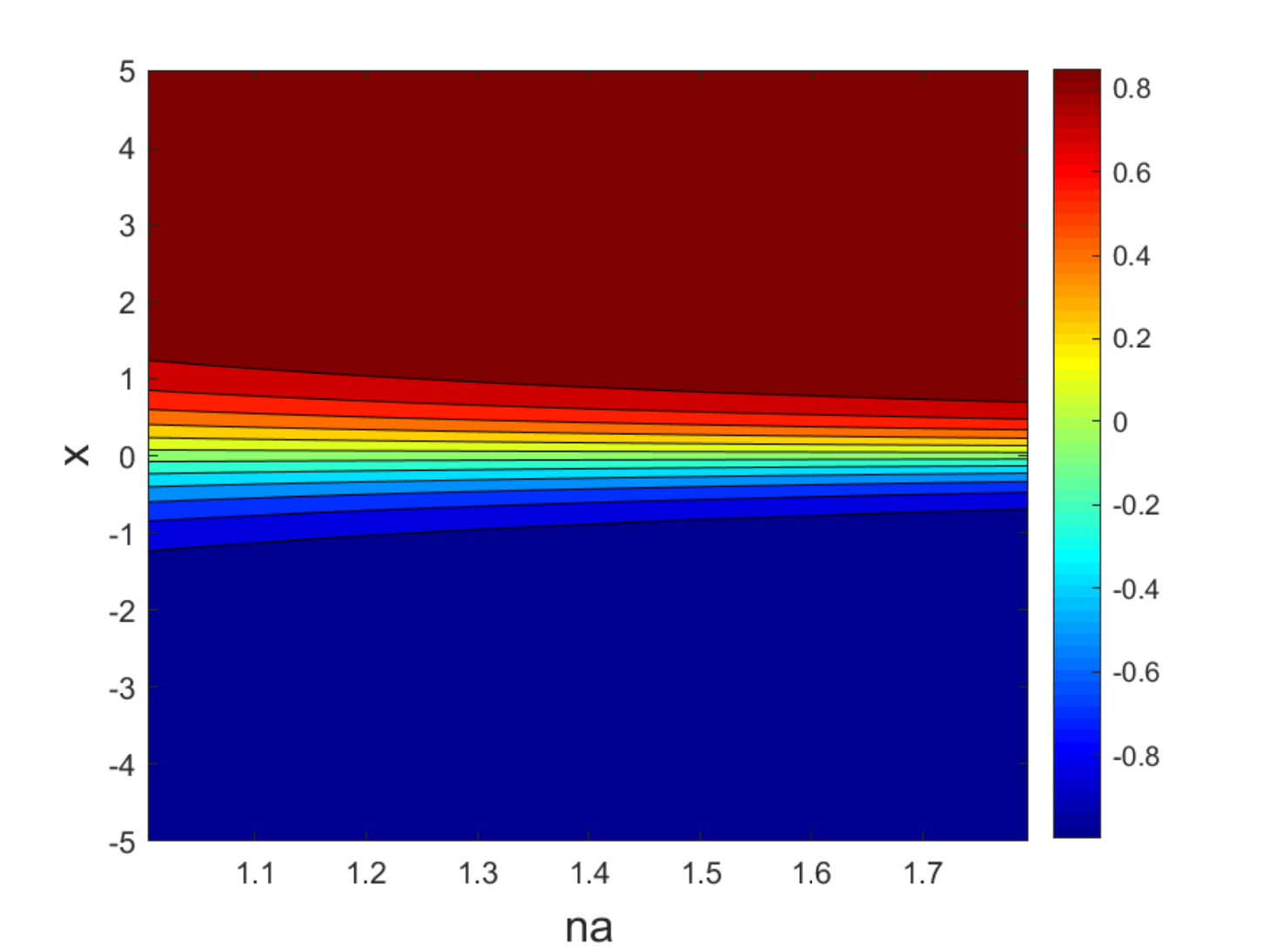}
\includegraphics[scale=0.54]{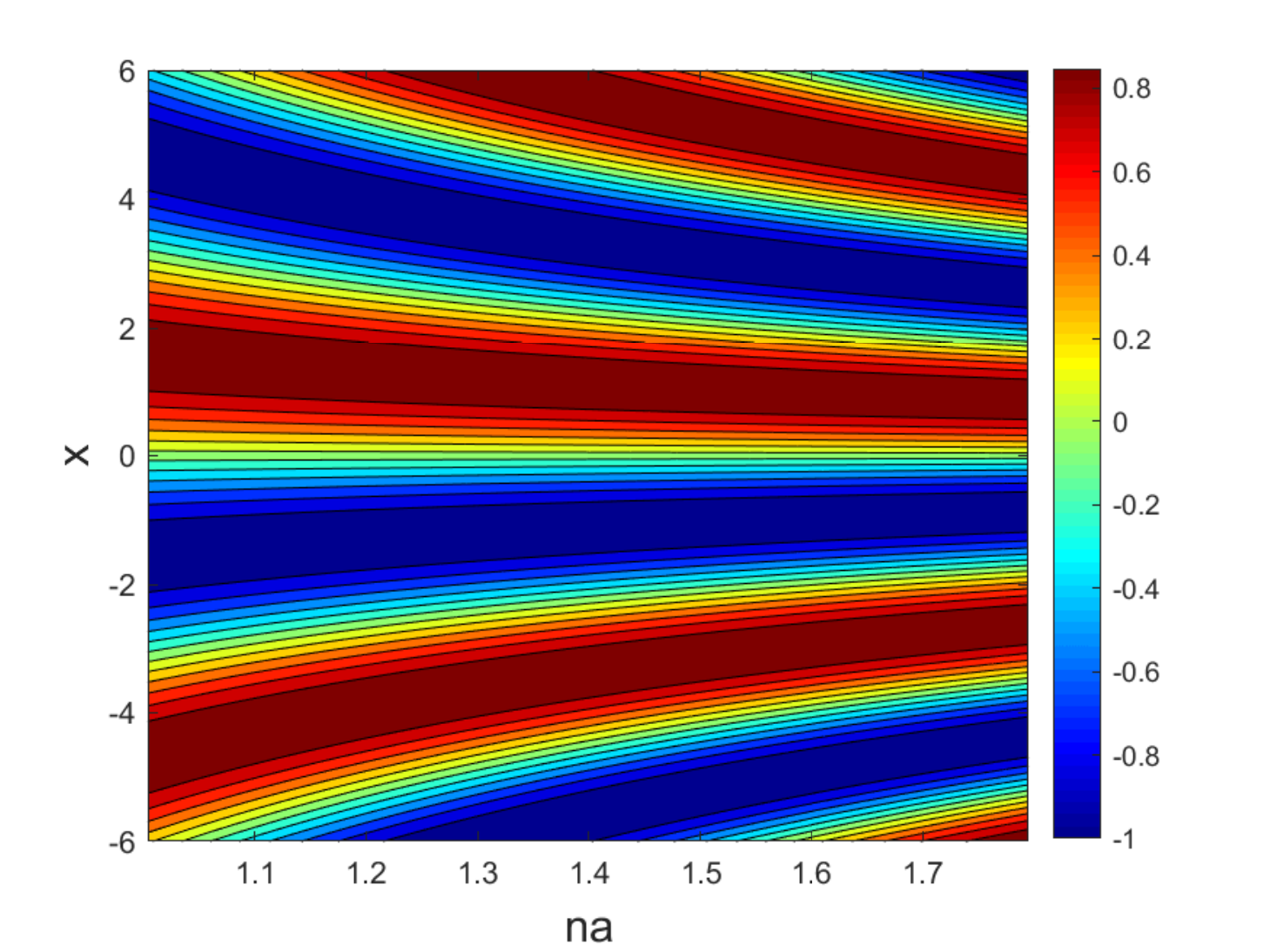}
\includegraphics[scale=0.54]{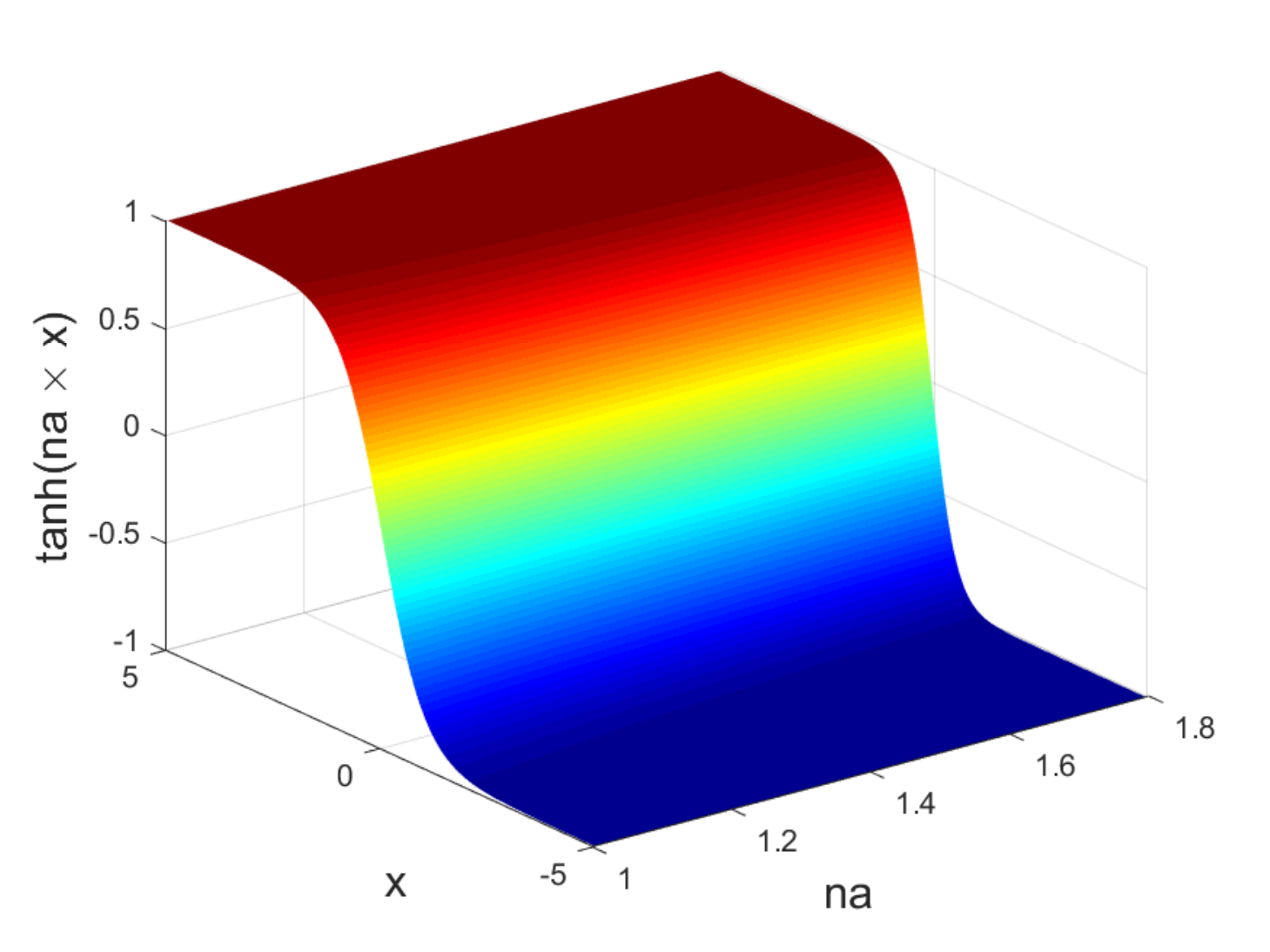}
\includegraphics[scale=0.54]{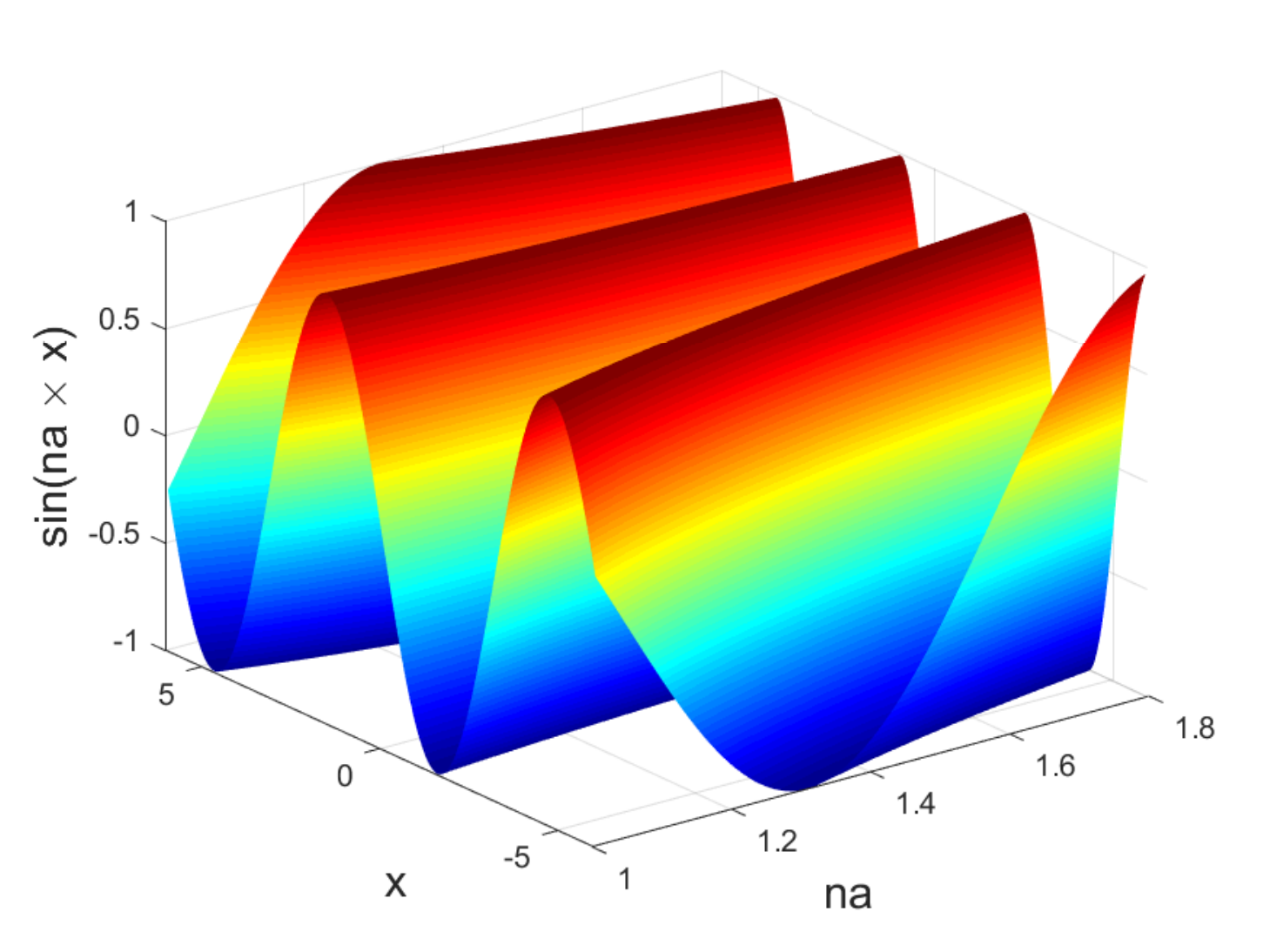}
\caption{Burgers equation: (Top row) Initial and final \textit{tanh} (left) and \textit{sin} (right) activation functions, (middle row) \textit{tanh} (left) and \textit{sin} (right) activation plane which shows the variation of activation function with $a$ and (bottom row) shows corresponding activation surfaces.}
\label{fig:BurAF}
\end{figure}

The introduction of a scalable hyper-parameter in the activation function dynamically changes the topology of the loss function thereby achieving faster convergence towards global minima. 
Figure \ref{fig:BurAF} (top row) shows the initial and final optimized \textit{'tanh'} (left) and \textit{'sin'} (right) activation functions. The corresponding \textit{activation planes ($na-x$ plane)} and \textit{activation surfaces} are given by the middle and bottom rows, respectively. In both cases we observe that the gradient of the activation function increases from its initial stage in the direction of large values of $na$, which contributes towards the fast learning process of the neural network.

\subsubsection{Initialization of scaled hyper-parameter and effect of large scaling factor}
Initialization of the scaled hyper-parameter can be done in various ways as long as such value does not cause divergence of the loss.
In this work the scaled hyper-parameter is initialized as $na=1, ~\forall n \geq 1$.
Although, increase in scaling factor speeds up the convergence rate, at the same time the parameter $a$ becomes more sensitive. This can be seen from the oscillations in the loss function as well as in the values of $a$, figure \ref{fig:BurMSE} (right). The reason behind this is that the SGD optimization algorithm becomes very sensitive, hence we should not use a large scaling factor which may eventually cause the solution to diverge. 
To overcome this difficulty we can think of adding a regularization term in the loss function. Although it partially suppresses the oscillations in the loss function, the other fatal effect of this addition is deterioration of convergence rate, which also reduces the accuracy of the solution. We can also add this regularization term after some iterations (instead of initially), which may nullify the effect of slower convergence rate but still we may not get rid of oscillations fully. Moreover, finding the correct value of regularization weight is still based on trial and error analysis. 

\subsection{Klein-Gordon equation}
The nonlinear Klein-Gordon equation is a second-order hyperbolic partial differential equation arising in many scientific fields like soliton dynamics and condensed matter physics \cite{CAU}, solid state physics, quantum field theory and nonlinear optics \cite{Waz}, nonlinear wave equations \cite{DOD} \textit{etc}. The inhomogeneous Klein-Gordon equation is given by
\begin{equation}\label{2Dkg}
u_{tt} + \alpha \Delta u + N(u) = h(x,t), \ \ x \in [-1,~1],\,\, t>0,
\end{equation}
where initial conditions and boundary condition are extracted from the exact solution given by 
$u(x,t) =x\cos(t)$. 
$\Delta$ is a Laplacian operator and $N(u) = \beta u +\gamma u^k$ is the nonlinear term with quadratic nonlinearity ($k=2$) and cubic non-linearity ($k=3$); $\alpha, \beta, \gamma$ are constants.
\begin{figure}[htpb]
\centering
\includegraphics[trim=0cm 3cm 0cm 0cm, clip=true, scale=0.99, angle = 0]{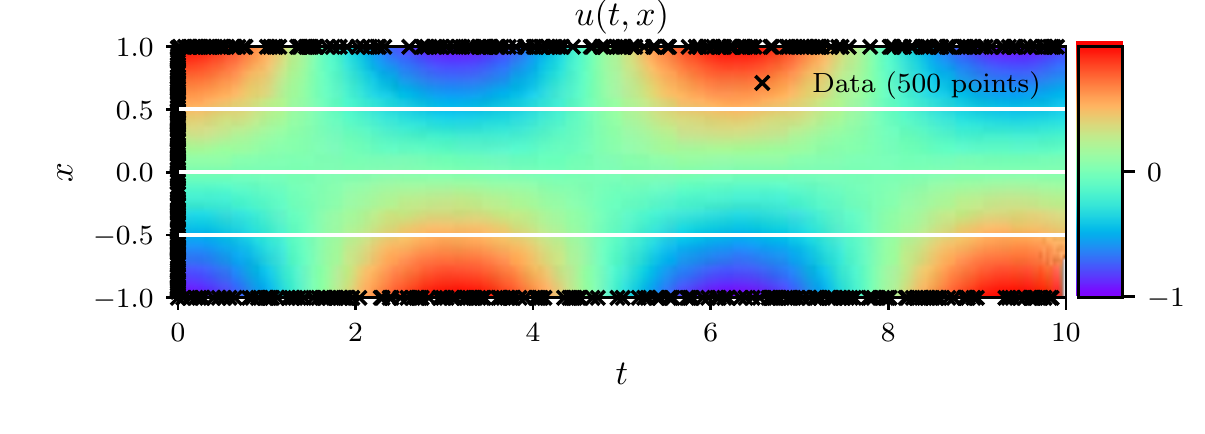}
\caption{Solution of the Klein-Gordon equation (after 1400 iterations) given by PINN using the adaptive activation function with variable $a, n = 5$. The three white horizontal lines indicate the locations where the exact solution is compared with the solution given by PINN in figure \ref{fig:KG333}.}
\label{fig:KG33}
\end{figure}
\begin{figure}
\centering
\includegraphics[trim=0cm 0cm 0cm 4cm, clip=true, scale=1, angle = 0]{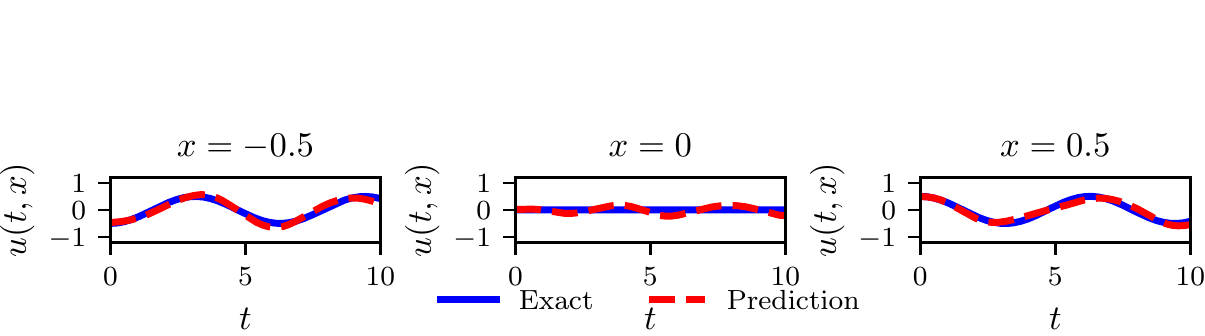}
\includegraphics[trim=0cm 0cm 0cm 4cm, clip=true, scale=1, angle = 0]{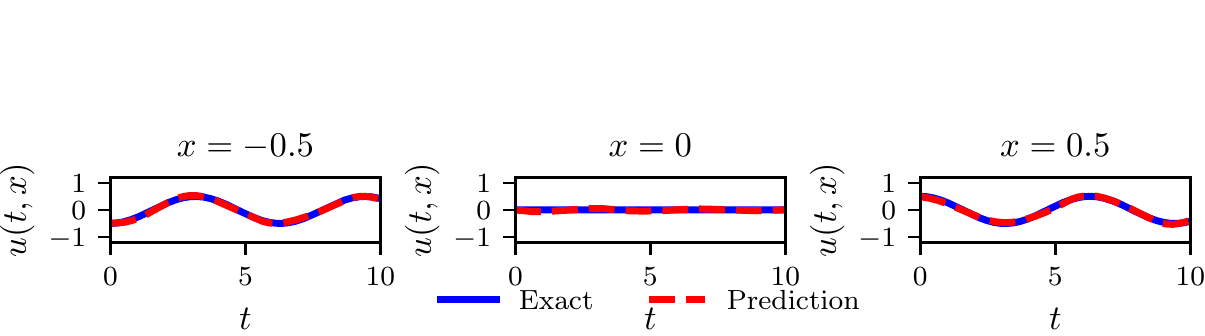}
\includegraphics[trim=0cm 0cm 0cm 4cm, clip=true, scale=1, angle = 0]{KG_T1N5.pdf}
\caption{Klein-Gordon equation: Comparison of the exact solution with the solution given by PINN using $a = 1$ (top), variable $a, n = 1$ (middle) and variable $a, n = 5$ (bottom) obtained after 1400 iterations.}
\label{fig:KG333}
\end{figure}
\begin{figure} 
\centering
\includegraphics[scale=0.65]{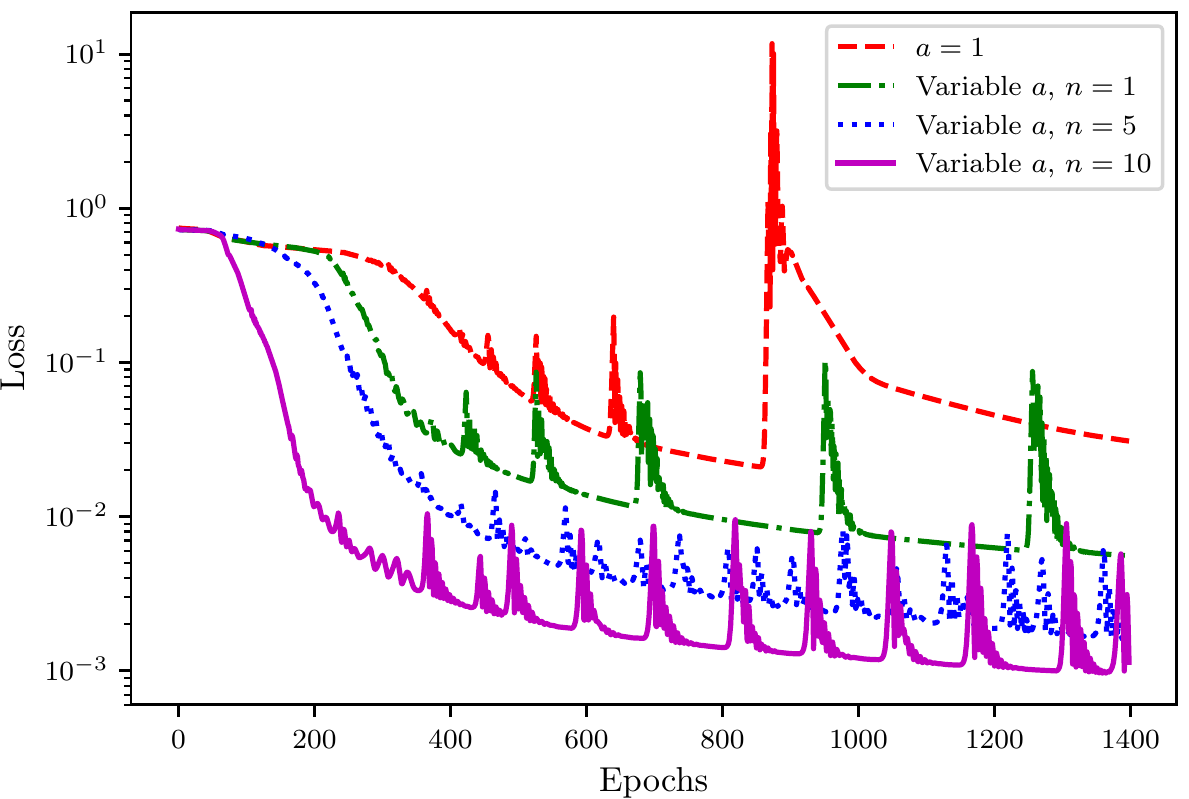}
\includegraphics[scale=0.65]{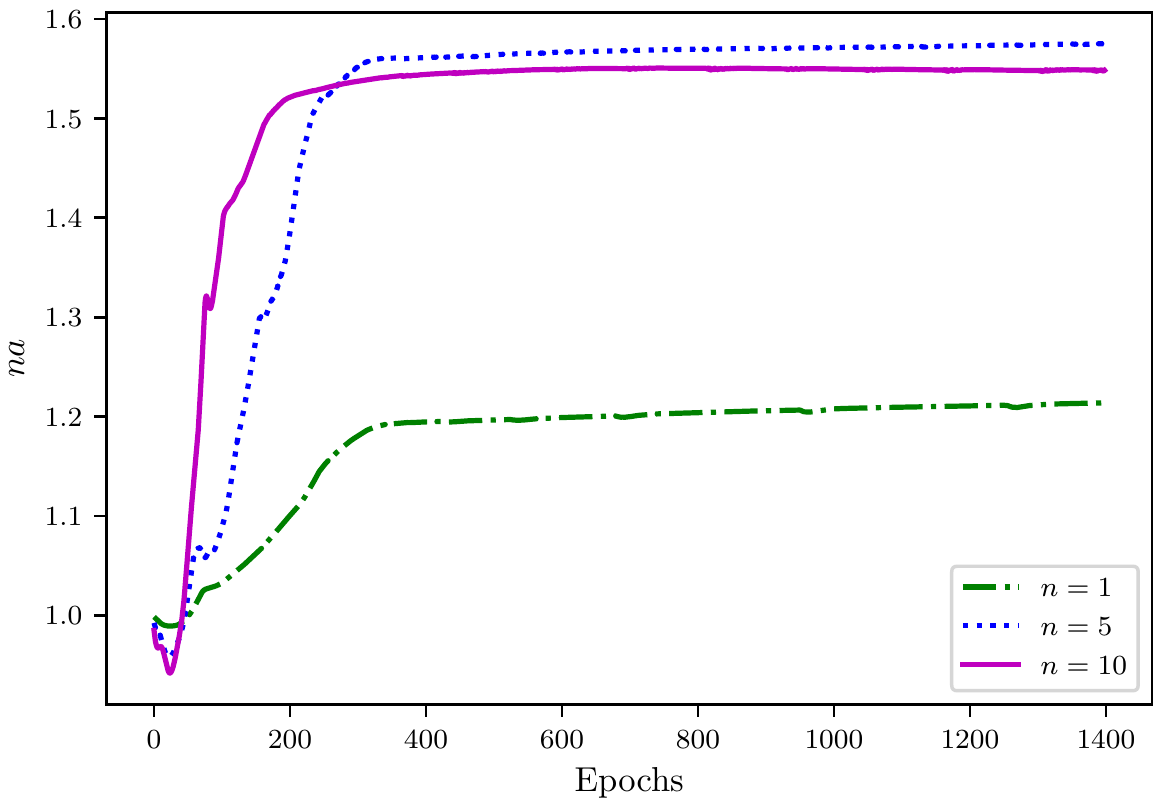}
\caption{Klein-Gordon equation: Loss vs. epochs for the fixed and variable $a$ with different values of $n$ (left) and corresponding variation in $na$ with epochs (right).}
\label{fig:KGmse}
\end{figure}

Now, let us consider the one-dimensional test case where the computational domain is $[-1, \,1] $, the initial conditions are $f(x) = x,~ g(x) = 0$ with $\alpha = -1, \beta = 0, \gamma = 1$,  $k=2$ and $h(x,t) = -x\cos(t) +x^2\cos^2(t)$.
The number of training data points on the boundary is 500, the number of residual training points is 10000 and $$\mathcal{F}\coloneqq u_{tt} - \Delta u_{NN} -N(u_{NN})$$
The neural network architecture used for the computation consists of two hidden layers with 40 neurons in each layer.
Figure \ref{fig:KG33} shows the contour plot of the solution of Klein-Gordon equation and figure \ref{fig:KG333} shows the comparison of exact and PINN solutions with fixed activation function (top) and adaptive activation function with variable $a$ and $n = 1$ (middle) and $n = 5$ (bottom). There is no difference in solution for $n = 5$ and $10$, hence plotted the solution for $n=5$ only. In these figures we can see that the solution is improving with increasing $n$. 
Figure \ref{fig:KGmse} (left) shows the loss function with epochs. Again, convergence is faster for the adaptive activation function with increasing $n$. Also, the value of $a$ converges to optimized value, which is nearly 1.548 with increasing scaling factor as shown in figure \ref{fig:KGmse} (right).
\begin{figure} [htpb] 
\centering
\includegraphics[scale=0.55]{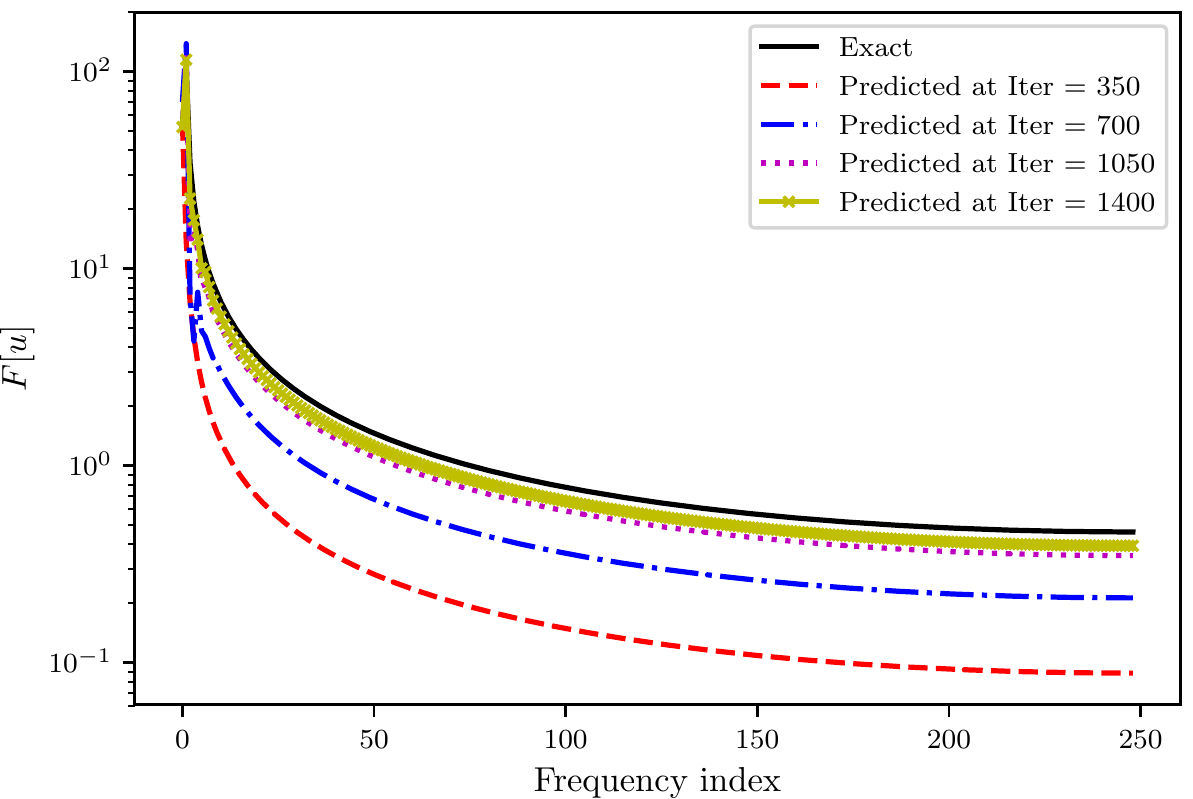}
\includegraphics[scale=0.55]{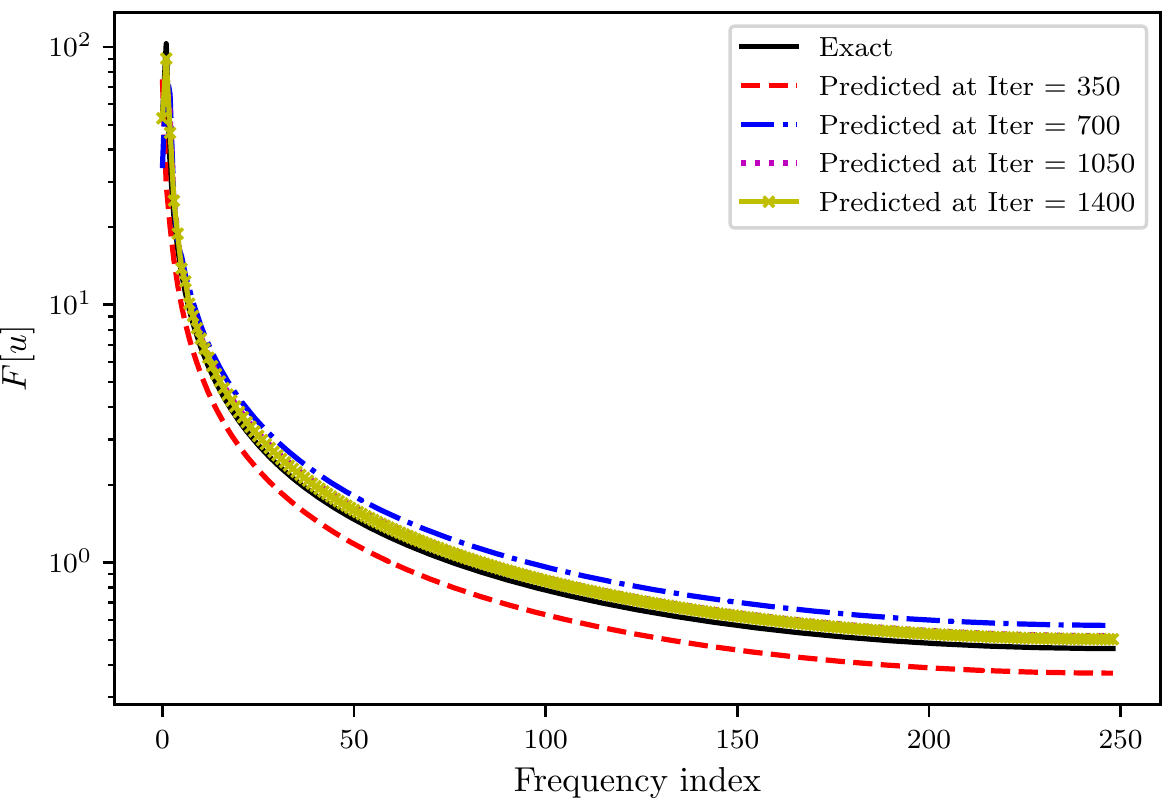}

\includegraphics[scale=0.55]{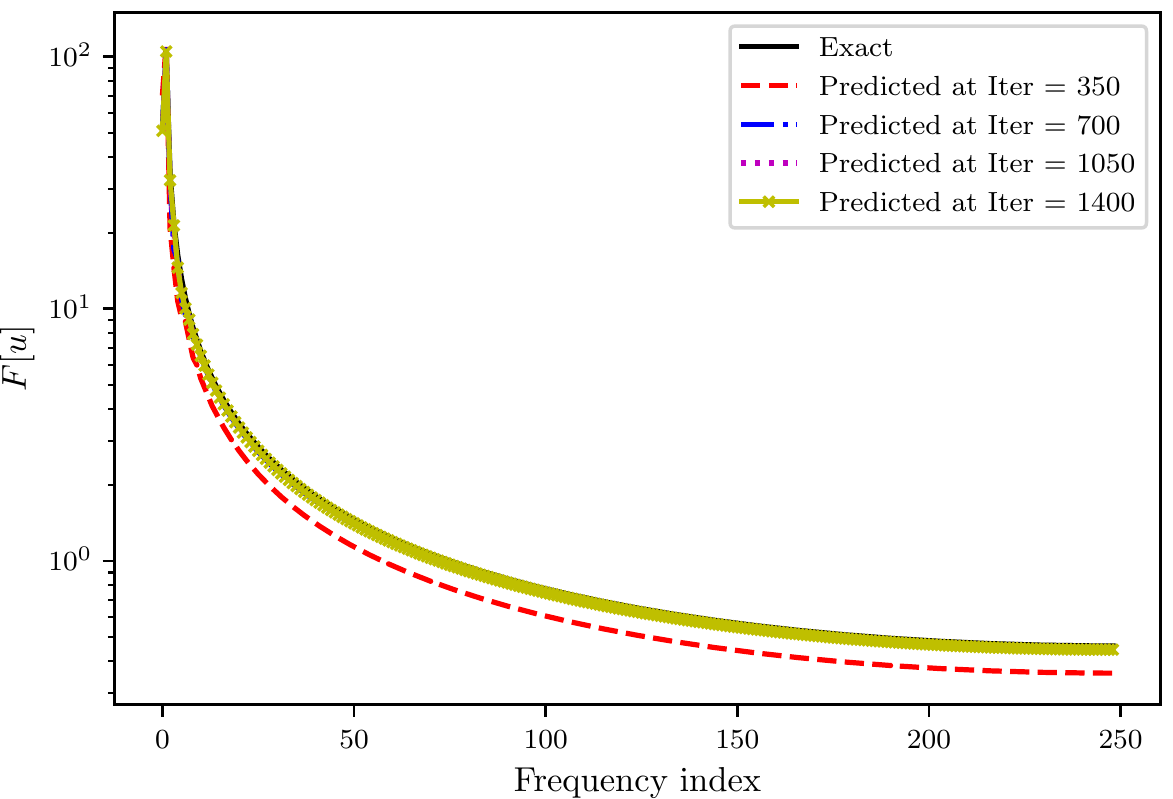}
\includegraphics[scale=0.55]{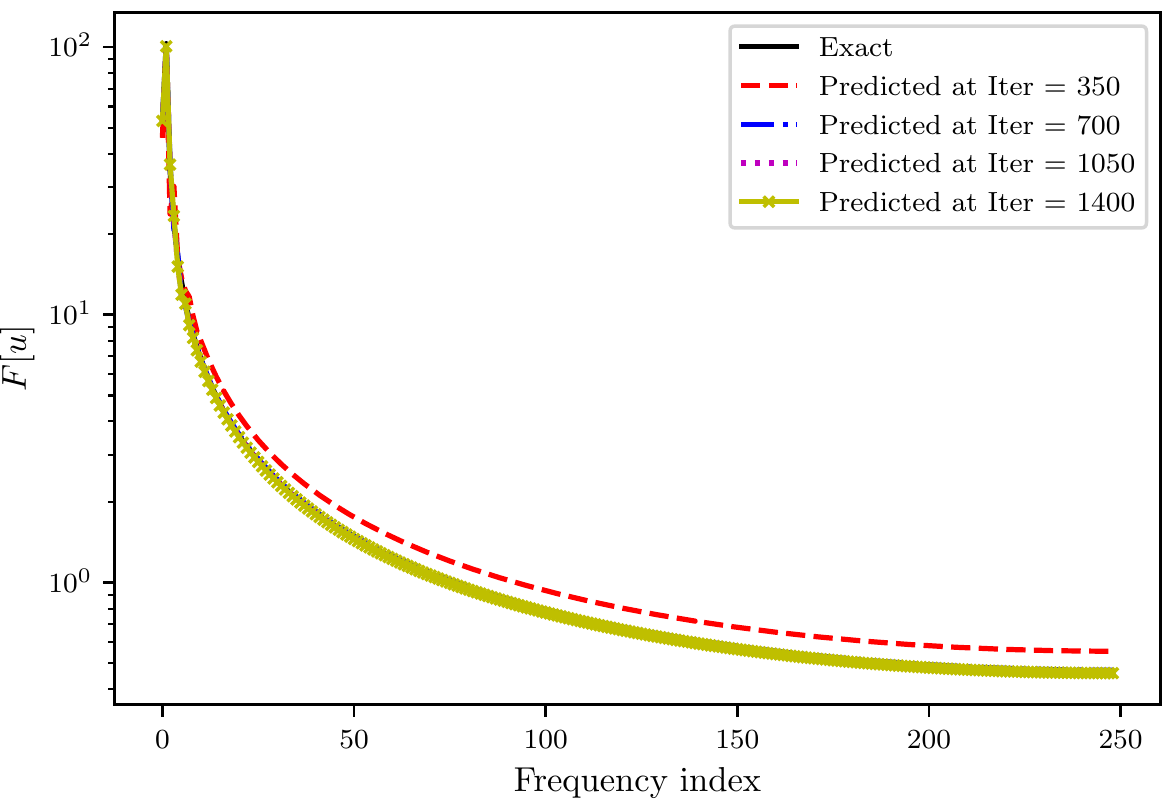}
\caption{Comparison of solution of the Klein-Gordon equation in frequency domain with the fixed (1st row) and variable $a, n = 5$ (2nd row) \textit{'tanh'} activation function. First column shows the frequencies in the solution at $x = -0.5$ whereas second column shows at $x = 0.5$.}
\label{fig:kgT}
\end{figure}
\begin{table}[htpb]
\begin{center}
\small \begin{tabular}{|c|c|c|c|c|} \hline 
 &  $a = 1$&
 Variable $a, (n = 1)$&Variable $a, (n = 5)$ & Variable $a, (n = 10)$
 \\ \hline
 Relative $L_2$ error & 1.953597e-01& 1.026246e-01 & 9.528848e-02 & 9.064256e-02 
 \\
 \hline 
 \end{tabular}
 \vspace{0.2 cm}
\caption{Klein-Gordon equation: Relative $L_2$ error after 1400 iterations with different values of $a$ with clean data.}\label{Table2}
\end{center}
\end{table}

We plotted the solution at $x = -0.5$ and $0.5$ (column-wise) in frequency domain and compared the results of fixed (top row) and adaptive activation (bottom row) functions as shown in figure \ref{fig:kgT}. The adaptive activation function captures all frequencies contained in the solution much faster than the fixed activation function. In both cases, all frequencies are captured in 400 iterations (approximately) by the adaptive activation whereas the fixed activation takes more than 1500 iterations.   

\subsection{Helmholtz equation}
The Helmholtz equation is one of the fundamental equations of mathematical physics arising in many physical problems like vibrating membranes, acoustics, electromagnetism equations, \textit{etc}, see book by Sommerfeld \cite{Som} for more details. In two dimensions it is given by
\begin{equation}\label{2DHel}
\Delta u + k^2 (u) = q(x,y), \ \ (x,y) \in [-1,~1]^2
\end{equation}
with homogeneous Dirichlet boundary conditions. The forcing term is given by
$$q(x,y) = 2\pi\cos(\pi~y)\sin(\pi~x)+ 2\pi\cos(\pi~x)\sin(\pi~y) +\
               (x+y)\sin(\pi~x)\sin(\pi~y) - 2\pi^2(x+y)\sin(\pi~x)\sin(\pi~y)$$
The exact solution for $k=1$ is
$$ u(x,y) = (x+y) \sin(\pi~x)\sin(\pi~y).$$

\begin{figure} [htpb] 
\centering
\includegraphics[scale=0.65]{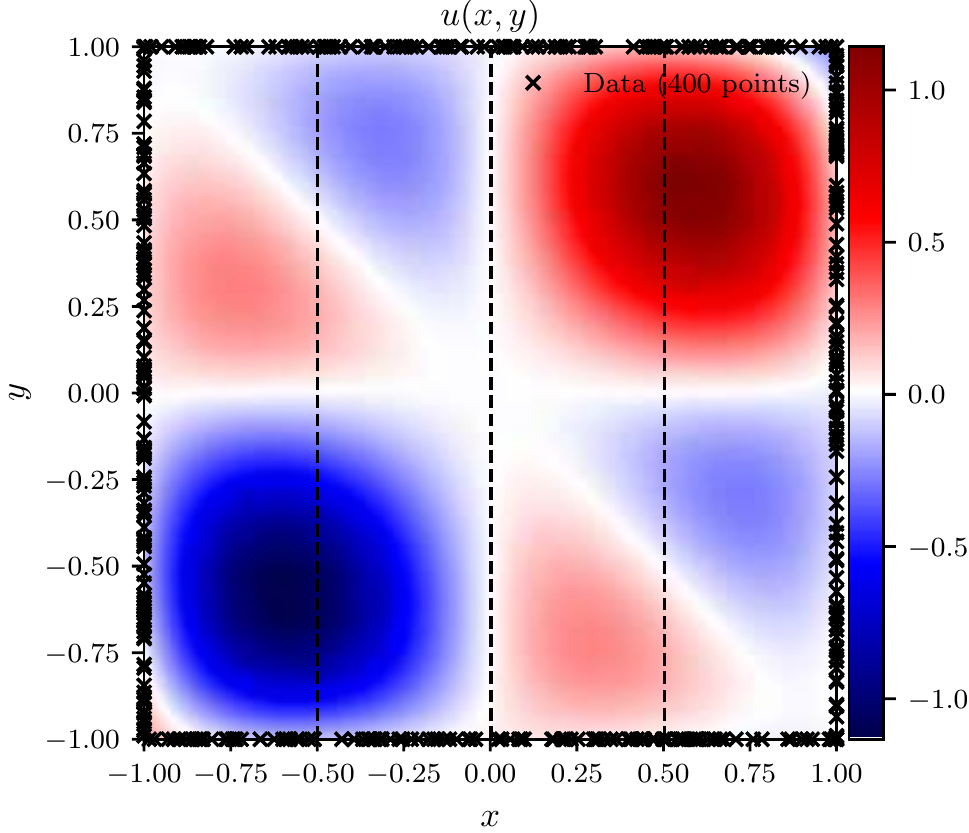}

\includegraphics[scale=0.75]{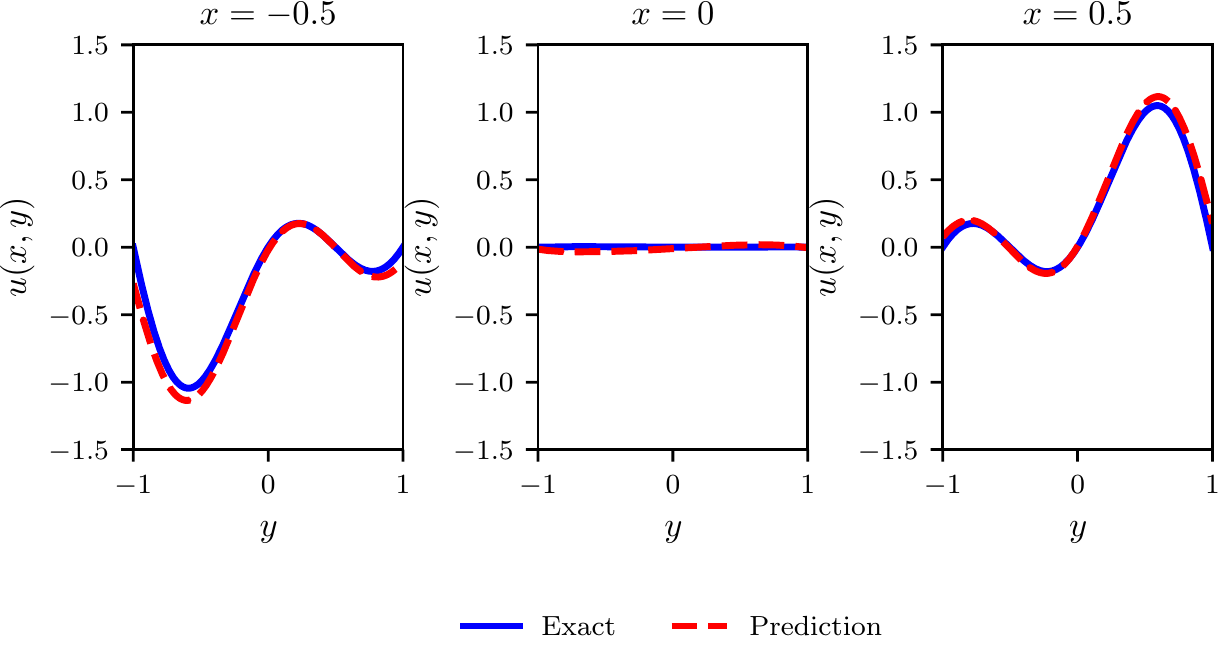}
\includegraphics[scale=0.75]{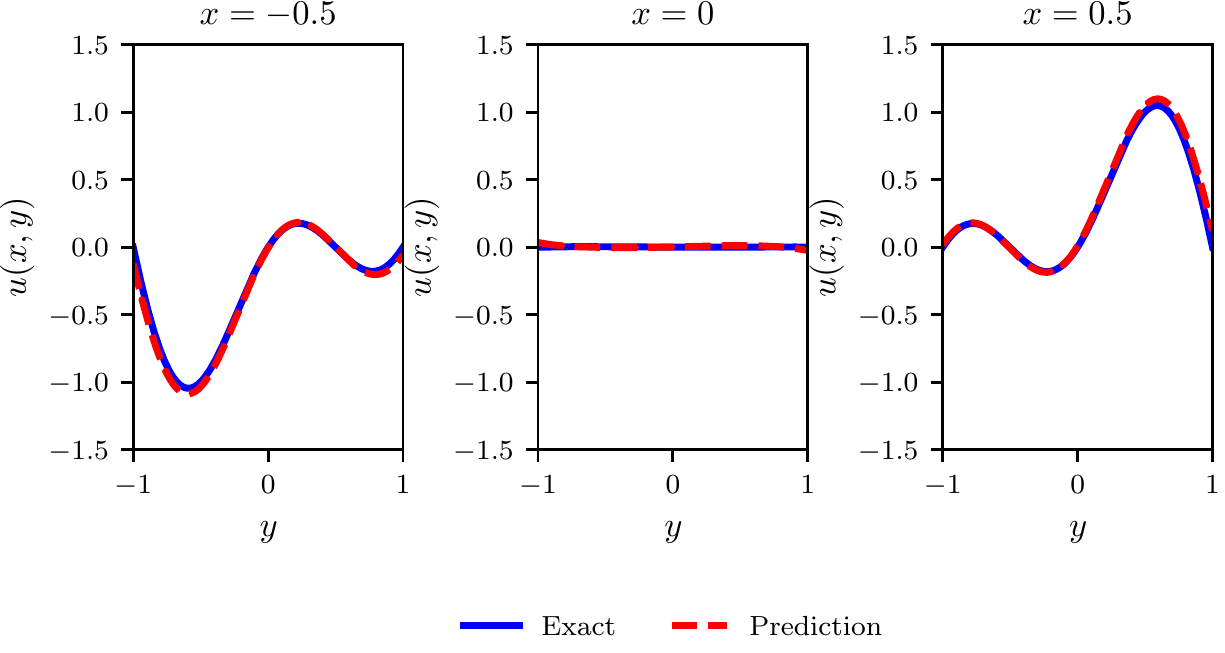}
\caption{Contour plot (top) shows the solution of Helmholtz equation using the adaptive activation function. The middle and bottom rows compare the PINN solution with exact solution using the fixed (middle) and variable $a, n = 10$ (bottom) 'tanh' activation function, respectively after 3600 iterations.}
\label{fig:Helm1}
\end{figure}
\begin{figure} [htpb] 
\centering
\includegraphics[scale=0.54]{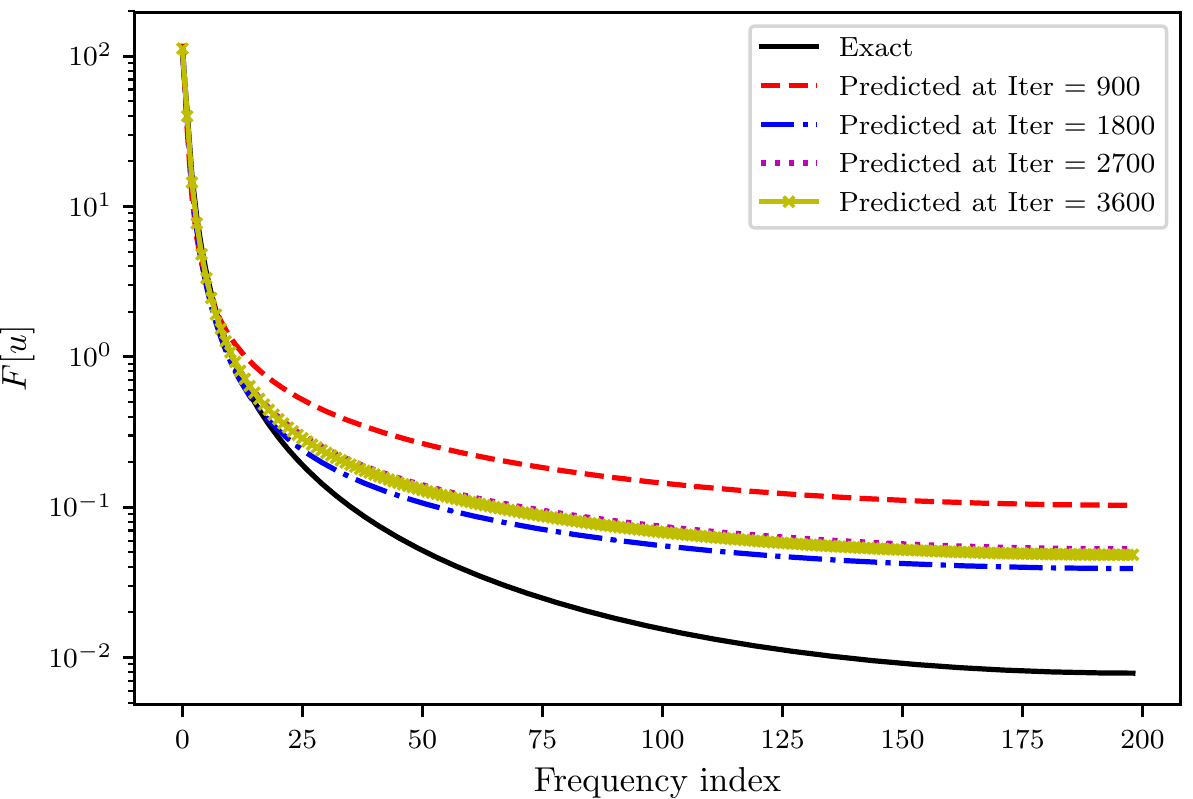}
\includegraphics[scale=0.54]{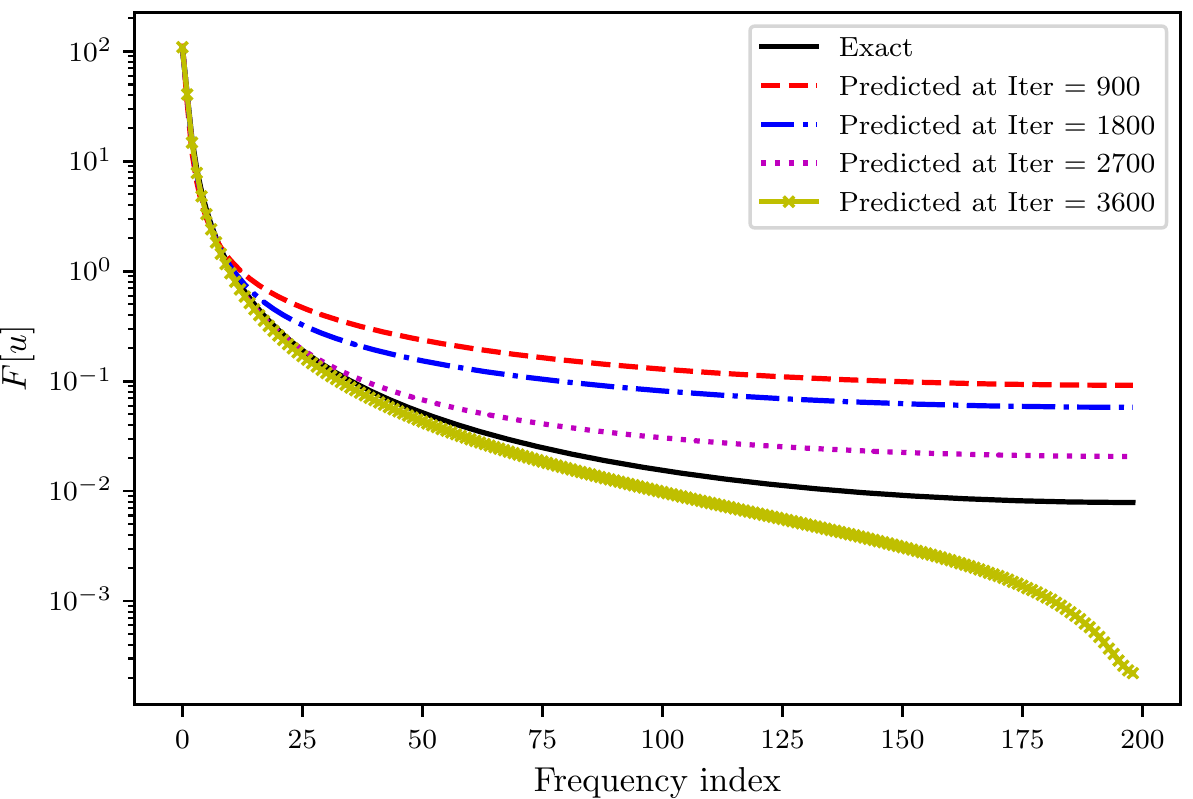}

\includegraphics[scale=0.54]{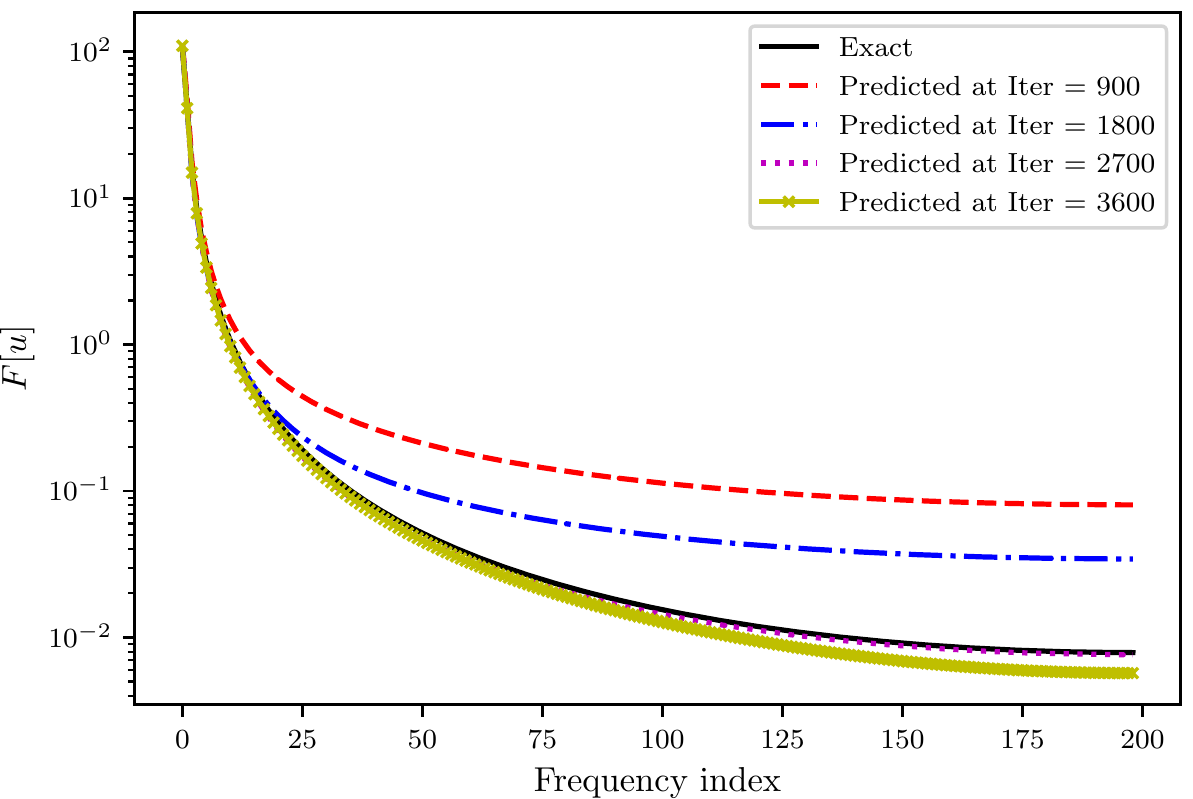}
\includegraphics[scale=0.54]{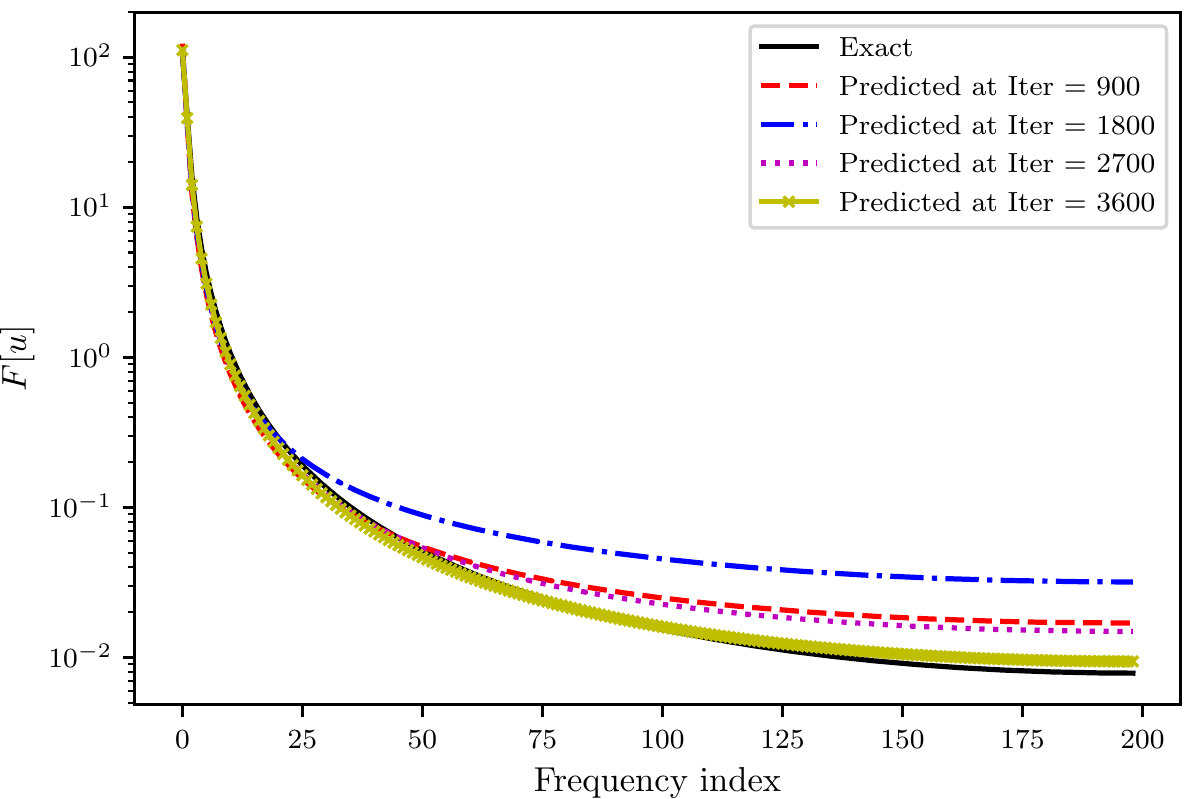}
\caption{Comparison of solution of Helmholtz equation in frequency domain using fixed (1st row) and variable $a, n = 10$ (2nd row) 'tanh' activation function. First column shows the frequencies in the solution at $x = -0.5$ location whereas second column shows at $x = 0.5$ location. }
\label{fig:Helm2}
\end{figure}

\begin{figure} [htpb] 
\centering
\includegraphics[scale=0.65]{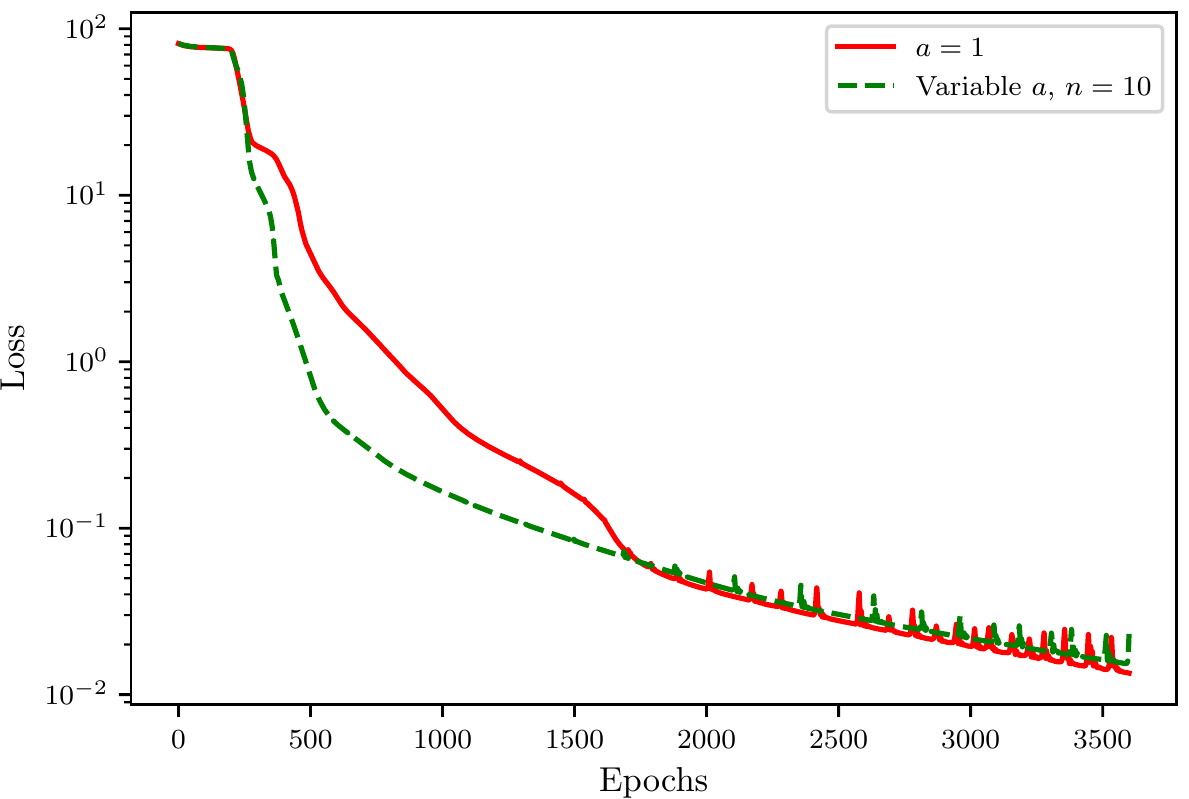}
\includegraphics[scale=0.65]{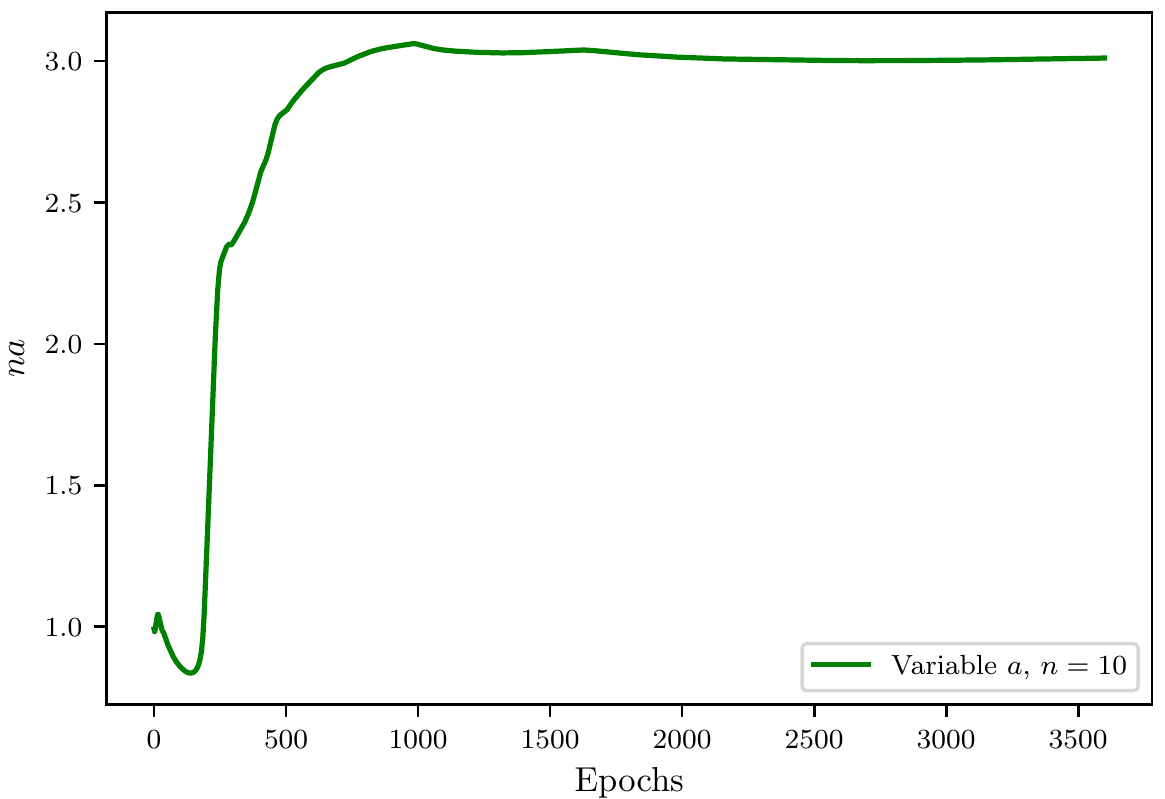}
\caption{Helmholtz equation: (Left) loss function variation with epochs for fixed $(a=1)$ and (right) adaptive activation ($n=10$) functions with variation in $a$.}
\label{fig:Helm3}
\end{figure}
In this case, the number of hidden layers is two and in each layer 40 neurons are used. The residual training points are 16000 whereas the $N_u$ points are 400.
Figure \ref{fig:Helm1} shows the comparison of exact solution with that of PINN solution using fixed and adaptive $(n=10)$ activation functions at three different location given by the black dash line in the contour plot. The relative $L_2$ error calculated at the end of 3600th iteration is 1.0591e-1 using the fixed activation function and 7.1945e-2 in the case of adaptive activation. We can also see the solution in the frequency domain given by figure \ref{fig:Helm2} using fixed (1st row) and variable $a, n = 10$ (2nd row) \textit{'tanh'} activation function. We can observe that the adaptive activation function captures frequencies faster than the fixed one in both locations $x = -0.5$ (first column) and $x = 0.5$ (second column). Finally, figure \ref{fig:Helm3} (left) shows the loss function comparison where we see the fast convergence of the adaptive activation as compared to the fixed activation, and figure \ref{fig:Helm3} (right) shows the variation of $a$ with number of iterations where the optimal value of $a$ is close to 3.

\subsection{Inverse problem for two-dimensional sine-Gordon equation}
An inverse problem is formulated as estimating the function $u \in \mathcal{U}$ from the data $s \in \mathcal{S}$  where $s = T(u) + noise$. Here $\mathcal{U, S}$ are topological vector spaces and $T:\mathcal{U}\rightarrow \mathcal{S}$ gives a mapping of a given function which gives rise to data in the absence of noise.
Machine learning when applied to inverse problems can be framed as the problem of reconstructing a nonlinear mapping $T^*: \mathcal{S}\rightarrow \mathcal{U}$ such that it satisfies the pseudo-inverse property as
$T^*(s) \approx u$ whenever data $s$ is related to $u$. An important step in machine learning approaches is to parameterize the pseudo-inverse operator, and then the learning phase refers to choosing optimal parameters using some training data by minimizing a suitably defined loss function which gives the learned pseudo-inverse operator $T^*_{\tilde{\boldsymbol{\Theta}}}$. The training data are \textit{iid} realizations of a $\mathcal{U}\times \mathcal{S}$-valued random variable $(u,s)$ with known probability density function. This data is rich enough to allow a machine learning scheme to identify the structure of nonlinear mapping $T$ given by the governing physical laws.

There are various approches proposed in the literature for a data-driven discovery of governing differential equations, for example, Raissi and Karniadakis \cite{RK} as well as Raissi et al. \cite{RK3} in the context of Gaussian process, Rudy et al \cite{Rud} proposed a sparse regression which is based on library of candidate terms and sparse model selection to select the important terms involved in the governing equation and recently by Berg and Nystr\"{o}m \cite{Berg}.
One of the efficient machine learning based approach for solving an inverse problem given by Raissi, et al. \cite{RK1} is the data-driven discovery of partial differential equations by writing the governing equation 
as
\begin{equation}\label{invP}
u_{tt} +  \mathcal{N}[u;\lambda] = 0
\end{equation}
where $\mathcal{N}[\cdot]$ contains parameterized linear/nonlinear terms. The network's job is to identify the unknown parameters $\lambda = \{\lambda_1,\lambda_2,\cdots \}$ as well as to obtain a qualitatively accurate reconstruction of the given solution. Here, we shall consider a two-dimensional sine-Gordon test case to solve the inverse problem where $N(u) = \phi ~\sin(u)$ . In this test case $\phi = -1$ and the domain is $-7 \leq x ,y \leq 7$. The initial conditions are given as
\begin{align*}
f(x,y) & = 4 \,\tan^{-1} (\text{exp}(x+y)) 
\\g(x,y) & = \frac{4 (\text{exp}(x+y))}{1+ (\text{exp}(2x+2y))}
\end{align*}
and boundary conditions are 
\begin{align*}
\frac{\partial u}{\partial x} & = \frac{4 (\text{exp}(x+y+t))}{(\text{exp}(2t)+ \text{exp}(2x+2y))} \ \ \text{for}\, x = \pm 7 \,\,\text{and}\,\,  y \in [-7, \,7], \,\, t\geq 0
\\ \frac{\partial u}{\partial y} & = \frac{4 (\text{exp}(x+y+t))}{(\text{exp}(2t)+ \text{exp}(2x+2y))} \ \  \text{for}\, y = \pm 7 \,\,\text{and}\,\, x \in [-7, \,7], \,\, t\geq 0
\end{align*}
which has the analytical solution
$$ u(x,y,t) = 4 \,\tan^{-1} (\text{exp}(x+y-t)). $$
\begin{figure} [htpb] 
\centering
\includegraphics[scale=0.42]{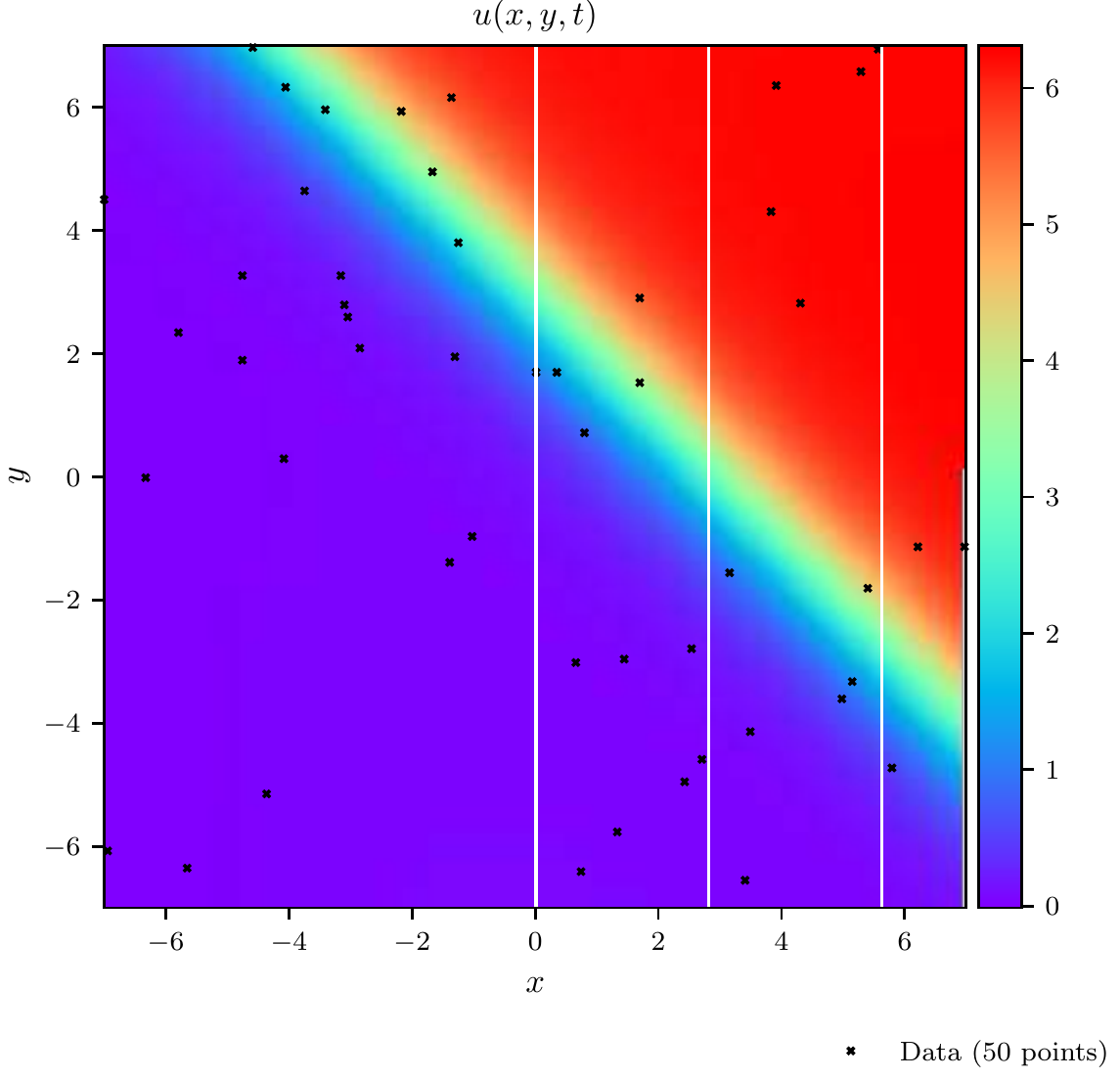}
\includegraphics[scale=0.42]{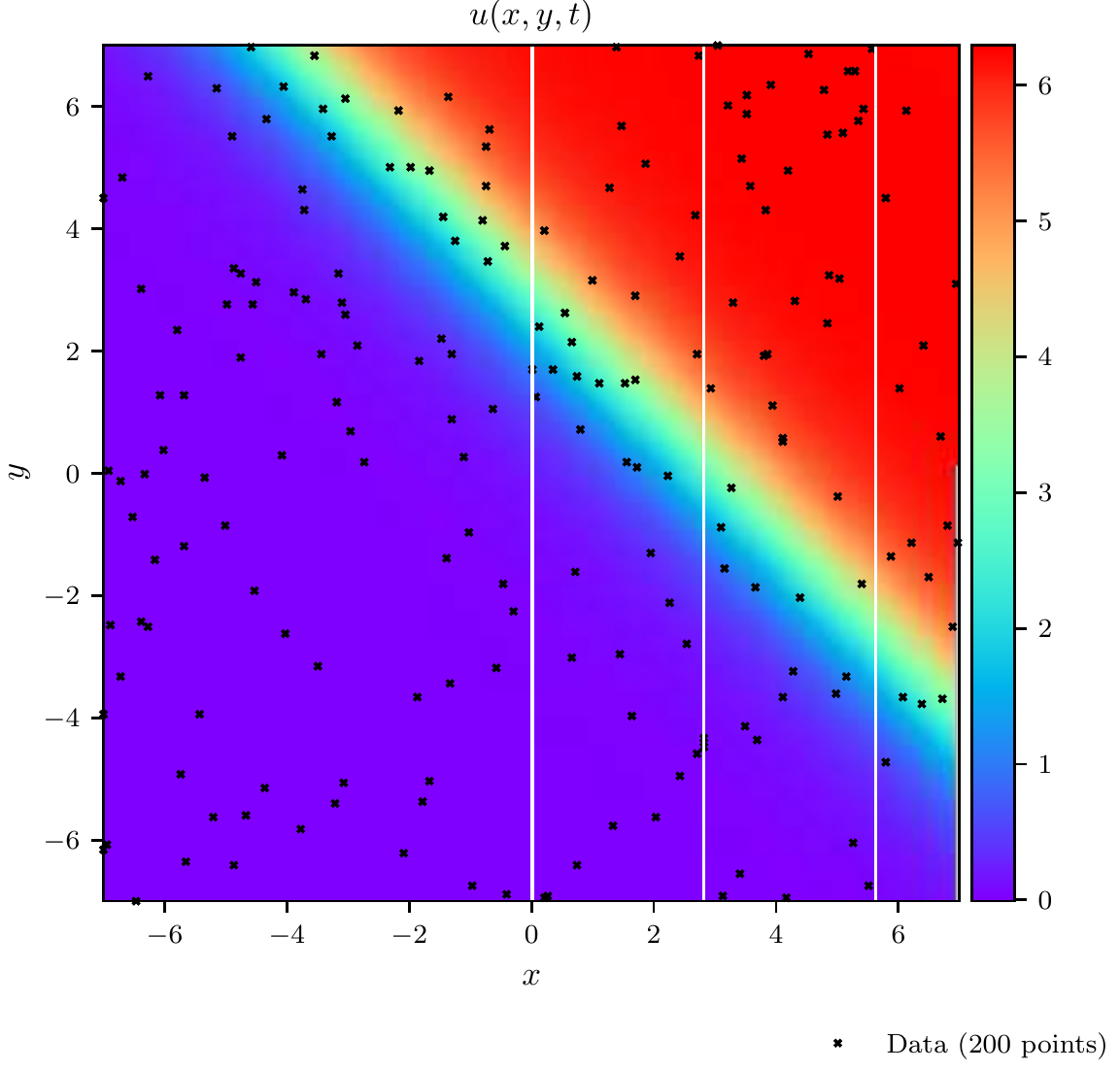}
\includegraphics[scale=0.42]{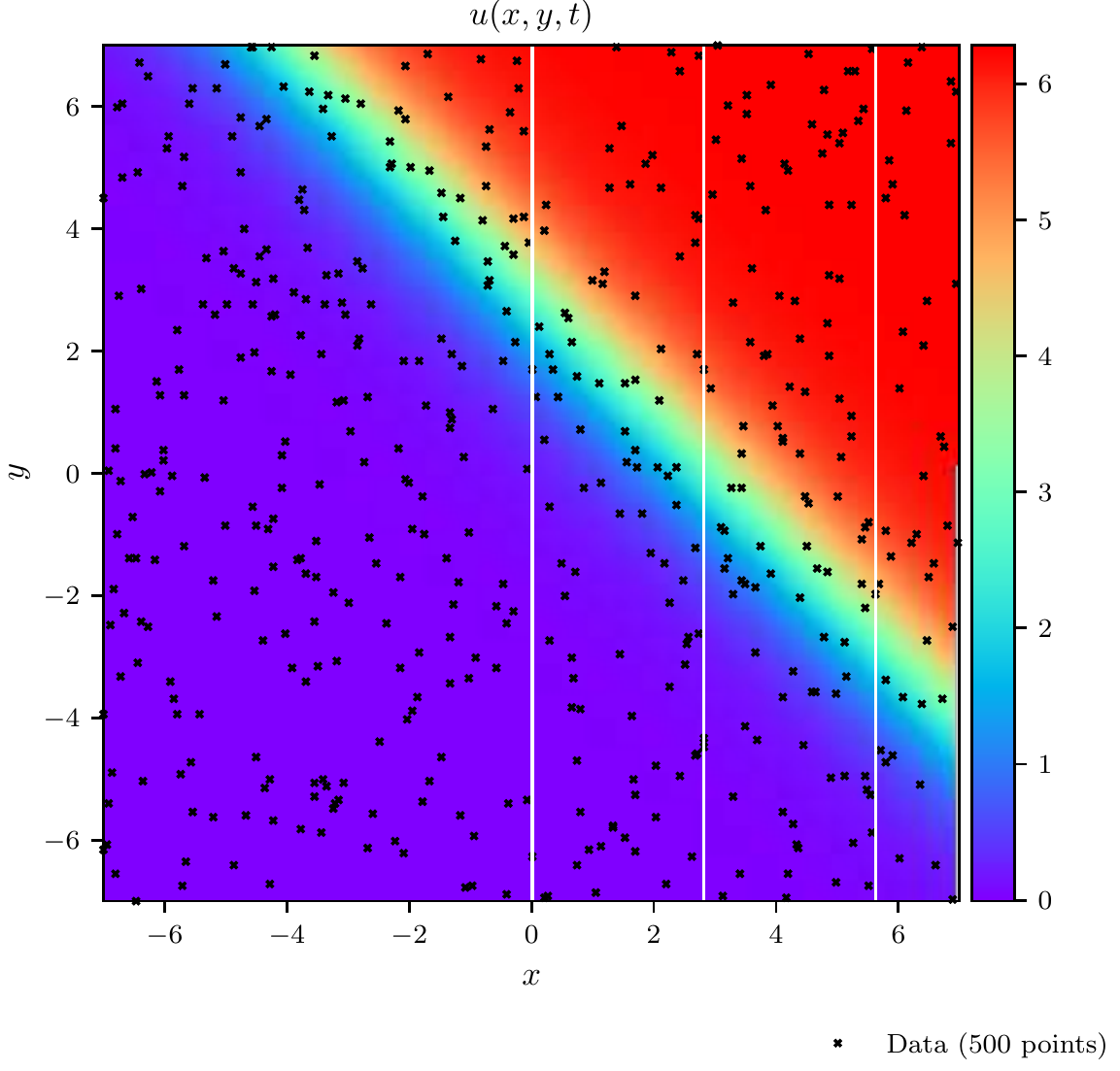}

\includegraphics[scale=0.72]{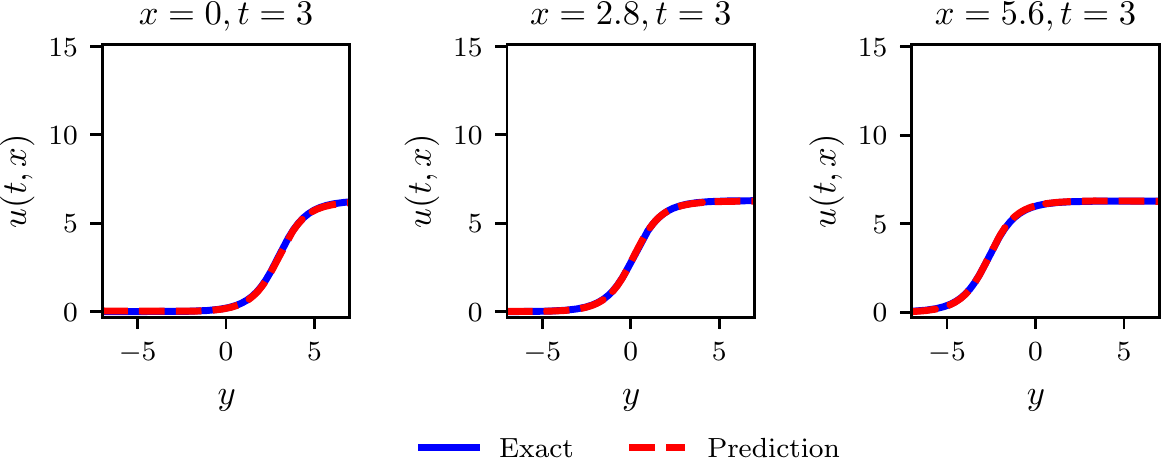}
\includegraphics[scale=0.72]{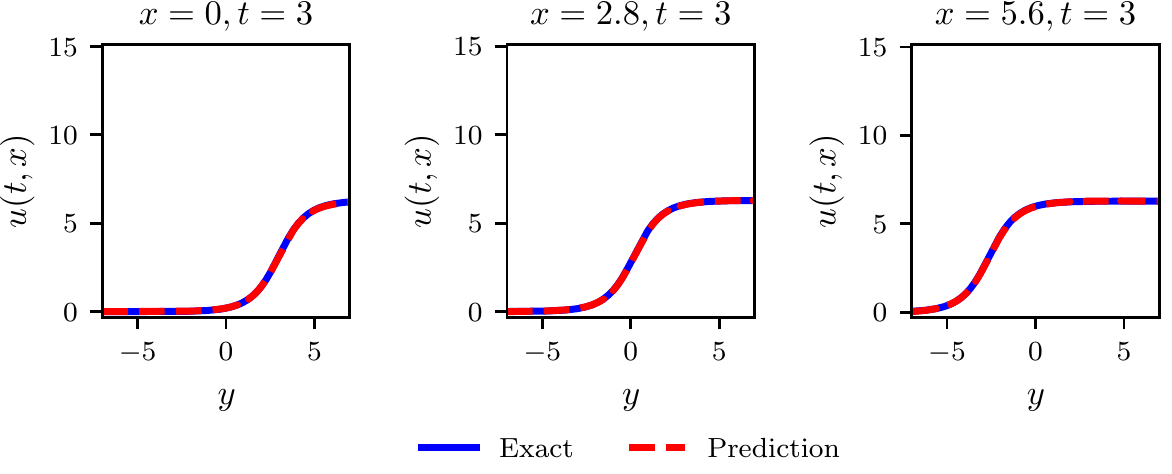}
\includegraphics[scale=0.72]{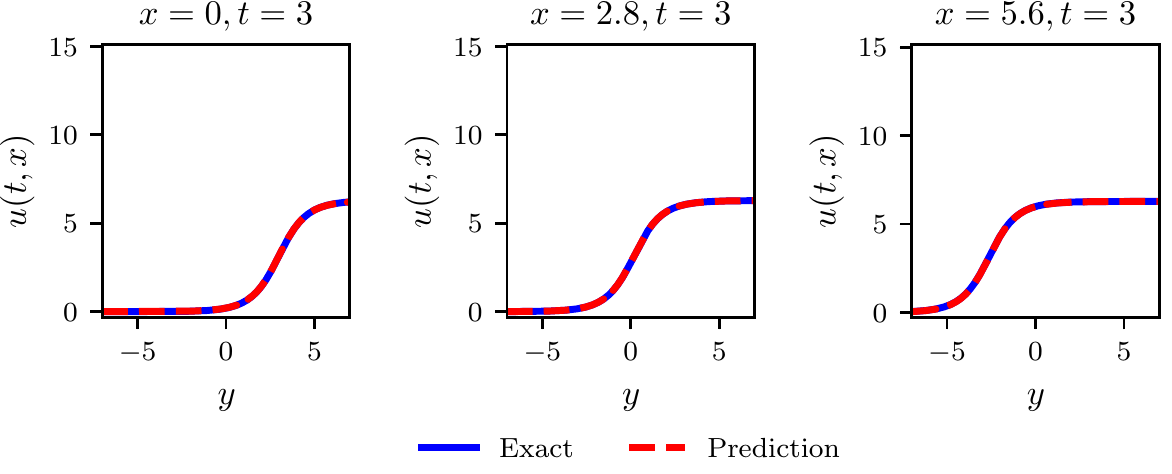}
\caption{Contour plots of the PINN solution for two-dimensional sine-Gordon equation with adaptive activation function ($n = 5$) using $N_u = 50, 200$ and 500 (top row). The second, third and fourth rows compare the PINN solution with exact one using $N_u = 50, 200$ and 500, respectively.}
\label{fig:sGidenO}
\end{figure}
In the case of the inverse problem, the loss function is the same as given by equation \eqref{loss}
but the MSE is given by
\begin{align*}
MSE_{\mathcal{F}} &=  \frac{1}{N_u}\sum_{i=1}^{N_u} |\mathcal{F}(x^i_u,y^i_u,t^i_u; \lambda_1,\lambda_2,\cdots)|^2, ~~~~~ MSE_u = \frac{1}{N_u}\sum_{i=1}^{N_u} |u^i - u(x^i_u,y^i_u,t^i_u)|^2.
\end{align*}
Here $\{x_u^i,y_u^i, t^i_u\}_{i=1}^{N_u}$ denotes the training data points from the boundary as well as from inside the domain. The loss $MSE_u$ corresponds to training data on the solution $u(x,y,t)$ whereas $MSE_{\mathcal{F}}$ enforces the governing equation on the same set of training data points.
For our analysis, 50, 200 and 500 training data points, which are randomly selected points are used.
The neural network architecture consists of four hidden layers with 20 neurons in each layer.

\begin{table}[h!]
\begin{center}
\begin{tabular}{| *{10}{c|} }
    \hline
    & \multicolumn{3}{c|}{$N_u = 50$}
            & \multicolumn{3}{c|}{$N_u = 200$}
                    & \multicolumn{3}{c|}{$N_u = 500$} \\
    \hline
    
    & Clean & 1\% noise& 2\% noise& Clean  & 1\% noise& 2\% noise& Clean & 1\% noise& 2\% noise \\ \hline
$\lambda_1$  & 3.28550&1.439983&3.218234   &    2.50646&0.831884&3.420329   &   1.04115 &   0.297201  &   1.896238\\
    \hline
$\lambda_2$   & 1.15104&2.333641&2.443504    &   2.04701&2.199578&1.212937  &   1.24181  &   1.380813  & 0.351852   \\
    \hline
$\lambda_3$   &0.50759&0.120240 &0.495362     &      2.61289&1.729769&3.965372      &  0.64331     &0.680488       &  1.020026    \\
    \hline
$\lambda_4$   &1.08709&1.030558 &1.178467       &    1.77241&1.530927&1.755536      &  0.92602     &1.200277       &   1.376885    \\
    \hline
\end{tabular}
\vspace{0.2 cm}
\caption{Sine-Gordon equation: Percentage relative $L_2$ error in $\lambda_1=1, \lambda_2=1, \lambda_3=1$ and $\lambda_4=4$ for different $N_u$ points using the adaptive activation function after 3500 iterations.}\label{Table301}
\end{center}
\end{table}

\begin{table}[h!]
\begin{center}
\begin{tabular}{| *{7}{c|} }
    \hline
    & \multicolumn{3}{c|}{$N_u = 500$ (fixed activation)}
                    & \multicolumn{3}{c|}{$N_u = 500$ (adaptive activation)} \\
    \hline
    
    & Clean & 1\% noise& 2\% noise & Clean & 1\% noise& 2\% noise \\ \hline
$\lambda_1$  &     6.25785&10.22437& 7.876128   &   1.04115 &   0.297201  &   1.896238\\
    \hline
$\lambda_2$   &       6.22453&7.21347&5.720103  &   1.24181  &   1.380813  & 0.351852   \\
    \hline
$\lambda_3$   &          8.99409&3.74553&9.965611      &  0.64331     &0.680488       &  1.020026    \\
    \hline
$\lambda_4$   &         3.55075&4.54000&3.60425     &  0.92602     &1.200277       &   1.376885    \\
    \hline
\end{tabular}
\vspace{0.2 cm}
\caption{Sine-Gordon equation: Comparison of percentage relative $L_2$ error in $\lambda_1=1, \lambda_2=1, \lambda_3=1$ and $\lambda_4=4$ obtained after 3500 iterations for the fixed and the adaptive activation functions using $N_u = 500$ points.}\label{Table302}
\end{center}
\end{table}

\begin{table}[h!]
\begin{center}
\small \begin{tabular}{|c|c|} \hline
 Correct PDE & $4~s(x,y) - u_{xx}-u_{yy}+ sin(u) = 0$
\\ \hline
Identified PDE (Clean data)& $(3.85797)~s(x,y) - (0.93743)~u_{xx}-(0.93776)~u_{yy}+ (0.9100591)~ \sin(u) = 0$
\\ \hline 
Identified PDE (1 \% noise)& $(3.81840)~s(x,y) - (0.89776)~u_{xx}-(0.92790)~u_{yy}+ (0.9625447)~ \sin(u) = 0$
\\ \hline
Identified PDE (2 \% noise)& $(3.85583)~s(x,y) - (0.92124)~u_{xx}-(0.94280)~u_{yy}+ (0.9003439)~ \sin(u) = 0$
\\ \hline
 \end{tabular}
\vspace{0.2 cm}
\caption{Sine-Gordon equation: Identification of two-dimensional sine-Gordon equation with the \textit{fixed} activation function after 3500 iterations.}\label{Table3}
\end{center}
\end{table}

\begin{table}[h!]
\begin{center}
\small \begin{tabular}{|c|c|} \hline
 Correct PDE & $4~s(x,y) - u_{xx}-u_{yy}+ sin(u) = 0$
\\ \hline
Identified PDE (Clean data)& $(3.96296)~s(x,y) - (1.01041)~u_{xx}-(0.98759)~u_{yy}+ (1.03064331)~ \sin(u) = 0$
\\ \hline 
Identified PDE (1 \% noise)& $(3.95199)~s(x,y) - (1.00297)~u_{xx}-(0.98619)~u_{yy}+ (0.9931951)~ \sin(u) = 0$
\\ \hline
Identified PDE (2 \% noise)& $(3.94492)~s(x,y) - (0.98104)~u_{xx}-(0.99648)~u_{yy}+ (0.9897997)~ \sin(u) = 0$
\\ \hline
 \end{tabular}
\vspace{0.2 cm}
\caption{Sine-Gordon equation: Identification of two-dimensional sine-Gordon equation using the \textit{adaptive} activation function with scaling parameter $n = 5$ after 3500 iterations.}\label{Table4}
\end{center}
\end{table}

\begin{figure} [htpb] 
\centering
\includegraphics[scale=0.66]{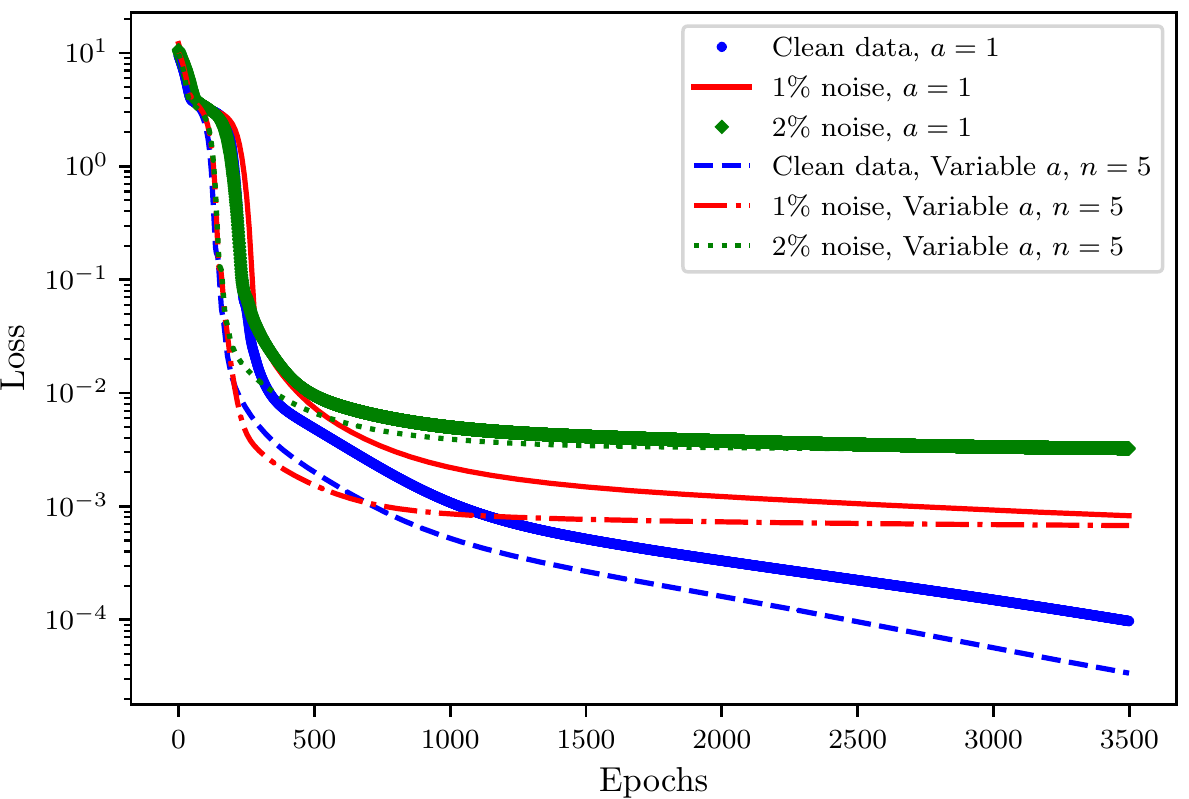}
\includegraphics[scale=0.66]{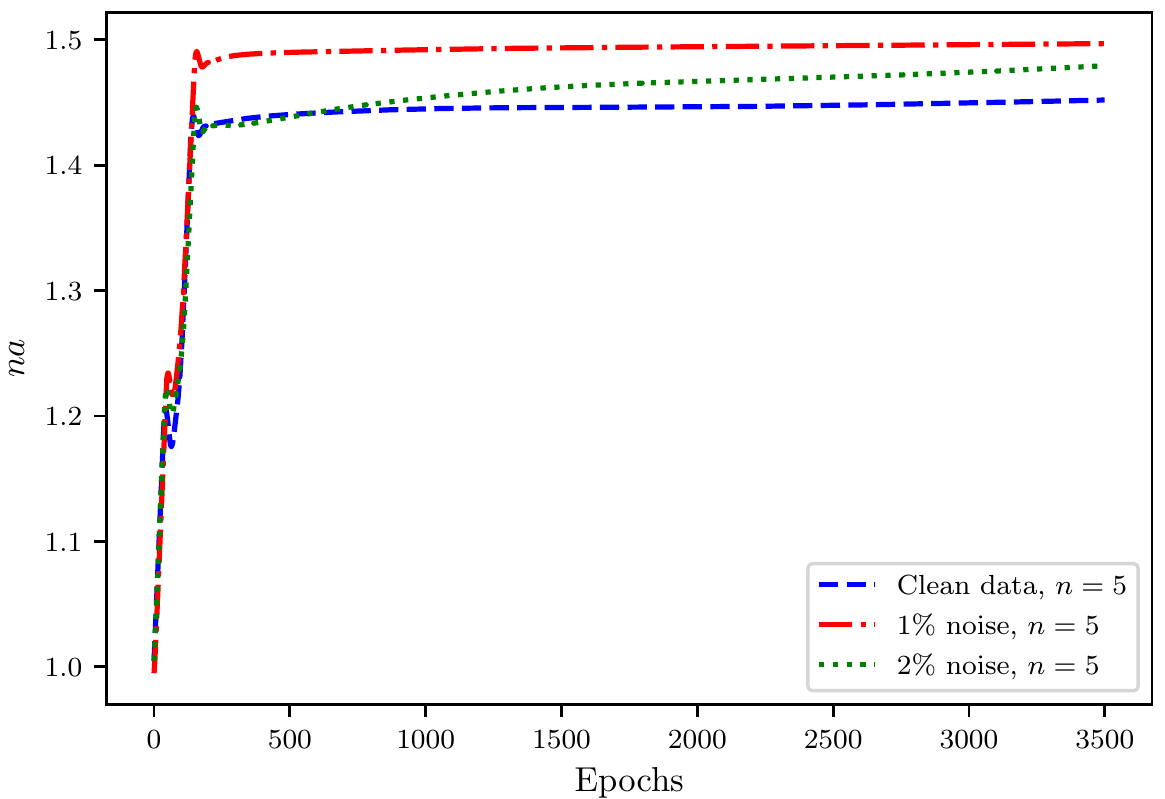}
\caption{Sine-Gordon equation: Loss function (left) and variation of $na$ (right) for clean data as well as data with noise using $N_u = 500$.}
\label{fig:sGiden}
\end{figure}
\begin{table}[h!]
\begin{center}
\small \begin{tabular}{|c|c|c|c|} \hline
 &   Clean data&
1 \% error&
2 \% error
 \\ \hline
Fixed activation function & 1.923454e-03&
1.996804e-03&
2.396984e-03 
 \\
 \hline 
Adaptive activation function with $n = 5$ & 9.93412e-04&
1.420345e-03&
1.986455e-03 
 \\
 \hline
 \end{tabular}
\vspace{0.2 cm}
\caption{Sine-Gordon equation: Comparison of relative $L_2$ error in the solution obtained after 3500 iterations for the fixed and the adaptable activation functions.}\label{Table5}
\end{center}
\end{table}

For careful scrutinization of the performance of the proposed approach, we vary the noise level in the training data. Given the noisy measurements of the data $(x^i,y^i, t^i, u^i)$, we are interested in learning the parameters $\lambda = \{\lambda_1,\lambda_2,\cdots \}$.
We define $\mathcal{F}$ as
$$\mathcal{F}\coloneqq \lambda_4 ~s(x,y) - \lambda_1u_{xx} -\lambda_2u_{yy} + \lambda_3 \sin(u)$$
where the function $s(x,y)$ is given by
$$s(x,y) = \frac{e^{9+x+y}-e^{3+3(x+y)}}{\left(e^6+e^{2(x+y)}\right)^2}$$ 
and $\lambda_1 = \lambda_2 = \lambda_3 = 1, \lambda_4 = 4$.

Table \ref{Table301} shows the percentage relative $L_2$ error in $\lambda$'s for adaptive activation function using 50, 200 and 500 training data points. We observe that the error decreases with increase in training points. Moreover, in some $\lambda$'s the error in the solution with $1\%$ noise is slightly less than that of clean data. Figure \ref{fig:sGidenO} shows the solution of PINN using adaptive activation function for different $N_u$ points, which accurately approximates the solution in the domain. Also, comparison of the PINN solution with the exact solution is considered at $x = 0, 2.8$ and 5.6 locations. Table \ref{Table302} shows the comparison of percentage relative $L_2$ error in $\lambda$'s using fixed and adaptive activation function after 3500 iterations. In the case of a fixed activation the maximum error is almost 9\% in clean data whereas it is just 1.24\% in adaptive activation function. This shows that the adaptive activation increases the accuracy of the solution.
Table \ref{Table3} shows the correct PDE along with the identified PDEs with clean data, 1\% and 2\% noise using the fixed activation function whereas table \ref{Table4} shows the corresponding PDEs using the adaptive activation function, which is more accurate than its fixed counterpart. From the results we observe that the neural network is able to correctly identify the unknown parameters involved in the governing equations with very good accuracy even in the presence of noise in the training data.
To quantify the accuracy we have given the relative $L_2$ error in table \ref{Table5}, which is again small for the adaptive activation function based PINN. Finally, the loss function using fixed and adaptive activation function is plotted for the three different data sets as shown in figure \ref{fig:sGiden}, which shows that the loss given by adaptive activation function based PINN converges faster.

\section{Conclusions}
Increasing the performance of deep learning algorithm is important in order to design fast and accurate machine learning techniques. By introducing the scalable hyper-parameter in the activation function, not only convergence of the neural network increases but also better accuracy is obtained. Thus, we can achieve a better performance of the neural network by the introduction of such parameter. To support our claim, various forward and inverse problems are solved using deep neural networks and physics-informed neural networks with smooth solution (given by Klein-Gordon equation, Helmholtz equation) as well as high gradient solution (given by the Burgers equation) in one and two dimensions. In all cases, it is shown that the decay in loss function is faster in the case of adaptive activation function, and correspondingly the relative $L_2$ error in solution is shown to be small in the proposed approach. In order to investigate the performance of such scalable hyper-parameter based adaptive activation function, the neural network solution is plotted in a frequency domain, revealing the capturing of the frequencies in the solution faster than fixed activation function. The proposed approach can be used both in PINNs as well as in standard neural networks and is a promising and simple approach to increase the efficiency, robustness and accuracy of the neural network based approximation of nonlinear functions as well as the solution of partial differential equations, especially for forward problems.

\section*{Acknowledgement}
This work was supported by the Department of Energy PhILMs grant DE-SC0019453, and by the DAPRA-AIRA grant HR00111990025.


\begin{thebibliography}{}

\bibitem{ARP}
D. Arpit, et al., A closer look at memorization in deep networks, arXiv preprint arXiv:1706.05394, 2017.

\bibitem{ARB}
A.R. Barron, Universal approximation bounds for superpositions of a sigmoidal function. IEEE Trans. Inform. Theory, 39(3), 930-945, 1993.

\bibitem{Bes}
C. Basdevant, et al., Spectral and finite difference solution of the Burgers equation, Comput. Fluids, 14 (1986) 23-41.


\bibitem{Bat}
H. Bateman, Some recent researches on the motion of fluids, Monthly Weather Review, 43(4), 163-170, 1915.

\bibitem{AD}
A.G. Baydin, B.A. Pearlmutter, A.A. Radul, J.M. Siskind, Automatic differentiation in machine learning: a survey, Journal of Machine Learning Research, 18 (2018) 1-43.

\bibitem{Ben}
Y. Bengio, Simard P. and Frasconi P., Learning long-term dependencies with gradient descent is difficult, IEEE Transactions on Neural Networks, 5(2), 157-166, 1994.

\bibitem{Berg}
J. Berg, K. Nystr\"{o}m , Data-driven discovery of PDEs in complex datasets, J. Comput. Phys. 384 (2019) 239-252.

\bibitem{Bur}
J.M. Burgers,A mathematical model illustrating the theory of turbulence. In advances in applied mechanics, Vol. 1, pp. 171-199, 1948.

\bibitem{CAU}
P.J. Caudrey, I.C. Eilbeck, J.D. Gibbon, The sine-Gordon equation as a model classical field theory,  Nuovo Cimento 25 (1975) 497-511.

\bibitem{GC}
G. Cybenko, Approximations by superpositions of sigmoidal functions, Approximation theory and its applications, 9(3), 17-28, 1989.

\bibitem{DOD}
R.K. Dodd, I.C. Eilbeck, J.D. Gibbon, H.C. Morris, Solitons and Nonlinear Wave Equations, Academic, London, 1982.
\bibitem{DZS}
K. Duraisamy, Z.J. Zhang, A.P. Singh, New approaches in turbulence and transition modeling using data-driven techniques, AIAA paper 2015-1284, 2015.

\bibitem{MiRa}
M. Dushkoff, R. Ptucha, Adaptive Activation Functions for Deep Networks, Electronic Imaging, Computational Imaging XIV, pp. 1-5(5).

\bibitem{Hay}
S.S. Haykin, Neural Networks: A comprehensive foundation, Prentice Hall, 1999.


\bibitem{ADAM}
D. P. Kingma, J. L. Ba, ADAM: A method for stochastic optimization,  arXiv:1412.6980v9, 2017.

\bibitem{LLF}
I.E. Lagaris, A. Likas, D.I. Fotiadis, Artifical neural network for solving ordinary and partial differential equations, IEEE Trans. Neural Networks, 9(5) 987-1000, 1998.

\bibitem{LLR}
B. Li, Y. Li and X. Rong, The extreme learning machine learning algorithm with tunable activation function, Neural Comput \& Applie (2013) 22: 531-539.

\bibitem{Min}
M. Minsky, S. Papert, Perceptrons, MIT Press, 1969.



\bibitem{OW}
H. Owhadi, Bayesian numerical homogenization, Multiscale Model. Simul. 13, 812-828, 2015.

\bibitem{PD}
E. J. Parish, K. Duraisamy, A paradigm for data-driven predictive modeling using field inversion and machine learning, J. Comput. Phys. 305, 758-774, 2016.

\bibitem{Qian}
S. Qian, et al, Adaptive activation functions in convolutional neural networks, Neurocomputing
Volume 272, 10 January 2018, Pages 204-212.

\bibitem{RAH}
N. Rahaman, et al., On the spectral bias of deep neural networks, arXiv preprint arXiv:1806.08734, 2018.


\bibitem{RK}
M. Raissi, G.E. Karniadakis, Hidden physics models: machine learning of nonlinear partial differential equations. J. Comput. Phys., 357, 125-141, 2018.

\bibitem{RK0}
M. Raissi, P. Perdikaris, G.E. Karniadakis, Numerical Gaussian processes for time-dependent and nonlinear partial differential equations. SIAM J. Sci. Comput. 40, A172-A198, 2018.

\bibitem{RK1}
M. Raissi, P. Perdikaris, G.E. Karniadakis, Physics-informed neural network: A deep learning framework for solving forward and inverse problems involving nonlinear partial differential equations. J. Comput. Phys., 378, 686-707, 2019. 

\bibitem{RK2}
M. Raissi, P. Perdikaris, G.E. Karniadakis, Inferring solutions of differential equations using noisy multi-fidelity data, J. Comput. Phys. 335 (2017) 736-746.





\bibitem{RK3}
M. Raissi, P. Perdikaris, G.E. Karniadakis, Machine learning of linear differential equations using Gaussian processes, J. Comput. Phys., 348, 683-693, 2017.

\bibitem{VIV}
M. Raissi, Z. Wang, M.S. Triantafyllou, G.E. Karniadakis, Deep learning of vortex-induced vibrations, J. Fluid Mech. (2019), vol. 861, pp. 119-137.



\bibitem{Rud}
S. Ruder, An overview of gradient descent optimization algorithms, arXiv:1609.04747v2, 2017.

\bibitem{Rud}
S.H. Rudy, et al., Data-driven discovery of partial differential equations, Sci. Adv. 3(4), 2017.

\bibitem{Som}
A. Sommerfeld, Partial Differential Equations in Physics, New York: Academic Press, 1949.

\bibitem{SWCC}
Y. Shen, B. Wang, F. Chen and L. Cheng, A new multi-output neural model with tunable activation function and its applications, Neural Processing Letters, 20: 85-104, 2004.

\bibitem{Wa}
J.-X Wang, et al.,  A comprehensive physics-informed machine learning framework for predictive turbulence modeling, arXiv:1701.07102.

\bibitem{Waz}
A. Wazwaz, New traveling wave solutions to the Boussinesq and the Klein-Gordon equations, Communications in Nonlinear Science and Numerical Simulation 13 (2008), 889-901.

\bibitem{WH}
G.B. Whitham, Linear and nonlinear waves, Vol. 42, John-Wiley \& Sons, 2011.

\bibitem{FPrin}
Z. Xu, Y. Zhang, T. Luo, Y. Xiao, Z. Ma, Frequency Principle: Fourier analysis sheds light on deep neural networks, arXiv:1901.06523v3, 2019.

\bibitem{DY}
Yarotsky D., Error bounds for approximations with deep ReLU networks, Neural Networks, 94, 103-114, 2017.

\bibitem{Yu}
C. Yu, et al.,An adaptive activation function for multilayer feedforward neural networks,  2002 IEEE Region 10 Conference on Computers, Communications, Control and Power Engineering. TENCOM '02. Proceedings.

\bibitem{ZD}
Z.J. Zhang, K. Duraisamy, Machine learning methods for data-driven turbulence modeling, AIAA paper 2015- 2460, 2015.



\end{thebibliography}
\end{document}